%% file: 00-BennettsLiang-ANZIAMJ-arXiv.tex
\documentclass{article}

\usepackage{arXiv}

\usepackage[utf8]{inputenc} 
\usepackage[T1]{fontenc}    
\usepackage{hyperref}       
\usepackage{url}            
\usepackage{booktabs}       
\usepackage{amsfonts}       
\usepackage{nicefrac}       
\usepackage{microtype}      
\usepackage{lipsum}		
\usepackage{graphicx}
\usepackage{doi}
\usepackage{bm}

\usepackage[numbers,sort&compress]{natbib}
\title{Modelling dynamic strains on ice shelves resulting from flexural and extensional motions forced by ocean wave packets}

\author{Luke G.~Bennetts \\
	University of Melbourne, Australia \\
	\texttt{luke.bennetts@unimelb.edu.au} 
	\And
	Jie~Liang \\
	University of Adelaide, Australia 
}

\date{}

\renewcommand{\shorttitle}{Ice shelf strains forced by ocean wave packets}

\hypersetup{
pdftitle={Ice shelf strains forced by ocean wave packets},
pdfsubject={q-bio.NC, q-bio.QM},
pdfauthor={Luke G.~Bennetts},
pdfkeywords={First keyword, Second keyword, More},
}

\input{arxiv-luke.tex}

\begin{document}
\maketitle

 \begin{abstract}
 	The transient response of an ice shelf to an incident wave packet from the open ocean is studied with a model that allows for extensional waves in the ice shelf, in addition to the standard flexural waves.
	Results are given for strains imposed on the ice shelf by the incident packet, over a range of peak periods in the swell regime and a range of packet widths.
	In spite of the large difference in speeds of the extensional and flexural waves, it is shown that there is generally an interval of time during which they interact, and the coherent phases of the interactions generate the greatest ice shelf strain magnitudes.
	The findings indicate that incorporating extensional waves into models is potentially important for predicting the response of Antarctic ice shelves to swell, in support of previous findings based on frequency-domain analysis.  
 \end{abstract}


%
\section{Introduction}

Antarctic ice shelves are the floating extensions of the Antarctic Ice Sheet onto the Southern Ocean. 
They play an important role in the dynamics of the Antarctic Ice Sheet \cite{noble2020sensitivity} and the Southern Ocean \cite{bennetts2024closing}, and are related to deep uncertainties in projections of the future climate \cite{portner2019ocean}.
They are typically hundreds of metres thick and cover areas up to hundreds of thousands of square kilometres, which enclose cavities of seawater \cite{fretwell2013bedmap2}.
They have a natural cycle of slowly gaining mass due to the inflow of tributary glaciers, and sudden{ly losing}  mass due to calving at {their} shelf front{s} ({their} seaward termin{i}), resulting from the propagation of fractures forced by internal forces (glacial flow) and the external environment \cite{benn2007calving}.
Climate change has potentially disrupted the mass balance between inflow and calving, as ice shelves are becoming thinner and weaker, such that the current rate of calving is unlikely to be compensated by glacial inflow \cite{greene2022antarctic}.

There are numerous observations of ocean waves propagating through Antarctic ice shelves \cite{bennetts2025wavesinice}, initially serendipitously \cite{thiel1960gravimetric} and subsequently with dedicated campaigns of increasing scale \cite{squire1994observations,cathles2009seismic,chen2018ocean}.
Waves in ice shelves have been measured over a broad spectrum of wave periods, including swell (10--30\,s), infragravity waves (50--300\,s) and very long period waves, such as tsunamis ($>300$\,s) \cite{chen2018ocean}.
Swell tends  to be detected on ice shelves during austral summer only \cite{baker2019seasonal}, as swell attenuates as it travels through the vast extent of sea ice surrounding Antarctica during winter \cite{golden2020modeling}.
In contrast, longer waves are detected year-round on ice shelves \cite{baker2019seasonal}.

Dynamic ice shelf flexure associated with ocean waves has been proposed as driver of ice shelf calving \cite{holdsworth1978iceberg}. 
The concept is supported by two observations of calving events on the Sulzberger Ice Shelf over a decade apart that both coincided with the arrival of tsunamis \cite{brunt2011antarctic,zhao2024long}.
It is also supported by the timing of multiple large-scale calving events on other Antarctic ice shelves, which followed prolonged reductions in regional sea ice that allowed energetic swell to reach the shelf fronts \cite{massom2018antarctic,teder2025large}.
It is likely that swell will become an increasingly important driver of ice shelf calving now that Antarctic sea ice appears to have entered a phase of retreat \cite{hobbs2024observational},
particularly for ice shelves that are thinning and weakening, for which swell is expected to be the dominant driver of calving  \cite{bassis2024stability}.

Mathematical models have been developed to predict ice shelf flexure in response to incident ocean waves, where the flexure is usually quantified in terms of a flexural strain (or stress).
It is standard to model the ice shelf as a Kirchhoff ``thin plate'', on the assumption that the shelf supports wavelengths much greater than its thickness.
The standard model involves incident waves from the open ocean interacting with an ice shelf of uniform thickness and semi-infinite length, is two dimensional, i.e., invariant in one (horizontal) spatial direction, and is studied in the frequency domain.
Early versions of the model imposed zero draught of the ice shelf to facilitate the solution method, and, hence, free-edge conditions were applied at the shelf front \cite{fox1991coupling,balmforth1999ocean}.
Thus, the ocean and ice shelf are coupled only along the lower surface of the ice shelf, which forces flexural waves in the ice shelf, under the thin-plate assumption combined with the standard assumptions of linear water waves.
The flexural waves are linked to wave motion in the underlying water cavity, and the coupled hydroelastic waves are  known as flexural--gravity waves owing to the two restoring forces involved \cite{bennetts2025wavesinice}.

Solution methods have been developed to accommodate an Archimedean ice shelf draught \cite{kalyanaraman2019shallow,papathanasiou2019resonant}, along with varying ice shelf thickness and seabed profiles \cite{ilyas2018time,meylan2021swell,bennetts2021complex}. 
However, the free-edge conditions have been incorrectly retained, such that water--ice coupling at the shelf front is overlooked and the draught only introduces the reduction in water depth between the open ocean and sub-shelf water cavity.
Numerical models have also been developed, in which, in contrast to the use of a thin-plate model for the ice shelf, the variation through the ice depth is not imposed  \cite{sergienko2010elastic,konovalov2019ice,kalyanaraman2020coupled,abrahams2023ocean}.
\citeauthor{abrahams2023ocean}~\cite{abrahams2023ocean} produced numerical time-domain simulations, for a model in which the  2D linear elasticity equations are used {for the} ice shelf and water--ice coupling occurs at the shelf front, to show that incident ocean waves generate extensional (Lamb) waves in the shelf, in addition to flexural waves, consistent with the most recent observations \cite{chen2018ocean}.

\citeauthor{bennetts2024thin}~\cite{bennetts2024thin} used a variational formulation of a model involving 2D linear elasticity for the ice shelf to derive a thin-plate model of the floating ice shelf, in which the ice shelf supports both flexural and extensional waves.
The flexural waves are forced by water--ice coupling at the shelf front and at the lower ice shelf surface. 
Hence, they are coupled directly to the sub-shelf water cavity and propagate as flexural--gravity waves.
The extensional waves are forced by water--ice coupling at the shelf front only and propagate along the shelf{,} without direct coupling to the sub-shelf water cavity, as noted by \citeauthor{abrahams2023ocean}~\cite{abrahams2023ocean}. 
\citeauthor{bennetts2024thin}~\cite{bennetts2024thin} used model outputs in the frequency domain to deduce that extensional waves make a major contribution to ice-shelf strains in response to incident swell,  
but that their contribution to strain is negligible in response to incident waves with longer periods.
The implication is that models that do not incorporate extensional waves under estimate the impact of swell on ice shelves.
However, frequency-domain results overlook the separation of the flexural and extensional waves in the ice shelf that would occur in response to transient forcing, due to the extensional waves propagating far faster than the flexural waves.
The separation is evident in \citeauthor{abrahams2023ocean}~\cite{abrahams2023ocean}'s time-domain simulations, and cognate phenomena occur in other water wave problems involving multiple wave modes, e.g., \cite{smith2020wiener}.

In this article, we analyse time-domain simulations for \citeauthor{bennetts2024thin}~\cite{bennetts2024thin}'s thin-plate model, in which the ice shelf is forced by an incident wave packet from the open ocean.
We produce time-domain solutions as superpositions of frequency-domain solutions, thus making the computations efficient enough to conduct an analysis over a parameter space of incident wave packet peak periods and widths, as well as different ice shelf thickness values.
We focus on the swell regime, interactions between extensional and flexural waves, and their relative contributions to strains experienced by the ice shelf.

\section{Mathematical model}\label{sec:math_model}

\begin{figure} 
\centering
\setlength{\figurewidth}{0.8\textwidth}
\setlength{\figureheight}{0.8\textwidth}
\input{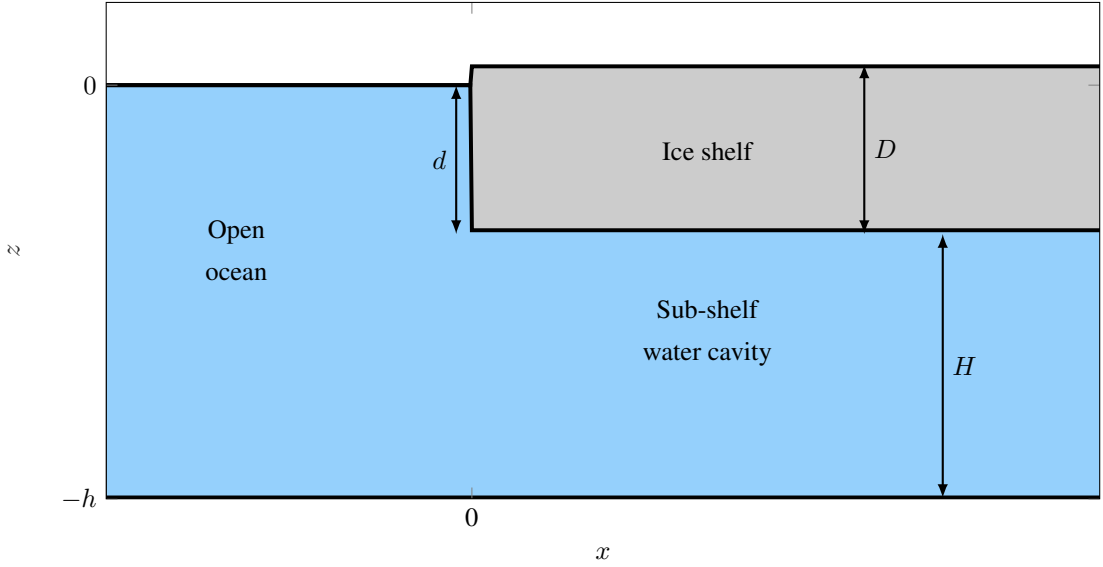}
\caption{Schematic (not to scale) of the equilibrium geometry.}
    \label{fig:diagram}
\end{figure}

Consider a two-dimensional water domain of finite depth and infinite horizontal extent.
Locations in the water are defined by the Cartesian coordinate $(x,z)$, where $x$ and $z$ denote the horizontal and vertical locations, respectively.
The water domain is bounded below by a flat impermeable seabed at $z=-h$.
In the absence of an ice cover, it is bounded above by a free surface, which is at $z=0$ when in equilibrium (Fig.~\ref{fig:diagram}).

A floating ice shelf of uniform thickness, $D$, occupies the water surface for $x>0$.
It has an Archimedean draught, such that its lower surface is at $z=-d$ in equilibrium, where  
\begin{equation}
    d = \frac{\rho_{\text{i}}\,D}{\rho_{\text{w}}},
\end{equation}
in which $\rho_{\text{i}}=922.5$\,kg\,m$^{-3}$ and $\rho_{\text{w}}=1025$\,kg\,m$^{-3}$ are the ice and water densities, respectively {\citep[where the ice shelf density is, e.g., approximately at the top end of the values given for the Amery Ice Shelf by][]{fricker2001distribution}}, so that $\rho_{\text{i}}\,/\,\rho_{\text{w}}=0.9$.
Therefore, the ice shelf encloses a water cavity of equilibrium depth $H\equiv{}h-d$ (Fig.~\ref{fig:diagram}).
The water domain for $x<0$ is referred to as the open ocean.

Water motion is modelled using linear potential-flow theory, assuming that the water is inviscid and incompressible, the motion is irrotational, and amplitudes are much smaller than {the} wavelengths involved.
The water displacement field {(rather the the velocity field)} is defined as the gradient of a scalar potential, $\Phi(x,z,t)$, where $t$ denotes time.
The displacement potential satisfies Laplace's equation, 
\begin{equation}
     \bm{\nabla}^2\,\Phi = 0,
\end{equation}
throughout the linearised water domain (open water and sub-shelf water cavity), plus a no-normal flow condition on the seabed, such that 
\begin{equation}
    \Phi_{z} = 0
    \mathfor x\in\mathbb{R}
    \mathand z=-h.
\end{equation}

The ice shelf is modelled as an elastic solid, with horizontal and vertical displacements, $U(x,z,t)$ and $W(x,z,t)$, respectively.
The displacements satisfy the equations of linear elasticity,
\begin{align}\label{eqs:full_field}
\rho_{\tn{i}}\,U_{tt} =\sigma_{11,x} + \sigma_{12,z}
\mathand
\rho_{\tn{i}}\,W_{tt} = \sigma_{12,x} + \sigma_{22,z}-\rho_{\tn{i}}\,g
\end{align}
for $x>0$ and $-d<z<D-d$.
In \eqref{eqs:full_field}, $g=9.81$\,m\,s$^{-2}$ is gravitational acceleration, and $\sigma_{ij}$ ($i,j=1,2$) are the components of the Cauchy stress tensor.
The ice shelf motion is coupled to the water motion via kinematic and dynamic conditions along its lower surface ($x>0$ and $z=-d$) and along the submerged portion of the shelf front ($x=0$ and $-d<z<0$) \cite{bennetts2024thin}.
Standard conditions are satisfied along its boundaries in contact with the atmosphere \cite{bennetts2024thin}.

Motions are excited by a rightwards propagating incident wave packet from the open ocean.
Following \cite{bennetts2021complex},
the incident packet is defined via the free surface elevation in the open ocean, $\eta(x,t) = \Phi_{z}(x,0{,t})$ {($x<0$)}, such that the incident packet creates the {surface} elevation
\begin{equation}{\label{Eq:packet}}
\eta(x,t)=\eta_{\text{inc}}(x,t)\equiv\real\left\{\int_0^{\infty} {A}(k)\,\varphi_{\text{inc}}(x)\,{\rm e}^{-\ci\,\omega\,t} 
\,\text{d} k\right\},  
\end{equation}
where $\varphi_{\text{inc}}={\rm e}^{\ci\,k\,x}$ and ${A}$ is the Gaussian
\begin{equation}\label{Amplitude}
  {A}(k) = A_0\,{\rm e}^{-(k-k_{\text{c}})^2/(2\,\sigma^2)-\ci\,k\,x_{\text{c}}}{,}  
\end{equation}
{which is chosen for convenience rather than based on observations.}
Here, $A_{0}$ is the packet amplitude ({in} wavenumber space), $k_{\text{c}}$ is its central wavenumber, $\sigma$ is its width (i.e., standard deviation), and $x_{\text{c}}$ is the   physical location of the packet centre at $t=0$.
For the results presented in \S\,\ref{sec:results}, the incident packet is separated from the ice shelf at $t=0$ by setting $x_{\text{c}}=-3\,/\,\sigma$. 
{Note that Eq.~\eqref{Eq:packet} sets the wavenumber $k$ as the spectral parameter for the problem, such that the} angular frequency, $\omega$, is treated as a function of  $k$, via the standard open water dispersion relation {\eqref{eq:dispersionrelation_openwater}.} 

In the absence of an ice shelf, the incident wave packet energy across the spatial domain, $\mathcal{E}(t)$, does not change as the packet propagates, and, hence,  $\mathcal{E}(t)=\mathcal{E}(0)\equiv\mathcal{E}_{0}$.
Noting that the energy at $t=0$ {is} proportional to $|\eta(x,0)|^2$ \cite{jeschke2017water} and that  ${A}(k)$ is the Fourier transform of the initial wave profile, $\eta(x,0)$, Parseval's theorem implies
\begin{equation}\label{Eq:energy}
  \mathcal{E}_{0}\propto \int_{-\infty}^{\infty} |\eta(x,0)|^2\,\text{d}x =  \int_{0}^{\infty} |{A}(k)|^2\,\text{d}k.
\end{equation}
Substituting \eqref{Amplitude} into \eqref{Eq:energy} 
\begin{subequations}\label{Eq:energy_2}
\begin{align}
  \mathcal{E}_{0} &\propto \int_{0}^{\infty} |A_0\,{\rm e}^{-(k-k_c)^2/(2\,\sigma^2)}|^2\,\text{d}k\\
  &= A_0^2\,\int_{0}^{\infty} \,{\rm e}^{-(k-k_c)^2/\sigma^2}\,\text{d}k\\
  &\approx A_0^2\,\sigma\sqrt{\pi},
  \end{align}
\end{subequations}
where the final result follows from the assumption that $k_{c}\gg \sigma$.
Thus, different wave packets have the same energy, $\mathcal{E}_{0}$, if they share the same value of $A_{0}^{2}\,\sigma$.

\section{Approximate solution}\label{sec:depthaverage}

Following \cite{bennetts2024thin}, a thin-plate (depth-averaged) approximation is applied to the ice shelf motions, based on the 
plane stress and plane strain assumptions. 
The thin-plate approximation uses the anzatzes
\begin{subequations}
  \begin{align}
      U(x,z,t) &= \mathcal{U}(x,t) - (z+d-D/2) \,\mathcal{W}_{x}(x,t),\\
      \mathandR W(x,z,t) &=  \mathcal{W}(x,t),
  \end{align}  
\end{subequations}
where $\mathcal{U}$ and $\mathcal{W}$ represent the extensional and flexural motions, respectively. 

The thin-plate approximation is combined with a single-mode (depth-averaged) approximation  for the water motion \cite{bennetts2007multi}, which assumes the vertical profile through the water column is that of the propagating wave mode.
Thus, separate anzatzes are applied to the potential in the open ocean and sub-shelf water cavity regions, such that
\begin{subequations}\label{eqs:SMA_3}
   \begin{align}
       \Phi(x,z,t) = \frac{g}{\omega^2}\,\varphi(x)\frac{\cosh \{k\,(z+h)\}}{\cosh{k\,h}}\,{\rm e}^{-\ci\,\omega t} & \quad\text{for}\quad x<0,
   \\[4pt]
       \mathandR 
       \Phi(x,z,t) = \frac{g}{\omega^2}\,\psi(x) \frac{\cosh \{\kappa\,(z+h)\}}{\cosh{\kappa\,H}}\,{\rm e}^{-\ci\,\omega t} & \quad\text{for}\quad x>0.
   \end{align}
 \end{subequations}
The wavenumbers $k$ and $\kappa$ satisfy the dispersion relations 
\begin{subequations}\label{eqs:dispersion_relations}
\begin{align}\label{eq:dispersionrelation_openwater}
    k\,\tanh(k\,h) & = \frac{\omega^{2}}{g}
    \\
    \mathandR
    \{F\,\kappa^{4} - J\,\omega^{2}\,\kappa^{2} + 1 -m\,\omega^{2}\}\,\kappa\,\tanh(\kappa\,H) 
    & = \frac{\omega^{2}}{g},
\end{align}
\end{subequations}
where $F$, $J$ and $m$ are scaled versions of the ice shelf flexural rigidity, rotational inertia and mass, respectively.
They are defined as
\begin{equation}
    F = \frac{E\,D^{3}}{12\,(1-\nu^{2})\,\rho_{\text{w}}\,g},
    \quad
    J = \frac{\rho_{\text{i}}\,D^{3}}{12\,\rho_{w}\,g} 
    \mathand
    m = \frac{\rho_{\text{i}}\,D}{\rho_{w}\,g}{,}
\end{equation}
{where $E=10$\,GPa and $\nu=0.3$ are the Young's modulus and Poisson's ratio of the ice shelf.}
The single-mode approximation (\ref{eqs:SMA_3}) maps the problem to the frequency domain through the assumption that the water motions are at a prescribed angular frequency, $\omega$. 
A compatible mapping is applied to the ice shelf motions, such that
\begin{equation}\label{eq:iceshelf_freqmap}
    \mathcal{U}(x,t)=u(x)\,{\rm e}^{-\ci\,\omega t} \mathand \mathcal{W}(x,t)=w(x)\,{\rm e}^{-\ci\,\omega t}.
\end{equation}
The unknowns in the frequency domain are complex-valued functions, i.e., $\varphi$, $\psi$, $u$, $w\in\mathbb{C}$. 

The governing equations of the thin-plate/single-mode approximation are
\begin{subequations}\label{eqs:odes}
\begin{align}
    \varphi''+k^{2}\,\varphi & = 0
    \mathfor x<0,
    \\
    a\,\psi'' + b\,\psi + \frac{\omega^{2}}{g}\,w & = 0
    \mathfor x>0,
    \\
    F\,w'''' + J\,\omega^{2}\,w'' + (1-m\,\omega^{2})\,w - \psi & = 0
    \mathfor x>0,
    \\
    \mathandR
    G\,u'' + m\,\omega^{2}\,u & = 0
    \mathfor x>0,
\end{align}
\end{subequations}
where
\begin{equation}
    a = \int_{-h}^{-d}\frac{\cosh{^{2}} \{\kappa\,(z+h)\}}{\cosh{^{2}}(\kappa\,H)} \,\text{d}z
    ,\quad
    b = \kappa^{2}\,a -\kappa\,\tanh(\kappa\,H)
\end{equation}
and $G=E\,D\,/\,\{\rho_{\text{w}}\,g\,(1-\nu^{2})\}$.
The solutions are subject to the jump conditions
\begin{align}\label{eq:jump}
    a_{0}\,\varphi = a\,\psi
    \mathand
    a_{1}\,\varphi' = a_{0}\,\psi' + \frac{\omega^{2}}{g}\,\{v_{0}\,u - v_{1}\,w'\}
    \mathfor x=0,
\end{align}
and the shelf front conditions
    \begin{align}\label{eq:shelffront}
        F\,w'' + v_{1}\,\varphi = 0
        ,\quad
        F\,w''' + J\,\omega^{2}\,w' = 0
        \mathand
         G\,u' + v_{0}\,\varphi = 0
    \end{align}
{for $x=0$,} where
\begin{subequations}
\begin{align}
    a_{0} & = \int_{-h}^{0}\frac{\cosh{^{2}} \{k\,(z+h)\}}{\cosh{^{2}}(k\,h)} \,\text{d}z
    ,
\\
    a_{1} & = \int_{-h}^{-d}\,\frac{\cosh \{k\,(z+h)\}\,\cosh\{\kappa(z+h)\}}{\cosh(k\,h)\,\cosh(\kappa H)} \,\text{d}z,
\\
    v_{0} & = \int_{-d}^{0}\,\frac{\cosh \{k\,(z+h)\}}{\cosh(k\,h)} \,\text{d}z,
\\
    \mathandR 
    v_{1} & = \int_{-d}^{0}\,\frac{\cosh \{k\,(z+h)\}}{\cosh(k\,h)} \left(z+d-D/2\right)\,\text{d}z.
\end{align}  
\end{subequations}

The second-order ODE in the open water region (\ref{eqs:odes}a) can be solved up to unknown constants.
It is expressed as
\begin{equation}
    \varphi(x) = \varphi_{\text{inc}}(x) + R\,\e^{-\ci\,k\,x},
\end{equation}
where the second term on the right-hand side is the wave reflected by the ice shelf, which has amplitude $R\in\mathbb{C}$.
The system of ODEs in the ice shelf/water cavity region (\ref{eqs:odes}b--{d}) can also be solved to give
\begin{align}
   \psi(x) & = T_{\text{cav}} {\rm e}^{\ci \,\kappa\,x} +
            \sum_{j={1,2}} T_{\text{cav}}^{(-j)}\,{\rm e}^{\ci \,\kappa_{-j}\,x},\\
    w(x) &= T_{\text{flex}} {\rm e}^{\ci \,\kappa\,x} +
            \sum_{j={1,2}} T_{\text{flex}}^{(-j)}\,{\rm e}^{\ci \,\kappa_{-j}\,x},\\
    \mathandR u(x) &= T_{\text{ext}}\,{\rm e}^{\ci \,q\,x}. 
\end{align}
The wavenumbers $\kappa_{-j}\in\mathbb{R}+\ci\,\mathbb{R}_{+}$ ($j=1,2$) are the roots of the quartic equation {(Appendix~\ref{app:details})}
\begin{equation}\label{eq:kappaj}
    a\,(F\,\kappa_{-j}^{4}-J\,\omega^{2}\,\kappa_{-j}^{2}+1-m\,\omega^{2})
    +
    \{F\,(\kappa^{2}+\kappa_{-j}^{2})-J\,\omega^{2}\}\,\kappa\,\tanh(\kappa\,H)=0,
\end{equation}
which approximate complex-valued roots of (\ref{eqs:dispersion_relations}b), and are typically such that $\kappa_{-2}=-\overline{\kappa_{-1}}$.
The extensional wavenumber, $q\in\mathbb{R}_{+}$, is
\begin{equation}
    q = \omega\,\sqrt{\frac{\rho_{\text{i}}\,(1-\nu^2)}{E}}.
\end{equation}
The amplitude of the transmitted propagating wave in the cavity is related to the flexural wave in the shelf by {(Appendix~\ref{app:details})}
\begin{equation}\label{eq:tcav}
    T_{\text{cav}}=\frac{\omega^{2}}{{g\,}\kappa\,\tanh(\kappa\,H)}\,T_{\text{flex}}, 
\end{equation}
and similarly
\begin{equation}\label{eq:tcavj}
    T_{\text{cav}}^{(-j)}=
    \frac{{-\{}F\,(\kappa^{2}+\kappa_{-j}^{2})-J\,\omega^{2}{\}\,\kappa\,\tanh(\kappa\,H)}}{a}\,T_{\text{flex}}^{(-j)}
    \mathfor j=1,2.
\end{equation}
The jump conditions (\ref{eq:jump}) and shelf front conditions (\ref{eq:shelffront}) are applied to calculate the five unknown amplitudes, $R$, $T_{\text{flex}}$, $T_{\text{flex}}^{(-j)}$ ($j=1,2$) and $T_{\text{ext}}$, thus completing the frequency-domain solution. 

The scaling used in the single-mode approximation (\ref{eqs:SMA_3}) is such that the incident free surface displacement is identical to the incident potential, $\varphi_{\text{inc}}$.
Therefore, the time-domain solution is obtained as a superposition of frequency-domain solutions, such that, e.g.,
\begin{align}
    \eta(x,t)=\real\left\{\int_0^{\infty} {A}(k)\,\varphi(x)\,\e^{-\ci\,\omega\,t} 
    \,\text{d} k\right\}{\mathfor x<0.} 
\end{align}
{The} only non-zero component of the strain tensor is  \cite{bennetts2024thin}
\begin{align}\label{eq:strain}
     \epsilon(x,z,t)&=\real\left\{\int_0^{\infty} {A}(k)\,\varepsilon(x,z)\,\e^{-\ci\,\omega\,t} 
    \,\text{d} k\right\}
    \\
    \mathwhereR
    \varepsilon(x,z) &=
    u'-(z+d-D\,/\,2)\,w''
\end{align}
{for $x>0$ and $-d<z<D-d$.}
{I}t can be decomposed as $\epsilon = \epsilon_{\text{flex}}+\epsilon_{\text{ext}}$, where $\epsilon_{\text{flex}}$ and $ \epsilon_{\text{ext}}$ are the strains produced by the flexural and extensional waves, respectively.
They are
\begin{subequations}
\begin{align}
     \epsilon_{\text{flex}}(x,z,t)&=\real\left\{\int_0^{\infty} {A}(k)\,\varepsilon_{\text{flex}}(x,z)\,\e^{-\ci\,\omega\,t} 
    \,\text{d} k\right\},
    \\
    \mathandR
    \epsilon_{\text{ext}}(x,t)&=\real\left\{\int_0^{\infty} {A}(k)\, 
    \varepsilon_{\text{ext}}(x)\,\e^{-\ci\,\omega\,t} 
    \,\text{d} k\right\}
    ,
\end{align}
\end{subequations}
in which
\begin{equation}\label{eq:strain_flex_ext}
    \varepsilon_{\text{flex}}(x,z)=-(z+d-D\,/\,2)\,w''
    \mathand
    \varepsilon_{\text{ext}}(x) =u'.
\end{equation}

\section{Results}\label{sec:results}

\subsection{Frequency domain analysis}

\begin{figure}[h!] 
    \centering
    \setlength{\figurewidth}{0.7\textwidth}
   \setlength{\figureheight}{0.7\textwidth}
    \input{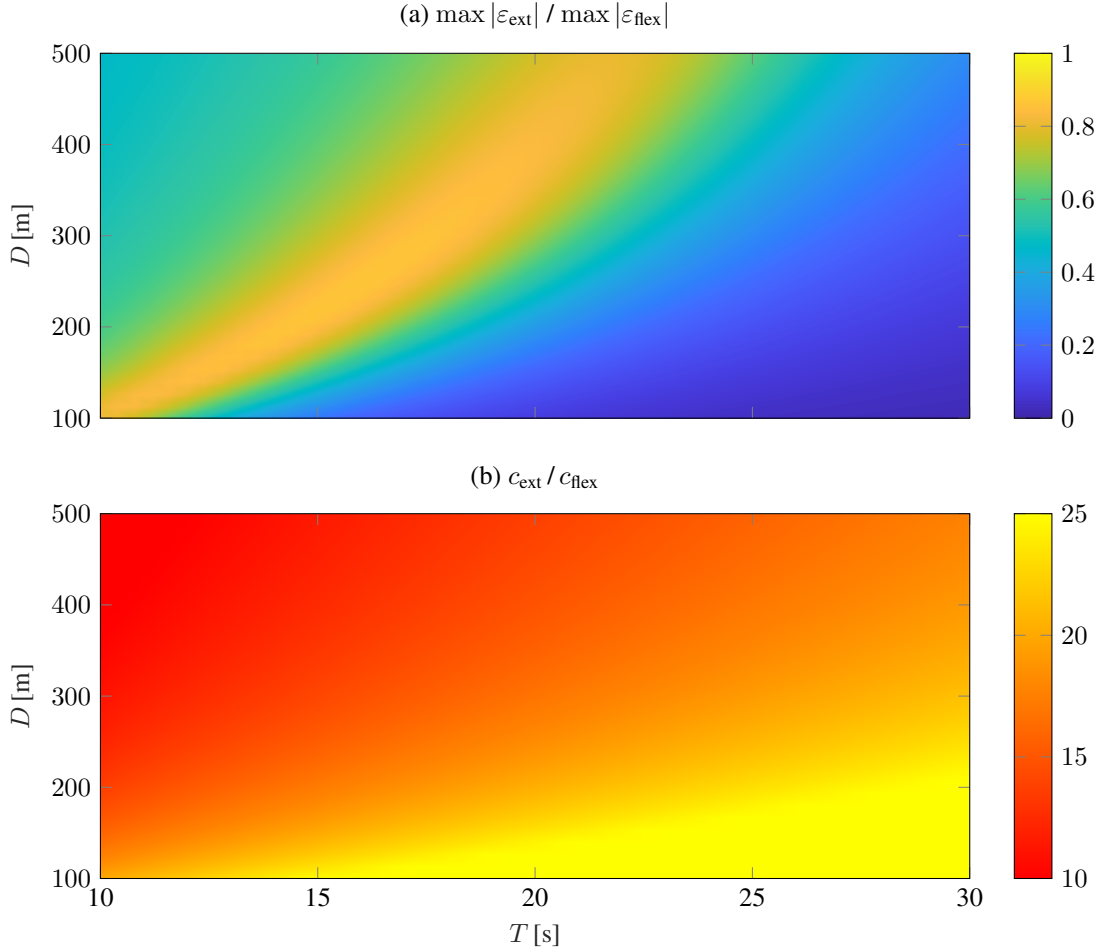}
\caption{Heatmaps showing ratios of (a)~maximum strain due to extensional waves to maximum strain due to flexural waves in the frequency domain, {$\max\vert\varepsilon_{\text{ext}}\vert\,/\,\max\vert\varepsilon_{\text{flex}}\vert$,} and (b)~phase speed of the extensional wave to that of the flexural wave{, $c_{\text{ext}}\,/\,c_{\text{flex}}$}, versus wave period{, $T$,} and ice shelf thickness{, $D$}.}
\label{fig:Heat_map}
\end{figure}

Fig.~\ref{fig:Heat_map}(a) shows the ratio of the maximum strain caused by extensional waves
to the maximum strain caused by flexural waves in the frequency domain,{~i.e., $\max\vert\varepsilon_{\text{ext}}\vert\,/\,\max\vert\varepsilon_{\text{flex}}\vert$ (\ref{eq:strain_flex_ext}),} as a function of ice shelf thickness, $D\in[100\text{\,m},500\text{\,m}]$, and wave period, $T\in[10\text{\,s},30\text{\,s}]$ where $T=2\,\pi\,/\,\omega$, i.e., the swell regime.
The water cavity depth is fixed at $H=800$\,m.
The maximum values are calculated over the ice shelf domain, {$x>0$ and $-d<z<D-d$,} noting that the extensional motions depend only on horizontal location, {$x$,} and the magnitudes of the flexural motions are symmetric about the ice shelf mid-plane{, $z=D\,/\,2-d$,} and attain maximum values at the upper/lower surfaces{, $z=-d$ and $D-d$}.
There is a (yellow) band of width 2--4\,s where the ratio is greatest (${\max\vert\varepsilon_{\text{ext}}\vert\,/\,\max\vert\varepsilon_{\text{flex}}\vert}\approx{}1$), 
i.e., the maximum strains due to extensional and flexural waves are almost identical.
The band moves from being centred at $T\approx{}10$\,s for $D=100$\,m to $T\approx{}20$\,s for $D=500$\,m.
On the shorter wave period side of the band, the ratio decreases slightly to no less than ${\max\vert\varepsilon_{\text{ext}}\vert\,/\,\max\vert\varepsilon_{\text{flex}}\vert=}0.5$, indicating that  strain{s} due to extensional waves remain comparable to those due to flexural waves. 
On the longer wave period side of the band, the ratio decreases sharply to ${\max\vert\varepsilon_{\text{ext}}\vert\,/\,\max\vert\varepsilon_{\text{flex}}\vert}\approx{}0$ (on the linear scale shown), indicating strains due to flexural waves are dominant. 

Fig.~\ref{fig:Heat_map}(b) shows the corresponding ratio of the phase speeds of the extensional and flexural waves, respectively
\begin{equation}
    c_{\text{ext}} = \frac{\omega}{q}
    \mathand
    c_{\text{flex}} = \frac{\omega}{\kappa}.    
\end{equation}
The ratio is greater than ten {($c_{\text{ext}}\,/\,c_{\text{flex}}>10$)} over the considered wave period and ice thickness region, 
which indicates the extensional waves are at least an order of magnitude faster than the flexural waves.
The ratio is ${c_{\text{ext}}\,/\,c_{\text{flex}}}\approx{}10$--15 in the wave period--ice thickness region where the strains due to extensional and flexural motions are comparable {($\max\vert\varepsilon_{\text{ext}}\vert\,/\,\max\vert\varepsilon_{\text{flex}}\vert\approx{}1$)}.
The ratio reaches up to ${c_{\text{ext}}\,/\,c_{\text{flex}}}\approx{}25$ for the longest wave periods and thinnest ice considered, which is the regime in which flexural motions dominate strain {($\max\vert\varepsilon_{\text{ext}}\vert\,/\,\max\vert\varepsilon_{\text{flex}}\vert\approx{}0$)}.


\subsection{Time domain: Case study}

\begin{figure}[h!] 
    \centering
    \setlength{\figurewidth}{0.8\textwidth}
    \setlength{\figureheight}{0.9\textwidth}
    \input{./tikz/Fig2_Strain_field.tex}
    \caption{Snapshots of the strain field{, $\epsilon(x,z,t)$ (\ref{eq:strain}),} in a $D=200$\,m-thick ice shelf, forced by  {a Gaussian} incident wave packet {(\ref{Eq:packet})} with a  peak period $T_{\text{peak}} = 15$\,s and width $\sigma = 2 \times 10^{-3}$\,m$^{-1}$.
    The free surface elevation of the incident packet, $\eta{(x,t)}=\eta_{\text{inc}}{(x,t)}$, is shown in the open ocean (black curve), and the displacement potential is shown in the sub-shelf water cavity{, $\Phi(x,z,t)$ (\ref{eqs:SMA_3}b)} (scaled by a factor $10^{-3}$ to match the colorbar for strain).
    The snapshots are at times (a)~$t = 20$\,s, (b)~$100$\,s, and (c)~$200$\,s.}
    \label{fig:Strain_field}
\end{figure}
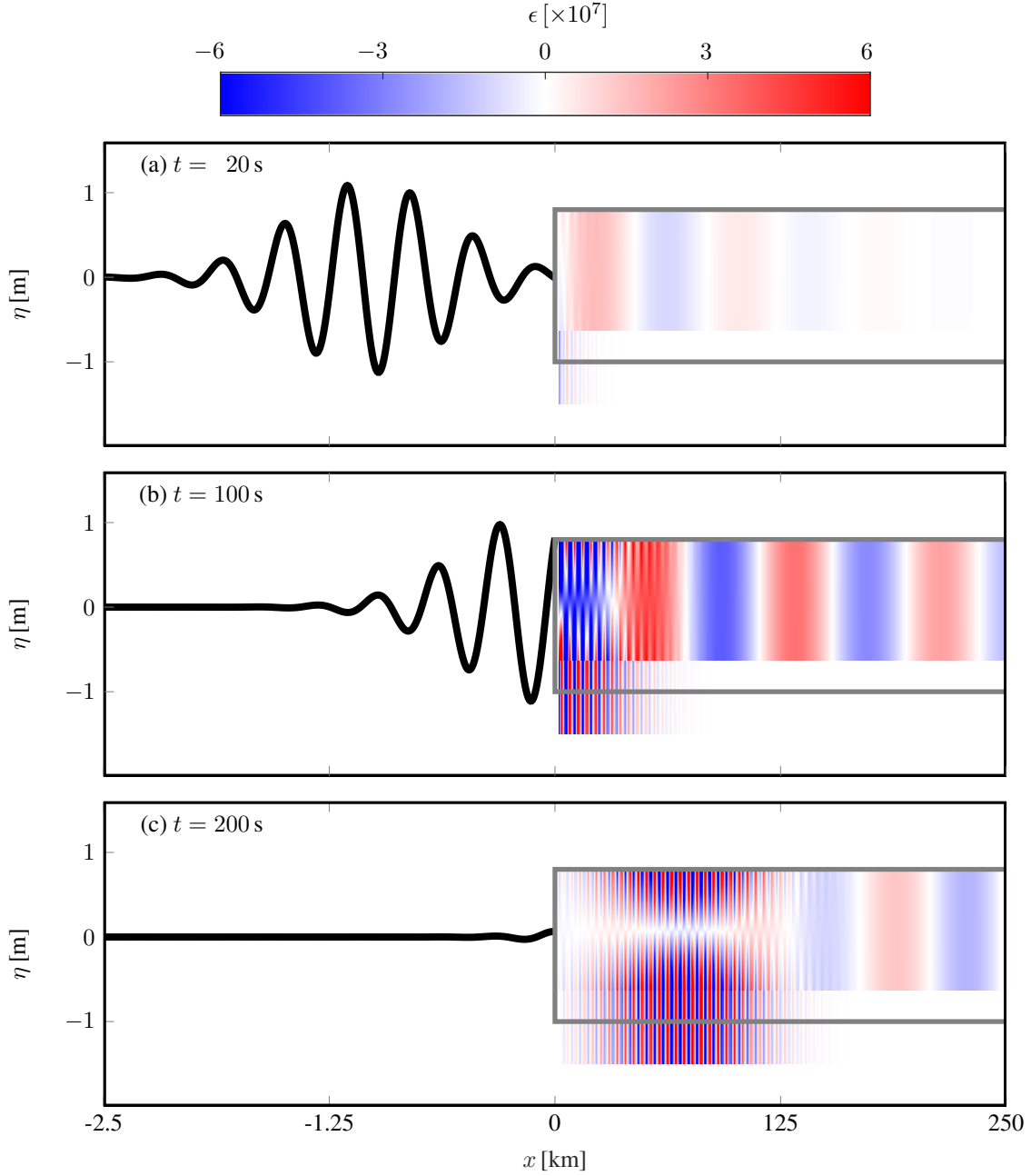

Consider  {a Gaussian} incident wave packet {(\ref{Eq:packet})} interacting with a $D=200$\,m thick ice shelf. 
The packet has peak wave period $T_{\text{peak}}=15$\,s, where $T_{\text{peak}}=2\,\pi\,/\,\omega_{\text{peak}}$ and $\omega_{\text{peak}}= {+}\sqrt{g\,k_{\text{c}}\,\tanh(k_{\text{c}}\,h)}$, such that the peak of the packet is where the strains due to extensional and flexural waves are of almost identical magnitude. 
The packet width is $\sigma=2\times10^{-3}$\,m$^{-1}$, which corresponds to a width of $\approx{}15$\,s in wave period, which is skewed towards longer periods around $T_{\text{peak}}$, i.e.,   $T_{\text{peak}}$ is not at the centre of the packet in terms of wave periods. 
   
When the leading edge of the incident wave packet interacts with the shelf front  at $t=20$\,s (Fig.~\ref{fig:Strain_field}a), it generates (i)~extensional {strain} waves in the ice shelf, {$\epsilon_{\text{ext}}$,} which are identified by the vertical bands that alternate between positive and negative strain, and propagate quickly along the shelf, such that they can be detected $>200$\,km from the shelf front, and (ii)~flexural {strain} waves, {$\epsilon_{\text{flex}}$,} which are coupled to motions in the sub-shelf water cavity, and are confined to $\approx{}30$\,km from the shelf front.  
When the peak of the packet reaches the shelf front  at $t=100$\,s (Fig.~\ref{fig:Strain_field}b), there is an interval of $\approx{}70$\,km from the shelf front where flexural and extensional waves co-exist, and with only extensional waves farther from the shelf front.
When the entire packet has interacted with the shelf  at $t=200$\,s (Fig.~\ref{fig:Strain_field}{c}), the extensional and flexural waves in the ice shelf have largely separated, due to the large difference in their celerities, with the flexural waves occupying up to $x\approx{}125$\,km and identified by the antisymmetric strains about mid-plane of the shelf.  

\begin{figure}[h!] 
    \centering
    \setlength{\figurewidth}{0.8\textwidth}
    \setlength{\figureheight}{0.9\textwidth}
    \input{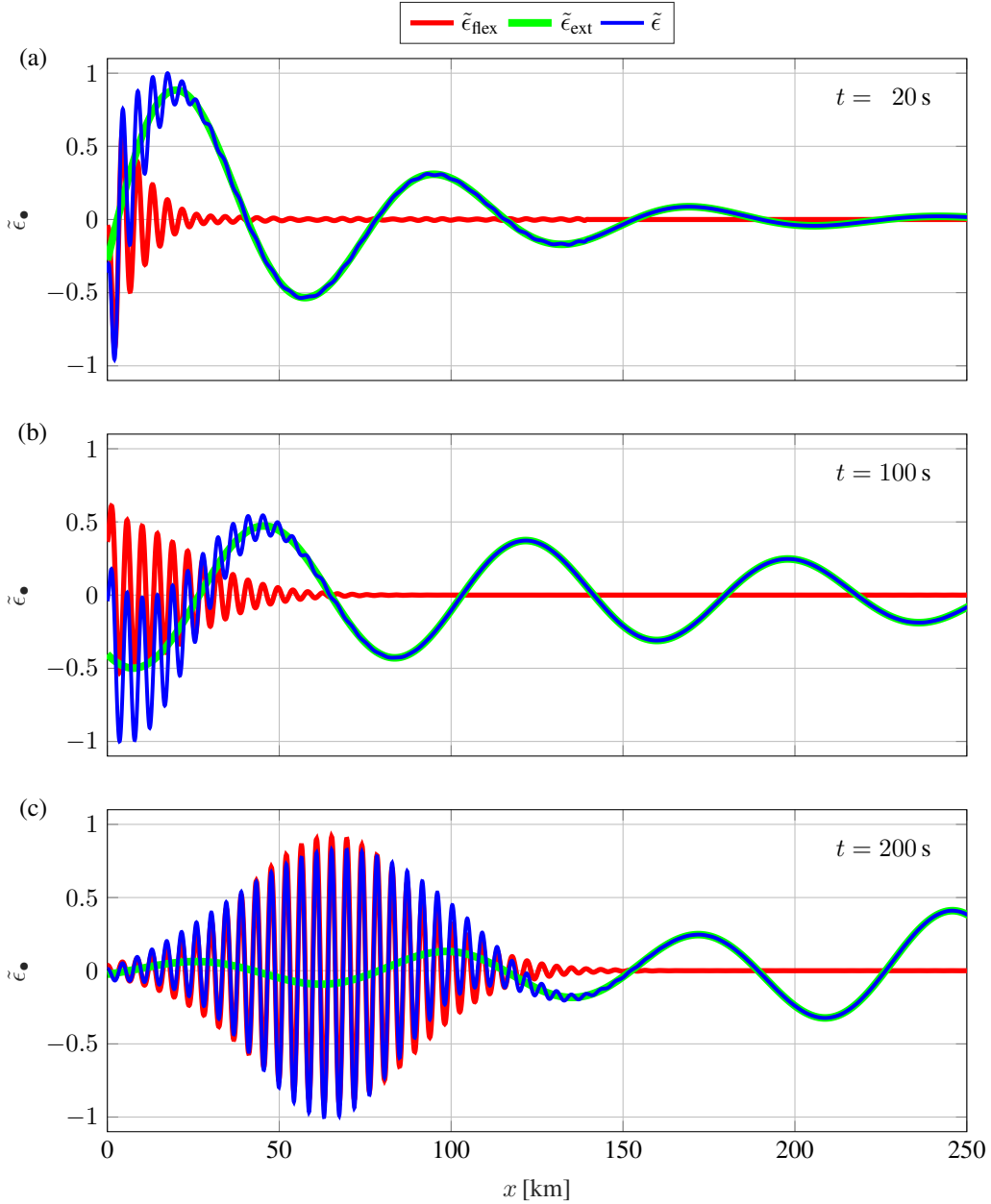}
    \caption{Snapshots of the  strain profile along the bottom of the ice shelf {(normalised by its maximum value at that instant of time; Eq.~\ref{eqs:strain_norm})},  
    $\tilde{\epsilon}(x,t)$ (blue curves), for the case shown in Fig.~\ref{fig:Strain_field}. 
    The corresponding contributions to the strain due to flexural waves, $\tilde{\epsilon}_{\text{flex}}(x,t)$ (red), and extensional waves, $\tilde{\epsilon}_{\text{ext}}(x,t)$ (green), are superimposed.}   
    \label{fig:Strain_wave_all}
\end{figure}

The strain{, $\epsilon$,} generally attains its maximum values along either its upper surface ($z=D-d$) or its lower surface ($z=-d$). 
Taking the lower surface for the purpose of illustration, {and using the normalised strains}  
\begin{subequations}\label{eqs:strain_norm}
\begin{align}
{\tilde{\epsilon}(x,t)} & {= \epsilon(x,{z=-d},t)\,/\,\max_{x}|\epsilon(x,{z=-d},t)|}	
\\[6pt]
{\tilde{\epsilon}_{\text{flex}}(x,t)} & {= \epsilon_{\text{flex}}(x,{z=-d},t)\,/\,\max_{x}|\epsilon(x,{z=-d},t)|}
\\[6pt]
{\mathandR
\tilde{\epsilon}_{\text{ext}}(x,t)} & {= \epsilon_{\text{ext}}(x,t)\,/\,\max_{x}|\epsilon(x,{z=-d},t)|,}
\end{align}
\end{subequations}
as expected (from Fig.~\ref{fig:Strain_wave_all}), there is a transition in the strain profile being dominated by the extensional waves during the early phase of the wave packet--ice shelf interaction (panel~a), to an increasing contribution from the flexural waves (b), to flexural waves dominating the strain over the first 100\,km (c).
In all cases, the strain magnitudes  reach their maximum values {($\vert\tilde{\epsilon}\vert=1$)} at locations where there are coherent interactions between the extensional and flexural waves.

\begin{figure}[h!] 
    \centering  
    \setlength{\figurewidth}{0.8\textwidth}
    \setlength{\figureheight}{0.9\textwidth}
    \input{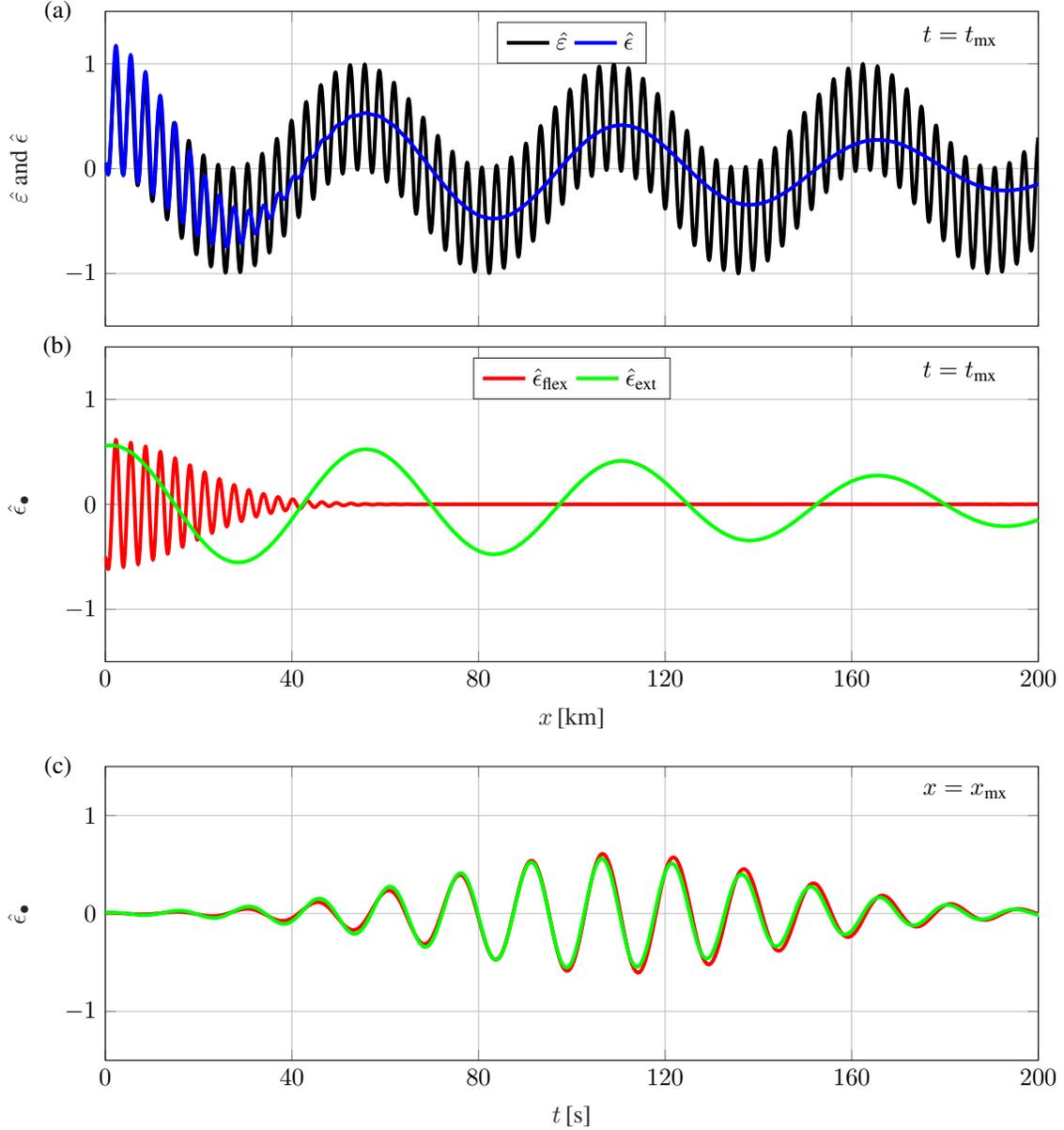}
    \caption{(a)~The strain profile relative to the maximum strain in the corresponding frequency-domain problem, 
    $\hat{\epsilon}(x,t)$ {(\ref{eqs:strains_relative_1}b)}, 
    at the time when the maximum strain magnitude is attained{, $t=t_{\max}$ (\ref{eq:tmax})} (blue curve), superimposed on the frequency-domain strain profile, $\hat{\varepsilon}(x)$ ({\ref{eqs:strains_relative_1}a; }black). 
    (b)~Corresponding strain profiles due to flexural waves, $\hat{\epsilon}_{\text{flex}}(x,t)$ ({\ref{eqs:strains_relative_2}a; }red), and extensional waves, $\hat{\epsilon}_{\text{ext}}(x,t)$ ({\ref{eqs:strains_relative_2}b; }green). 
    (c)~Time series of the strains due to flexural and extensional waves at the location where the maximum strain magnitude is attained{, $x=x_{\max}$ (\ref{eq:xmax})}.}
    \label{fig:Strain_wave_x}
\end{figure}

The time when the maximum strain magnitude is attained is 
\begin{equation}\label{eq:tmax}
    t_{\max}=\mathrm{argmax}\Big\{\max_{x}\vert\epsilon(x,z=-d,t)\vert\Big\}.
\end{equation}
The strain profile at $t=t_{\max}\approx{}109$\,s is similar to the strain profile in the frequency domain at the peak period $T_{\text{peak}}$/central wavenumber $k_{\text{c}}$ and corresponding amplitude, $A_{0}\equiv{A}(k_{\text{c}})${, as illustrated by similarity of the profiles}
\begin{subequations}\label{eqs:strains_relative_1}
\begin{align}
{\hat{\varepsilon}(x)} & { =\real\{\varepsilon(x,{z=-d}:k_{\text{c}})\}}
\\[6pt]
{\mathandR
\hat{\epsilon}(x,t)} & {= \epsilon(x,{z=-d},t)\,/\,\max_{x}|{A}_{0}\,\varepsilon(x,{z=-d}:\omega(k_{\text{c}}))|}	
\end{align}
\end{subequations}
{at $t=t_{\max}$}
(Fig.~\ref{fig:Strain_wave_x}a).
The main difference is that the short-wave component of the strain in the time domain exists only up to 30--40\,km from the shelf front.
Over this interval, the amplitudes of the strain due to extensional and flexural waves are almost identical{, as indicated by comparison of} (Fig.~\ref{fig:Strain_wave_x}b)
\begin{subequations}\label{eqs:strains_relative_2}
\begin{align}
{\hat{\epsilon}_{\text{flex}}(x,t)} & {= \epsilon_{\text{flex}}(x,{z=-d},t)\,/\,\max_{x}|{A}_{0}\,\varepsilon(x,{z=-d}:\omega(k_{\text{c}}))|}
\\[6pt]
{\mathandR
\hat{\epsilon}_{\text{ext}}(x,t)} & {= \epsilon_{\text{ext}}(x,t)\,/\,\max_{x}|{A}_{0}\,\varepsilon(x,{z=-d}:\omega(k_{\text{c}}))|.}
\end{align}
\end{subequations}

The corresponding location of maximum strain magnitude is
\begin{equation}\label{eq:xmax}
    x_{\max}=\mathrm{argmax}\Big\{\max_{t}\vert\epsilon(x,z=-d,t)\vert\Big\}.
\end{equation}
It is the location where the strains due to extensional and flexural waves are in phase with one another (Fig.~\ref{fig:Strain_wave_x}c).
At this location, the amplitudes of the two strains build up until the maximum strain is attained (at $t=t_{\max}$) and then decrease (Fig.~\ref{fig:Strain_wave_x}c).

\subsection{Time domain: Parameter analysis}
 
\begin{figure}[h!] 
    \centering
    \setlength{\figurewidth}{0.8\textwidth}
    \setlength{\figureheight}{0.9\textwidth}
    \input{./tikz/Fig4_Strain.tex}
    \caption{Maximum strain, $\hat{\epsilon}^{\max}$ {(\ref{eq:hatepsmax})}, of a $D=200$\,m-thick ice shelf, and corresponding strains due to flexural and extensional waves,  $\hat{\epsilon}^{\max}_{\text{flex}}$ and $\hat{\epsilon}^{\max}_{\text{ext}}$ {(\ref{eq:hatepsmax_flex_ext})}, respectively, as functions of (a)~peak period, for wave packet width $\sigma = 2\times 10^{-3}$\,m$^{-1}$, and (b)~wave packet width $\sigma$ for peak period  $T_{\text{peak}} = 15$\,s.}
 \label{fig:paramter_test}
\end{figure}
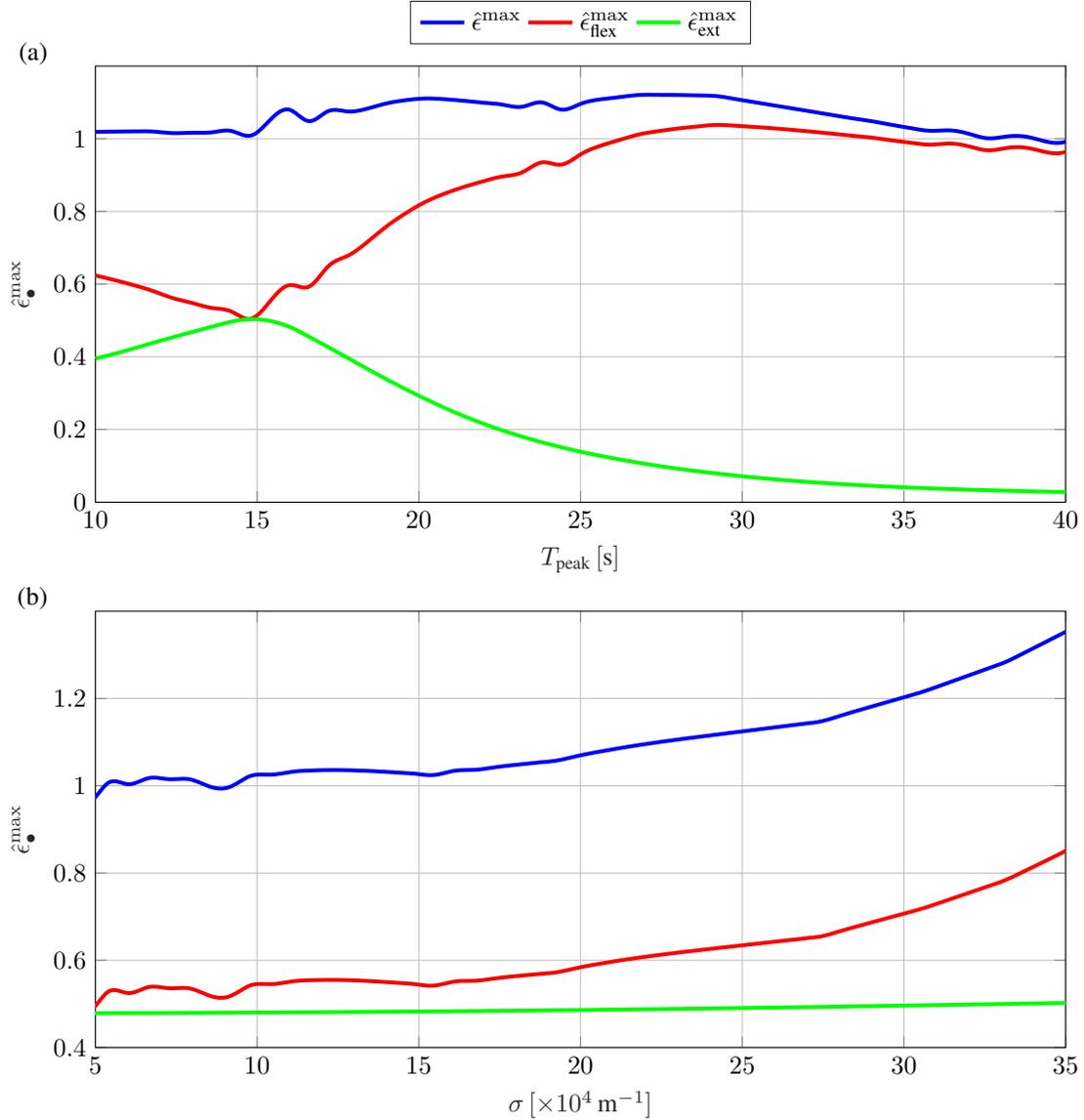

The maximum strain modulus {(maximum of $\vert\epsilon\vert$ with respect to time and space)}  along the lower surface of a $D=200$\,m-thick ice shelf, induced by an incident wave packet, is similar to that of the corresponding frequency-domain problem over peak periods from 10--40\,s, with the packet width held fixed  at $\sigma=2\times10^{-3}$\,m$^{-1}${, as shown by}
\begin{equation}\label{eq:hatepsmax}
{\hat{\epsilon}^{\max}\equiv \vert\hat{\epsilon}(x_{\max},t_{\max})\vert}
\end{equation}
{as a function of $T_{\text{peak}}$}
 (Fig.~\ref{fig:paramter_test}a).
The contribution  of flexural waves to the maximum strain is greater than that of extensional motions across the peak period range, {as shown by}
\begin{equation}\label{eq:hatepsmax_flex_ext}
{\hat{\epsilon}^{\max}_{\text{flex}}\equiv \vert\hat{\epsilon}_{\text{flex}}(x_{\max},t_{\max})\vert
\mathand
\hat{\epsilon}^{\max}_{\text{ext}}\equiv \vert \hat{\epsilon}_{\text{ext}}(x_{\max},t_{\max})\vert}
\end{equation}
{(Fig.~\ref{fig:paramter_test}a)}
except for around the 15\,s peak period where they are almost identical (cf.~Fig.~\ref{fig:Strain_wave_x}c).
The contribution to the maximum strain due to extensional waves attains its maximum around the 15\,s peak period. 
It decreases slightly as peak period decreases below $T_{\text{peak}}=15$\,s and strongly as peak period increases, such that flexural waves contribute $>90$\% of the maximum strain for peak periods greater than 27\,s.

\begin{figure}[h!] 
    \centering
     \setlength{\figurewidth}{0.8\textwidth}
    \setlength{\figureheight}{0.9\textwidth}
    \input{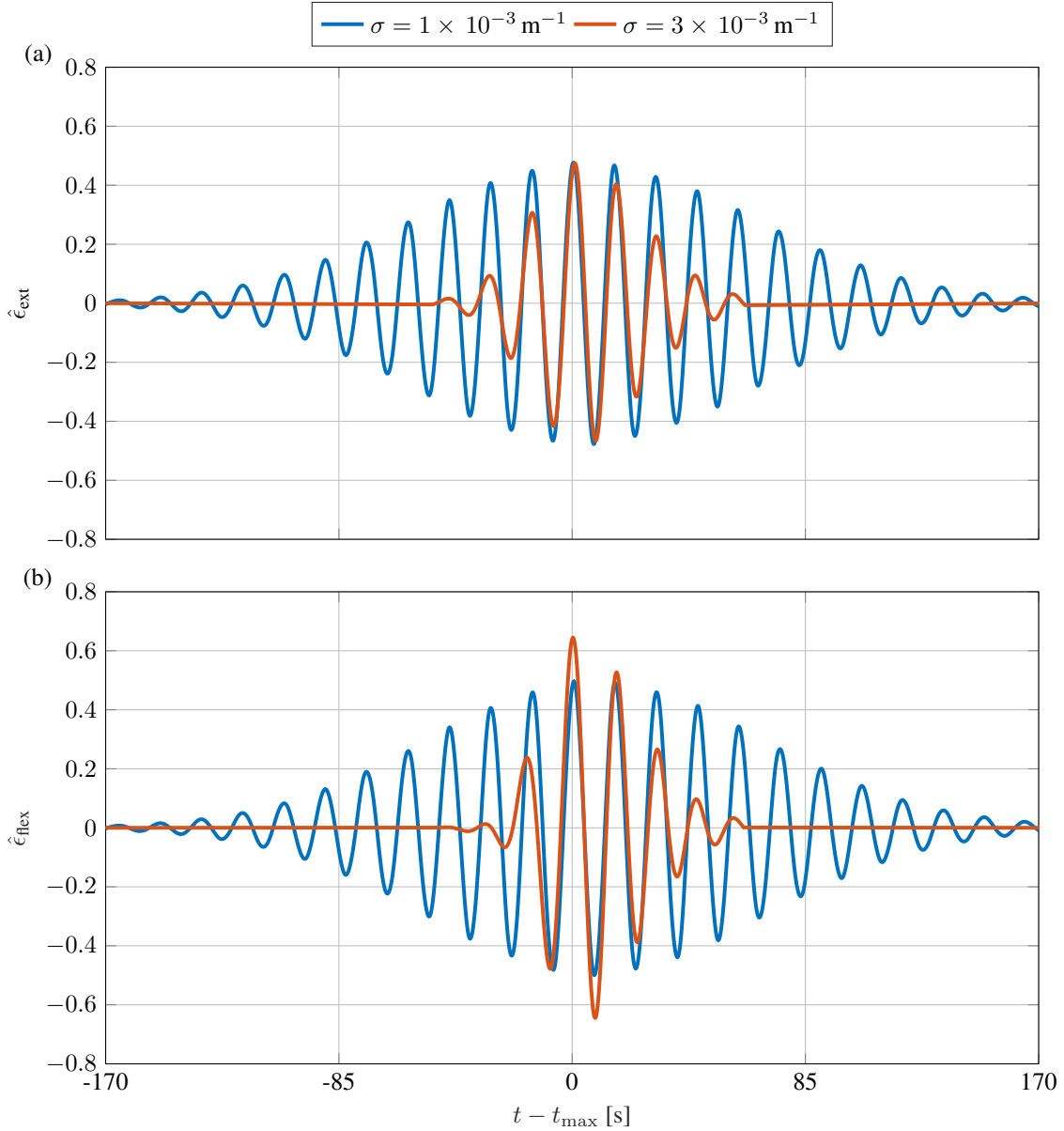}
    \caption{Time series of the strains  on a $D=200$\,m-thick ice shelf, due to (a)~extensional waves{, $\hat{\epsilon}_{\text{ext}}$ (\ref{eqs:strains_relative_2}b), } and (b)~flexural waves{, $\hat{\epsilon}_{\text{flex}}$ (\ref{eqs:strains_relative_2}a)}, at the location of the maximum strain, $x=x_{\text{mx}}$, caused by an incident wave packet of peak period $T_{\text{peak}}=15$\,s and packet width $\sigma = 1\times\,10^{-3}$\,m$^{-1}$ (blue curves) and $\sigma = 3\times\,10^{-3}$\,m$^{-1}$ (red).  }  
    \label{fig:Strain_different_width}
\end{figure}

 For a fixed peak period of $T_{\text{peak}}=15$\,s, the maximum strain {modulus} increases with respect to that of the corresponding frequency-domain problem {(i.e., $\hat{\epsilon}^{\max}$ increases)} as the packet width{, $\sigma$,} increases (Fig.~\ref{fig:paramter_test}b).
The contribution due to flexural waves{, $\hat{\epsilon}^{\max}_{\text{flex}}$,} also increases with increasing packet width, but the contribution due to extensional waves{, $\hat{\epsilon}^{\max}_{\text{ext}}$,} is insensitive to the packet width.
This is backed by time series of the strains due to flexural and extensional waves{, $\hat{\epsilon}_{\text{flex}}$ and $\hat{\epsilon}_{\text{ext}}$,} at the location of the maximum strain{, $x=x_{\max}$} (Fig.~\ref{fig:Strain_different_width}).
Increasing the packet width in the wavenumber (or frequency) domain, i.e., increasing $\sigma$, results in narrowing of the packet in the time domain.
The maximum value of {$\hat{\epsilon}_{\text{ext}}$} is unaffected by the packet width (Fig.~\ref{fig:Strain_different_width}a), noting that the amplitude $A_{0}$ scales with $1\,/\,\sqrt{\sigma}$ to maintain a consistent energy and, thus, the smaller width produces a greater maximum value for the unnormalised strains.
For the strain due to flexural waves, narrowing of the packet in the time domain is accompanied by an increase in {the peak of $\hat{\epsilon}_{\text{flex}}$} (Fig.~\ref{fig:Strain_different_width}b).

\begin{figure}[h!] 
    \centering
     \setlength{\figurewidth}{0.8\textwidth}
    \setlength{\figureheight}{0.9\textwidth}
    \input{./tikz/Fig5_boxplot_period.tex}
    \caption{{Statistics of $\max_{x}\hat{\epsilon}(x,t)$} versus peak period{,} with fixed packet width $\sigma = 2\times10^{-3}$\,m$^{-1}$, for ice shelf thickness (a)~$D = 100$\,m, (b)~$D=200$\,m and (c)~$D=400$\,m. 
    Median strains (blue curves) are shown at each peak period, along with interquartile ranges (boxes) and min--max values (whiskers) for subsets of peak periods, where statistics are for datasets over $0<x<50$\,km, from the time the incident packet peak reaches the shelf shelf front to the time is reaches $x=50$\,km, with data stored at spatial and temporal resolutions of 200\,m and 2\,s, respectively.} 
    \label{fig:boxplot_period}
\end{figure}
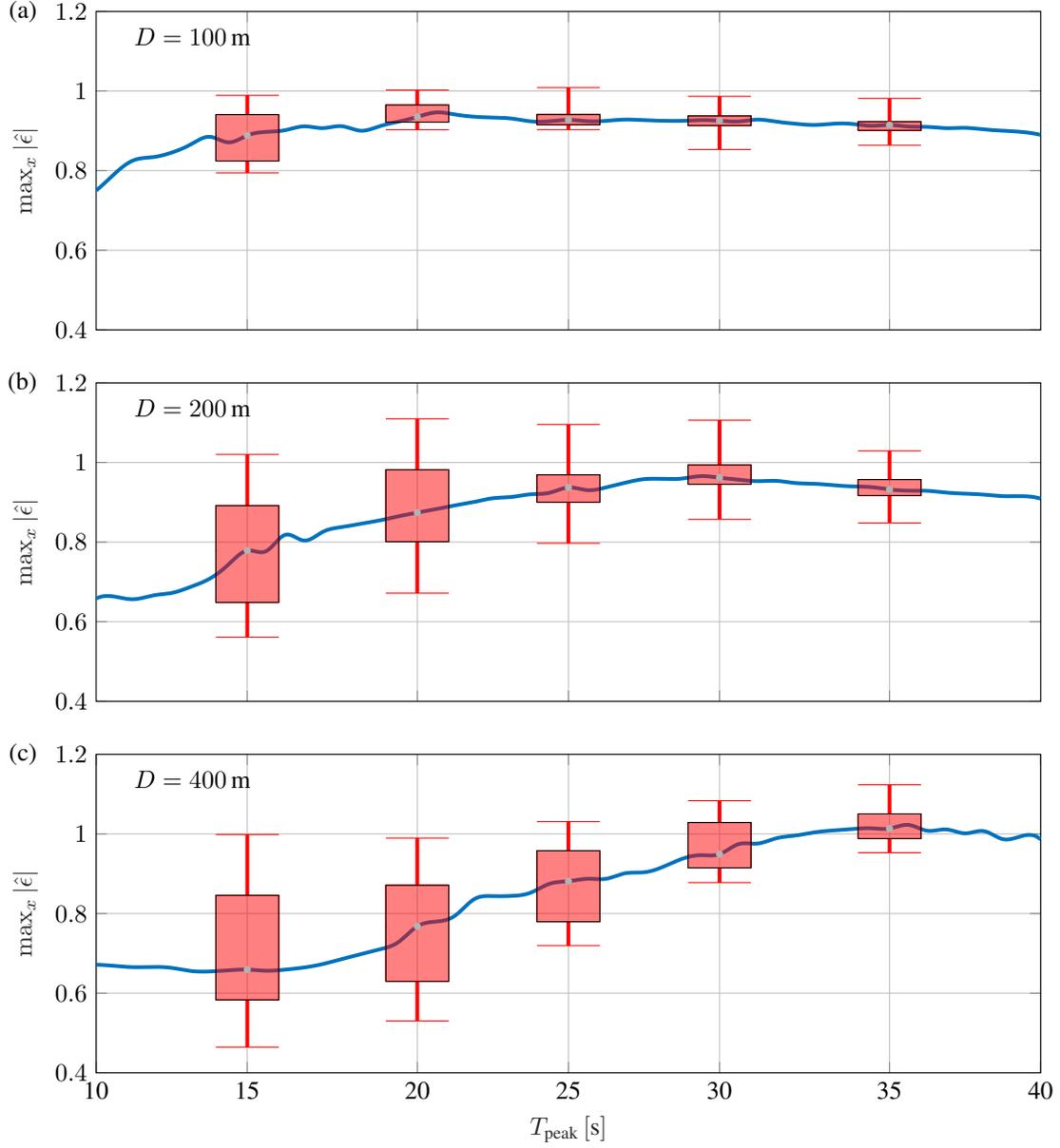

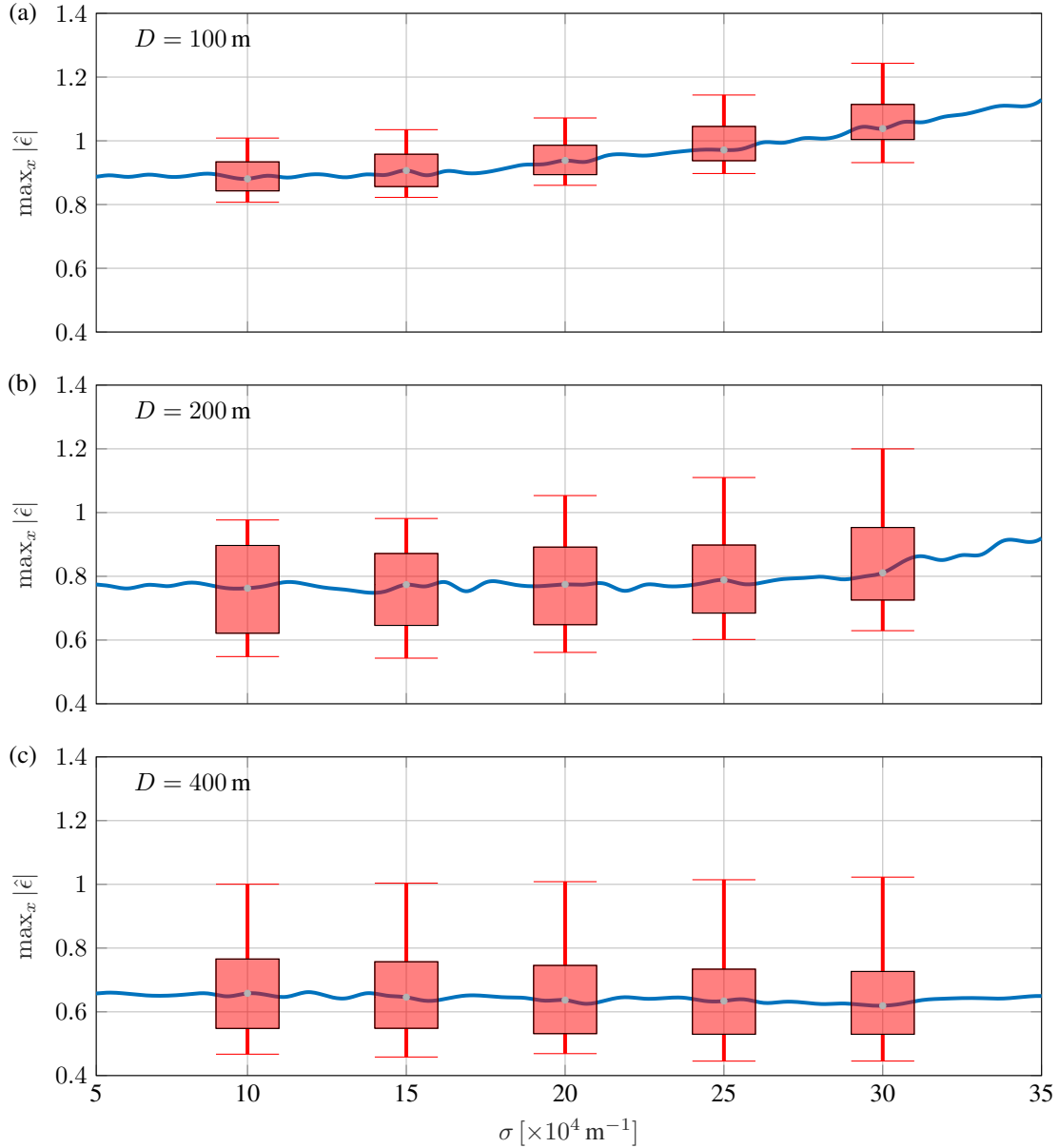
\begin{figure}[h!] 
    \centering
    \setlength{\figurewidth}{0.8\textwidth}
    \setlength{\figureheight}{0.9\textwidth}
    \input{./tikz/Fig5_boxplot_width.tex}
    \caption{As in Fig.~\ref{fig:boxplot_period} but versus wave packet width, with peak period $T_{\text{peak}} = 15$\,s.}
    \label{fig:boxplot_width}
\end{figure}

Extensional and flexural waves can interact coherently, which creates the maximum strains (e.g., Fig.~\ref{fig:Strain_wave_x}c), and incoherently, which reduces the strains in relation to the individual extensional and flexural components of the strain field.
Thus, for shorter peak periods, where strains have comparable contributions from flexural and extensional waves, there are relatively large spreads of 
\begin{equation}
{
\max_{x}\vert\hat{\epsilon}\vert 
\mathfor 0<x<50\text{\,km},
}
\end{equation}
{over a time interval for which the packet crosses that length of the ice shelf,}
whereas, at  longer periods, 
where strains are dominated by flexural waves,
distributions are far narrower  (Fig.~\ref{fig:boxplot_period}).
Thus, over the 10--40\,s peak period interval shown, the median strain magnitude tends to increase with peak period for shorter peak periods and be relatively insensitive to peak period for longer peak periods.
The transition between these regimes shifts to longer peak periods as the shelf thickness increases, 
as extensional waves make an appreciable contribution to strain up to longer periods for thicker ice shelves (cf.~Fig.~\ref{fig:Heat_map}a).
In contrast, there is little variation in the spread of the (normalised) strain magnitudes with respect to the incident packet width (Fig.~\ref{fig:boxplot_width}).

\section{Conclusions}

The flexural strain imposed on a floating ice shelf by an incident wave pack{et} from the open ocean has been studied using a mathematical model in which the ice shelf is modelled as a thin plate that supports extensional waves in addition to flexural waves.
In the wave period interval of interest (the swell regime), it was shown that the flexural and extensional waves generated in the ice shelf by the incident packet interact before they separate due to their different speeds.
Coherent interactions create the largest ice shelf strains, although these are accompanied by incoherent interactions that can reduce the strains.
The maximum strains are comparable to, and usually slightly exceed, those predicted by a corresponding frequency-domain problem.

The findings support the conclusions of \cite{bennetts2024thin} on the importance of incorporating extensional waves into models when making predictions of the strains imposed on ice shelves by ocean swell.
To be compatible with state-of-the-art predictions of the impacts of swell on Antarctic ice shelves \cite{bennetts2022modeling,tazhimbetov2023simulation,liang2024pan}, the model must be developed to incorporate spatial variations in the ice shelf geometry{. The spatial variations should include ice shelf thickening away from the shelf front, which causes flexural waves to attenuate away from the shelf front \citep{liang2024pan}, so that the models can then test if attenuation of the extensional waves is significantly weaker, as observed on the Ross Ice Shelf \citep{chen2018ocean}.} {Models with spatial variations can also be used to test the effects of crevasses on ice shelf strain, noting that models involving only flexural waves predict that crevasses create local amplifications in strain forced by swell}~\cite{bennetts2022modeling}.

\appendix

\section{Approximate solution: Additional details}\label{app:details}

Solutions of the governing equations for flexural waves in the ice shelf--cavity region (\ref{eqs:odes}b--c) are sought by writing
\begin{equation}\label{eq:exp_solns}
\psi(x)=c\,\e^{\ci\,\lambda}
\mathand
w(x)=\gamma\,\e^{\ci\,\lambda},
\end{equation}
where $\lambda$, $c$ and $\gamma$ are to be found.
Substituting (\ref{eq:exp_solns}) into (\ref{eqs:odes}b--c) gives
\begin{subequations}\label{eqs:eigensys}
\begin{align}
a\,(\kappa^{2}-\lambda^{2})\,c - \kappa\,\tanh(\kappa\,H)\,c + \frac{\omega^{2}}{g}\,\gamma & = 0
\\
\mathandR
\{F\,\lambda^{4} - J\,\omega^{2}\,\lambda^{2} + 1-m\,\omega^{2}\}\,\gamma - c
& = 0,
\end{align}
\end{subequations}
where $b=\kappa^{2}\,a-\kappa\,\tanh(\kappa\,H)$ has been used in (\ref{eqs:eigensys}a).
Nontrivial solutions, $(c,\gamma)$, of the system~(\ref{eqs:eigensys}a--b) exist if its determinant is zero, that is
\begin{equation}\label{eq:det}
\{ a\,(\kappa^{2}-\lambda^{2})-\kappa\,\tanh(\kappa\,H) \}
\,
\{F\,\lambda^{4} - J\,\omega^{2}\,\lambda^{2} + 1-m\,\omega^{2}\}
+
\frac{\omega^{2}}{g} = 0.
\end{equation}
Using the dispersion relation (\ref{eqs:dispersion_relations}b) to replace the final term on the left-hand side of (\ref{eq:det}) and rearranging, gives
\begin{equation}\label{eq:det2}
(\kappa^{2}-\lambda^{2})
\,
\Big[
a\,\{F\,\lambda^{4} - J\,\omega^{2}\,\lambda^{2} + 1-m\,\omega^{2}\}
+
\{F\,(\kappa^{2}+\lambda^{2}) - J\,\omega^{2}\}\,\kappa\,\tanh(\kappa\,H)
\Big]
=
0.
\end{equation}
Eq.~(\ref{eq:det2}) is satisfied if $\lambda=\pm\kappa$, in which case (\ref{eq:exp_solns}b) becomes
\begin{align}
\{F\,\kappa^{4} - J\,\omega^{2}\,\kappa^{2} + 1-m\,\omega^{2}\}\,\gamma - c
& = 0
\\
\Rightarrow\quad
\frac{\omega^{2}\,\gamma}{g\,\kappa\,\tanh(\kappa\,H)}
-c
& = 0,
\end{align}
from which (\ref{eq:tcav}) can be deduced.
If $\lambda\neq\pm\kappa$, then Eq.~(\ref{eq:det2}) requires the factor in square brackets on its left-hand side to be zero, i.e., (\ref{eq:kappaj}) holds with $\lambda=\kappa_{-j}$.
In this case, substituting (\ref{eq:kappaj}) with $\kappa_{-j}=\lambda$ into (\ref{eq:exp_solns}b) gives
\begin{equation}
\{F\,(\kappa^{2}+\lambda^{2}) - J\,\omega^{2}\}\,\kappa\,\tanh(\kappa\,H)\,\gamma + a\,c = 0,
\end{equation}
from which (\ref{eq:tcavj}) can be deduced.



\bibliography{bibli.bib}

\end{document}

%% file: arxiv-luke.tex
\usepackage{amsmath}

\input{definitions-luke.tex}
\input{tikz-luke.tex}
\input{colours-luke.tex}

%% file: definitions-luke.tex
\newcommand{\real}{\textrm{Re}}

\newcommand{\ci}{\mathrm{i}}   
\newcommand{\e}{\mathrm{e}}

\usepackage{enumerate}

\usepackage{overpic,rotating}

\newcommand{\tn}[1]{\textnormal{#1}}

\newcommand{\beq}{\begin{equation}}
\newcommand{\eeq}{\end{equation}}
\newcommand{\bea}{\begin{eqnarray}}
\newcommand{\eea}{\end{eqnarray}}
\newcommand{\bean}{\begin{eqnarray*}}
\newcommand{\eean}{\end{eqnarray*}}

\newlength{\mylinelength}
\newlength{\mydashlength}
\newlength{\mydashspace}
\newlength{\mychainlengthA}
\newlength{\mychainlengthB}
\newlength{\mychainspace}
\newlength{\mylinethickness}
\newcommand{\mathand}{\quad\textnormal{and}\quad}
\newcommand{\mathandR}{\textnormal{and}\quad}  
  
\newcommand{\mathfor}{\quad\textnormal{for}\quad}

\newcommand{\mathwhereR}{\textnormal{where}\quad}

\usepackage{pifont}
\ifdefined\showmylabels
 \usepackage[inline]{showlabels}
 
\fi

\usepackage[normalem]{ulem}
\usepackage{soul}
\usepackage{tcolorbox}

%% file: tikz-luke.tex
\usepackage{tikz,pgfplots}
\usetikzlibrary{calc,patterns,decorations.pathmorphing,decorations.markings,decorations.pathmorphing,angles,quotes,arrows,arrows.meta}
\tikzset{snake it/.style={decorate, decoration=snake}}

\newlength{\figurewidth}

\newlength{\figureheight}

%% file: colours-luke.tex
\usepackage{xcolor}
\definecolor{darkgreen}{rgb}{0,0.55,0}
\definecolor{midgreen}{rgb}{0,0.8,0.2}
\definecolor{magenta}{rgb}{1,0,1}
\definecolor{purple}{rgb}{0.5,0,0.5}
\definecolor{darkorange}{rgb}{1,0.55,0}
\definecolor{maroon}{rgb}{0.5,0,0}
\definecolor{olive}{rgb}{0.5,0.5,0}
\definecolor{midgrey}{rgb}{0.5,0.5,0.5}
\definecolor{lightgrey}{rgb}{0.75,0.75,0.75}
\definecolor{matlabblue}{rgb}{0,0.447,0.741}
\definecolor{matlabred}{rgb}{0.85,0.325,0.098}
\definecolor{lightblue}{rgb}{0,0.5,1}
\definecolor{darkgrey}{rgb}{0.25,0.25,0.25}
\definecolor{teal}{rgb}{0,0.5,0.5}
\definecolor{navy}{rgb}{0,0,0.5}
\definecolor{goldenrod}{rgb}{0.85,0.6,0.1}

%% file: tikz/Fig2_Strain_field.tex
%
\definecolor{mycolor1}{rgb}{0.85000,0.32500,0.09800}%
\begin{tikzpicture}


\begin{axis}[
name = Strain1,
 mark layer=axis tick labels,axis on top,
width=\figurewidth,
height=0.3\figureheight,
at={(0\figurewidth,0.65\figureheight)},
scale only axis,
point meta min=0,
point meta max=1,
xmin=-1,
xmax=1,
ymin=-2,
ymax=1.6,
xtick={-1,-0.5,0,0.5,1},
ytick={-1,0,1},
xticklabel=\empty,
ylabel={$\eta$\,[m]},
axis background/.style={fill=white},
axis y line*=left,
]
  \addplot [color=black, line width=3.0pt, forget plot]
  table[row sep=crcr]{%
-1	0.00280700319253898\\
-0.998	0.00262225649295487\\
-0.996	0.00239409096708535\\
-0.994	0.00212139210767092\\
-0.992	0.00180345795684145\\
-0.99	0.00144003968046975\\
-0.988	0.00103137968052633\\
-0.986	0.000578246612866211\\
-0.984	8.19666768889449e-05\\
-0.982	-0.000455549450023693\\
-0.98	-0.00103178464594618\\
-0.978	-0.00164359893592039\\
-0.976	-0.00228721920093961\\
-0.974	-0.00295823552482921\\
-0.972	-0.0036516046522374\\
-0.97	-0.00436166097558767\\
-0.968	-0.00508213540576629\\
-0.966	-0.0058061824106432\\
-0.964	-0.00652641542782272\\
-0.962	-0.00723495077382055\\
-0.96	-0.00792346008169084\\
-0.958	-0.00858323120381956\\
-0.956	-0.00920523741688661\\
-0.954	-0.00978021466276546\\
-0.952	-0.0102987464534333\\
-0.95	-0.0107513559606437\\
-0.948	-0.0111286047035416\\
-0.946	-0.0114211971404337\\
-0.944	-0.0116200903659921\\
-0.942	-0.0117166080133759\\
-0.94	-0.0117025573633976\\
-0.938	-0.0115703485711406\\
-0.936	-0.0113131148357879\\
-0.934	-0.0109248322628492\\
-0.932	-0.0104004381009347\\
-0.93	-0.00973594597877355\\
-0.928	-0.00892855672349258\\
-0.926	-0.00797676330932173\\
-0.924	-0.00688044846786506\\
-0.922	-0.0056409734877593\\
-0.92	-0.0042612567437604\\
-0.918	-0.002745840523631\\
-0.916	-0.00110094476631569\\
-0.914	0.000665493612960062\\
-0.912	0.00254379705550624\\
-0.91	0.00452254049784645\\
-0.908	0.00658856234998013\\
-0.906	0.00872699421579988\\
-0.904	0.0109213101104645\\
-0.902	0.0131533957660189\\
-0.9	0.015403638438011\\
-0.898	0.0176510374334552\\
-0.896	0.0198733353754747\\
-0.894	0.0220471700038251\\
-0.892	0.0241482460848108\\
-0.89	0.0261515267709634\\
-0.888	0.0280314435119353\\
-0.886	0.0297621233759144\\
-0.884	0.0313176323975906\\
-0.882	0.0326722333268912\\
-0.88	0.0338006559150252\\
-0.878	0.0346783776432989\\
-0.876	0.035281912578814\\
-0.874	0.0355891058320004\\
-0.872	0.0355794308971027\\
-0.87	0.0352342869808977\\
-0.868	0.0345372932699624\\
-0.866	0.0334745769554454\\
-0.864	0.0320350517291265\\
-0.862	0.0302106833881659\\
-0.86	0.0279967391405344\\
-0.858	0.0253920171909742\\
-0.856	0.0223990532101397\\
-0.854	0.0190243003491747\\
-0.852	0.0152782795594306\\
-0.85	0.0111756971135296\\
-0.848	0.00673552639995775\\
-0.846	0.0019810512791509\\
-0.844	-0.00306013145562811\\
-0.842	-0.00835615267509725\\
-0.84	-0.0138709563987907\\
-0.838	-0.0195644316299831\\
-0.836	-0.0253925916000773\\
-0.834	-0.0313078010475857\\
-0.832	-0.037259051693638\\
-0.83	-0.0431922855876505\\
-0.828	-0.0490507654868931\\
-0.826	-0.0547754909068559\\
-0.824	-0.0603056579400894\\
-0.822	-0.0655791603947932\\
-0.82	-0.0705331292562251\\
-0.818	-0.0751045069299539\\
-0.816	-0.0792306521917902\\
-0.814	-0.0828499712514075\\
-0.812	-0.0859025698414213\\
-0.81	-0.0883309207772947\\
-0.808	-0.0900805410026991\\
-0.806	-0.0911006717457067\\
-0.804	-0.0913449550701356\\
-0.802	-0.0907720998190927\\
-0.8	-0.0893465297202385\\
-0.798	-0.0870390062596787\\
-0.796	-0.0838272188385828\\
-0.794	-0.079696334708023\\
-0.792	-0.0746395012367101\\
-0.79	-0.068658293206495\\
-0.788	-0.0617630980538436\\
-0.786	-0.0539734322836377\\
-0.784	-0.0453181826750772\\
-0.782	-0.0358357663780969\\
-0.78	-0.025574204561146\\
-0.778	-0.0145911049153936\\
-0.776	-0.00295354904310299\\
-0.774	0.00926211844500089\\
-0.772	0.0219706014324413\\
-0.77	0.0350780657796569\\
-0.768	0.0484826551817134\\
-0.766	0.0620751087634965\\
-0.764	0.0757394787036009\\
-0.762	0.0893539450779188\\
-0.76	0.102791723995328\\
-0.758	0.115922063972198\\
-0.756	0.128611324371185\\
-0.754	0.140724128625478\\
-0.752	0.15212458389458\\
-0.75	0.162677557764825\\
-0.748	0.172250001629808\\
-0.746	0.180712309475651\\
-0.744	0.187939699965955\\
-0.742	0.193813608983754\\
-0.74	0.198223079154271\\
-0.738	0.201066132353949\\
-0.736	0.202251110818232\\
-0.734	0.20169797220205\\
-0.732	0.199339523831195\\
-0.73	0.195122581416461\\
-0.728	0.189009037691277\\
-0.726	0.180976826781385\\
-0.724	0.171020770624602\\
-0.722	0.15915329443041\\
-0.72	0.145404999002241\\
-0.718	0.129825078737012\\
-0.716	0.112481575262115\\
-0.714	0.0934614579632981\\
-0.712	0.0728705240892837\\
-0.71	0.0508331126805223\\
-0.708	0.0274916282482236\\
-0.706	0.0030058719121276\\
-0.704	-0.0224478204237901\\
-0.702	-0.0486776313361593\\
-0.7	-0.0754775030701982\\
-0.698	-0.102628561103319\\
-0.696	-0.129900712492925\\
-0.694	-0.157054409380325\\
-0.692	-0.183842566015405\\
-0.69	-0.210012615710811\\
-0.688	-0.235308692252582\\
-0.686	-0.259473918508822\\
-0.684	-0.282252783311151\\
-0.682	-0.303393586157032\\
-0.68	-0.322650927915342\\
-0.678	-0.339788224532401\\
-0.676	-0.35458021974955\\
-0.674	-0.366815472072989\\
-0.672	-0.376298790697037\\
-0.67	-0.382853594786198\\
-0.668	-0.386324170480117\\
-0.666	-0.386577800207175\\
-0.664	-0.383506739382643\\
-0.662	-0.377030016329211\\
-0.66	-0.367095032291425\\
-0.658	-0.353678939718419\\
-0.656	-0.336789778555546\\
-0.654	-0.316467352106526\\
-0.652	-0.292783826091677\\
-0.65	-0.265844036820133\\
-0.648	-0.235785496897192\\
-0.646	-0.202778089581642\\
-0.644	-0.167023445769427\\
-0.642	-0.128754000583822\\
-0.64	-0.0882317296709286\\
-0.638	-0.045746568503314\\
-0.636	-0.00161452125260897\\
-0.634	0.043824530928591\\
-0.632	0.0902093091079814\\
-0.63	0.137159886930019\\
-0.628	0.184280692660863\\
-0.626	0.231163740432837\\
-0.624	0.277392065328317\\
-0.622	0.322543334256695\\
-0.62	0.366193602109426\\
-0.618	0.407921180458103\\
-0.616	0.447310584118593\\
-0.614	0.48395651926718\\
-0.612	0.517467875487056\\
-0.61	0.547471683166766\\
-0.608	0.57361699708454\\
-0.606	0.595578666808784\\
-0.604	0.613060954736053\\
-0.602	0.62580096318104\\
-0.6	0.633571832931094\\
-0.598	0.636185677079498\\
-0.596	0.633496215751508\\
-0.594	0.625401079525087\\
-0.592	0.611843751910117\\
-0.59	0.592815124167154\\
-0.588	0.568354638996801\\
-0.586	0.538551003186957\\
-0.584	0.503542453137014\\
-0.582	0.463516561251706\\
-0.58	0.418709575475421\\
-0.578	0.369405288680431\\
-0.576	0.315933439186969\\
-0.574	0.258667648334745\\
-0.572	0.198022905697895\\
-0.57	0.134452617191059\\
-0.568	0.068445235904835\\
-0.566	0.000520499985915447\\
-0.564	-0.0687746938074249\\
-0.562	-0.138870748137522\\
-0.56	-0.209180118118867\\
-0.558	-0.279102319208403\\
-0.556	-0.348029146602423\\
-0.554	-0.415350059955972\\
-0.552	-0.48045768482374\\
-0.55	-0.542753380251266\\
-0.548	-0.601652820449454\\
-0.546	-0.656591537486863\\
-0.544	-0.707030371451177\\
-0.542	-0.752460774576741\\
-0.54	-0.79240991641672\\
-0.538	-0.8264455382586\\
-0.536	-0.854180506636916\\
-0.534	-0.875277017978308\\
-0.532	-0.889450409106604\\
-0.53	-0.896472531519305\\
-0.528	-0.896174650996144\\
-0.526	-0.888449838184346\\
-0.524	-0.873254820288048\\
-0.522	-0.85061126883075\\
-0.52	-0.820606503614829\\
-0.518	-0.783393598422636\\
-0.516	-0.739190879637567\\
-0.514	-0.688280814756023\\
-0.512	-0.631008293655293\\
-0.51	-0.567778311419222\\
-0.508	-0.499053067443284\\
-0.506	-0.4253485013828\\
-0.504	-0.34723029221225\\
-0.502	-0.265309352170509\\
-0.5	-0.180236852618097\\
-0.498	-0.0926988237722591\\
-0.496	-0.003410374860422\\
-0.494	0.0868904146078733\\
-0.492	0.177448878052356\\
-0.49	0.267500361923885\\
-0.488	0.356276982449976\\
-0.486	0.443014476695666\\
-0.484	0.526959084128712\\
-0.482	0.607374394014434\\
-0.48	0.68354809373488\\
-0.478	0.754798553537068\\
-0.476	0.820481184266699\\
-0.474	0.879994506331539\\
-0.472	0.932785870450176\\
-0.47	0.978356773658458\\
-0.468	1.01626771754236\\
-0.466	1.04614255971096\\
-0.464	1.06767231407951\\
-0.462	1.08061836055735\\
-0.46	1.08481503018106\\
-0.458	1.08017153754707\\
-0.456	1.06667323852326\\
-0.454	1.04438219759656\\
-0.452	1.01343705577879\\
-0.45	0.974052196682171\\
-0.448	0.926516215121247\\
-0.446	0.871189699331746\\
-0.444	0.808502344551783\\
-0.442	0.738949422218551\\
-0.44	0.663087635328725\\
-0.438	0.581530396528646\\
-0.436	0.494942571179471\\
-0.434	0.404034732924563\\
-0.432	0.309556984116925\\
-0.43	0.212292397793715\\
-0.428	0.113050141667752\\
-0.426	0.0126583478031236\\
-0.424	-0.088043205779896\\
-0.422	-0.188210533393618\\
-0.42	-0.287002930423404\\
-0.418	-0.383590442493433\\
-0.416	-0.477161245134223\\
-0.414	-0.56692886481829\\
-0.412	-0.652139173699821\\
-0.41	-0.732077092529073\\
-0.408	-0.806072938991151\\
-0.406	-0.873508362113985\\
-0.404	-0.933821807367795\\
-0.402	-0.986513461597689\\
-0.4	-1.03114963194695\\
-0.398	-1.06736651839009\\
-0.396	-1.09487334534673\\
-0.394	-1.11345482403069\\
-0.392	-1.12297292364051\\
-0.39	-1.12336793615358\\
-0.388	-1.1146588262776\\
-0.386	-1.09694286497367\\
-0.384	-1.07039455182408\\
-0.382	-1.0352638383077\\
-0.38	-0.991873670697385\\
-0.378	-0.940616877742264\\
-0.376	-0.881952434477627\\
-0.374	-0.816401139356775\\
-0.372	-0.744540747364066\\
-0.37	-0.667000606793985\\
-0.368	-0.58445585191838\\
-0.366	-0.497621207769924\\
-0.364	-0.4072444667066\\
-0.362	-0.314099699257776\\
-0.36	-0.218980263962003\\
-0.358	-0.122691682471036\\
-0.356	-0.0260444471015927\\
-0.354	0.0701531717391653\\
-0.352	0.16510225124533\\
-0.35	0.258020209081756\\
-0.348	0.348147599199984\\
-0.346	0.434754648693901\\
-0.344	0.517147476702613\\
-0.342	0.594673939838213\\
-0.34	0.666729052603568\\
-0.338	0.732759935726313\\
-0.336	0.792270250221429\\
-0.334	0.844824080253966\\
-0.332	0.89004923344993\\
-0.33	0.927639933138598\\
-0.328	0.957358883042951\\
-0.326	0.979038691104424\\
-0.324	0.992582645370716\\
-0.322	0.997964841127907\\
-0.32	0.995229664657905\\
-0.318	0.984490645087749\\
-0.316	0.965928691708757\\
-0.314	0.939789739823654\\
-0.312	0.906381833574004\\
-0.31	0.866071679257821\\
-0.308	0.819280707320978\\
-0.306	0.766480685453576\\
-0.304	0.708188929006087\\
-0.302	0.644963158227302\\
-0.3	0.57739605458986\\
-0.298	0.506109570688091\\
-0.296	0.431749049851412\\
-0.294	0.354977212705113\\
-0.292	0.27646806842525\\
-0.29	0.19690080837788\\
-0.288	0.116953739213089\\
-0.286	0.0372983113148789\\
-0.284	-0.0414067031917823\\
-0.282	-0.11852083088122\\
-0.28	-0.193427266601734\\
-0.278	-0.265537945997606\\
-0.276	-0.334298278322523\\
-0.274	-0.399191500783856\\
-0.272	-0.459742619872104\\
-0.27	-0.515521909617362\\
-0.268	-0.566147941425516\\
-0.266	-0.611290125030534\\
-0.264	-0.650670745103872\\
-0.262	-0.684066483135747\\
-0.26	-0.71130941929367\\
-0.258	-0.732287514020162\\
-0.256	-0.746944574103677\\
-0.254	-0.755279712796207\\
-0.252	-0.757346318211594\\
-0.25	-0.753250548676622\\
-0.248	-0.743149377882114\\
-0.246	-0.727248216556207\\
-0.244	-0.705798140923548\\
-0.242	-0.679092761392901\\
-0.24	-0.647464767706496\\
-0.238	-0.611282189166791\\
-0.236	-0.570944410514117\\
-0.234	-0.526877985550438\\
-0.232	-0.47953229168364\\
-0.23	-0.429375069201209\\
-0.228	-0.376887889274526\\
-0.226	-0.322561594452319\\
-0.224	-0.266891754735522\\
-0.222	-0.210374181251312\\
-0.22	-0.153500538080911\\
-0.218	-0.0967540909667476\\
-0.216	-0.0406056294559984\\
-0.214	0.0144904034413621\\
-0.212	0.0680995427595261\\
-0.21	0.119810665930708\\
-0.208	0.169239064879298\\
-0.206	0.216029204133352\\
-0.204	0.259857146748251\\
-0.202	0.300432633450931\\
-0.2	0.337500804083777\\
-0.198	0.370843554116954\\
-0.196	0.400280522669017\\
-0.194	0.425669712091686\\
-0.192	0.446907742700688\\
-0.19	0.463929749637279\\
-0.188	0.476708932093463\\
-0.186	0.485255768199013\\
-0.184	0.489616911724248\\
-0.182	0.489873789375574\\
-0.18	0.486140919830806\\
-0.178	0.478563977760937\\
-0.176	0.467317627900407\\
-0.174	0.452603155748216\\
-0.172	0.434645922700147\\
-0.17	0.413692674323789\\
-0.168	0.390008731092173\\
-0.166	0.363875091191105\\
-0.164	0.335585475015056\\
-0.162	0.305443340675276\\
-0.16	0.273758899272674\\
-0.158	0.240846157850834\\
-0.156	0.207020016857243\\
-0.154	0.17259344762176\\
-0.152	0.137874773830507\\
-0.15	0.103165079252464\\
-0.148	0.0687557620880516\\
-0.146	0.0349262542779391\\
-0.144	0.00194192196104807\\
-0.142	-0.029947838971584\\
-0.14	-0.0605113005901966\\
-0.138	-0.0895358846220443\\
-0.136	-0.116829421213647\\
-0.134	-0.14222118020553\\
-0.132	-0.165562672696359\\
-0.13	-0.186728222950927\\
-0.128	-0.205615312909403\\
-0.126	-0.222144703663768\\
-0.124	-0.23626034026333\\
-0.122	-0.247929048078191\\
-0.12	-0.257140030672956\\
-0.118	-0.263904180710359\\
-0.116	-0.268253216805159\\
-0.114	-0.270238660474269\\
-0.112	-0.269930668373217\\
-0.11	-0.267416735867417\\
-0.108	-0.262800288657177\\
-0.106	-0.256199179657643\\
-0.104	-0.247744108630802\\
-0.102	-0.237576982179809\\
-0.1	-0.225849231651707\\
-0.098	-0.212720106260002\\
-0.096	-0.198354958342197\\
-0.094	-0.182923537119071\\
-0.092	-0.166598306633334\\
-0.09	-0.149552802727499\\
-0.088	-0.131960042987314\\
-0.086	-0.113991002541752\\
-0.084	-0.095813167487498\\
-0.082	-0.0775891765098593\\
-0.08	-0.0594755600178331\\
-0.078	-0.041621584813536\\
-0.076	-0.0241682109901616\\
-0.074	-0.00724716641245469\\
-0.072	0.00901985720651757\\
-0.07	0.0245218839452945\\
-0.0679999999999999	0.0391592076441487\\
-0.0659999999999999	0.0528438200811277\\
-0.0639999999999999	0.0654997144424497\\
-0.0620000000000001	0.0770630663287704\\
-0.0600000000000001	0.0874822956230446\\
-0.0580000000000001	0.0967180135529846\\
-0.056	0.10474286020448\\
-0.054	0.111541238575974\\
-0.052	0.117108952003045\\
-0.05	0.121452752423827\\
-0.048	0.124589807497054\\
-0.046	0.126547095024315\\
-0.044	0.127360733466282\\
-0.042	0.127075257580345\\
-0.04	0.125742848346038\\
-0.038	0.123422526387763\\
-0.036	0.120179318055382\\
-0.034	0.116083403186633\\
-0.032	0.111209253356361\\
-0.03	0.105634769121914\\
-0.028	0.0994404244081662\\
-0.026	0.0927084257462573\\
-0.024	0.0855218935943116\\
-0.022	0.0779640724336146\\
-0.02	0.0701175757574203\\
-0.018	0.0620636714594154\\
-0.016	0.0538816124924617\\
-0.014	0.0456480170131271\\
-0.012	0.0374363015610793\\
-0.01	0.0293161701517833\\
-0.00800000000000001	0.0213531614930169\\
-0.00600000000000001	0.0136082558769751\\
-0.004	0.00613754265631919\\
-0.002	-0.00100805140989494\\
0	-0.00778297325070289\\
};
\node[anchor=north west] at (axis cs:-0.01,0.9) {\includegraphics[width=8.5cm]{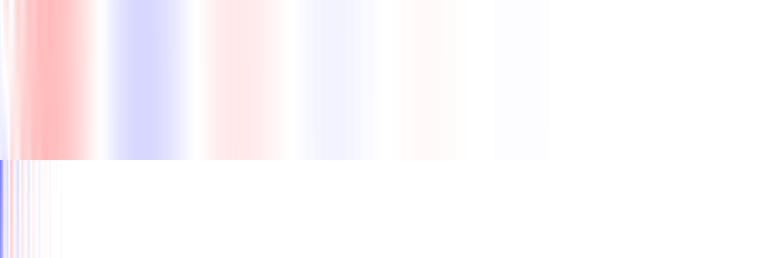}}; 
\draw[line width=2.0pt, draw=black!50] (axis cs:0,-1) rectangle (axis cs:1,0.8);
\draw[line width=2.0pt, draw=black] (axis cs:-1,-2) rectangle (axis cs:1,1.6);

\end{axis}

\begin{axis}[
  name = Strain2,
  mark layer=axis tick labels,axis on top,
  width=\figurewidth,
  height=0.3\figureheight,
  at={(0\figurewidth,0.325\figureheight)},
  scale only axis,
  point meta min=0,
  point meta max=1,
  xmin=-1,
  xmax=1,
  ymin=-2,
  ymax=1.6,
  ylabel={$\eta$\,[m]},
  xtick={-1,-0.5,0,0.5,1},
  ytick={-1,0,1},
xticklabel=\empty,
  axis background/.style={fill=white},
  axis y line*=left,
  ]
  \addplot [color=black, line width=3.0pt, forget plot]
  table[row sep=crcr]{%
-1	1.46903717080749e-05\\
-0.998	2.83400603390835e-06\\
-0.996	-9.66347318263438e-06\\
-0.994	-2.25203628343612e-05\\
-0.992	-3.54461825565006e-05\\
-0.99	-4.81467198500505e-05\\
-0.988	-6.03291636114024e-05\\
-0.986	-7.17072392462444e-05\\
-0.984	-8.2006258223906e-05\\
-0.982	-9.0967995961129e-05\\
-0.98	-9.83553144976874e-05\\
-0.978	-0.000103956450384277\\
-0.976	-0.000107588893535314\\
-0.974	-0.000109102789428903\\
-0.972	-0.000108383804794059\\
-0.97	-0.000105355405798199\\
-0.968	-9.99805074403177e-05\\
-0.966	-9.22624633559037e-05\\
-0.964	-8.22453762589769e-05\\
-0.962	-7.00137206763007e-05\\
-0.96	-5.56912812101463e-05\\
-0.958	-3.94394211763962e-05\\
-0.956	-2.14547078481891e-05\\
-0.954	-1.96593151152356e-06\\
-0.952	1.87694340504343e-05\\
-0.95	4.04692724121525e-05\\
-0.948	6.28313007433989e-05\\
-0.946	8.55379079117395e-05\\
-0.944	0.000108261274477858\\
-0.942	0.000130668689623915\\
-0.94	0.000152427976347816\\
-0.938	0.000173212934182028\\
-0.936	0.000192708708039281\\
-0.934	0.00021061699274128\\
-0.932	0.000226660985289685\\
-0.93	0.000240590000880045\\
-0.928	0.000252183674128392\\
-0.926	0.000261255673721386\\
-0.924	0.00026765686676098\\
-0.922	0.000271277878163026\\
-0.92	0.000272051000620648\\
-0.918	0.00026995142144791\\
-0.916	0.000264997744196335\\
-0.914	0.000257251794770227\\
-0.912	0.000246817713909603\\
-0.91	0.000233840350026057\\
-0.908	0.00021850297828568\\
-0.906	0.000201024383329956\\
-0.904	0.000181655353926041\\
-0.902	0.00016067464794192\\
-0.9	0.000138384495164028\\
-0.898	0.000115105713437834\\
-0.896	9.1172520382953e-05\\
-0.894	6.69271281807946e-05\\
-0.892	4.27142128177175e-05\\
-0.89	1.887535137993e-05\\
-0.888	-4.2564783186706e-06\\
-0.886	-2.6362242516267e-05\\
-0.884	-4.7141949978283e-05\\
-0.882	-6.63194176632164e-05\\
-0.88	-8.36466227563134e-05\\
-0.878	-9.8907567327127e-05\\
-0.876	-0.00011192159030427\\
-0.874	-0.000122546070999405\\
-0.872	-0.00013067847898107\\
-0.87	-0.000136257736499892\\
-0.868	-0.000139264871621098\\
-0.866	-0.000139722952689647\\
-0.864	-0.000137696307335359\\
-0.862	-0.000133289041825924\\
-0.86	-0.000126642888941469\\
-0.858	-0.000117934424417495\\
-0.856	-0.000107371703216965\\
-0.854	-9.51903772683769e-05\\
-0.852	-8.16493655190135e-05\\
-0.85	-6.7026155281462e-05\\
-0.848	-5.16118204257254e-05\\
-0.846	-3.57058471858993e-05\\
-0.844	-1.96108618761568e-05\\
-0.842	-3.62735661558317e-06\\
-0.84	1.19514906275819e-05\\
-0.838	2.68448070587762e-05\\
-0.836	4.07887358515134e-05\\
-0.834	5.35407804430351e-05\\
-0.832	6.48837027109163e-05\\
-0.83	7.46289035647278e-05\\
-0.828	8.26192239496453e-05\\
-0.826	8.8731115038864e-05\\
-0.824	9.28761380681555e-05\\
-0.822	9.50017669041266e-05\\
-0.82	9.50914795642017e-05\\
-0.818	9.31641385042338e-05\\
-0.816	8.9272673163309e-05\\
-0.814	8.35020918447071e-05\\
-0.812	7.59668632617347e-05\\
-0.81	6.68077206630817e-05\\
-0.808	5.61879533046815e-05\\
-0.806	4.42892607010899e-05\\
-0.804	3.13072545982283e-05\\
-0.802	1.7446701581095e-05\\
-0.8	2.91660558724848e-06\\
-0.798	-1.20747657770012e-05\\
-0.796	-2.73248045751483e-05\\
-0.794	-4.2641328160261e-05\\
-0.792	-5.78470810707482e-05\\
-0.79	-7.27838704333703e-05\\
-0.788	-8.73162374633645e-05\\
-0.786	-0.000101334574879498\\
-0.784	-0.000114757609248916\\
-0.782	-0.000127534178065582\\
-0.78	-0.000139644243931024\\
-0.778	-0.00015109910195266\\
-0.776	-0.000161940751617481\\
-0.774	-0.000172240420303873\\
-0.772	-0.000182096242341301\\
-0.77	-0.000191630114532558\\
-0.768	-0.000200983766337939\\
-0.766	-0.000210314099878937\\
-0.764	-0.000219787871503464\\
-0.762	-0.000229575802430293\\
-0.76	-0.000239846220638115\\
-0.758	-0.000250758349592177\\
-0.756	-0.000262455371039443\\
-0.754	-0.000275057399006571\\
-0.752	-0.000288654509893267\\
-0.75	-0.000303299979014814\\
-0.748	-0.000319003877000012\\
-0.746	-0.000335727179914271\\
-0.744	-0.0003533765447523\\
-0.742	-0.000371799897096116\\
-0.74	-0.000390782970046269\\
-0.738	-0.000410046923301326\\
-0.736	-0.000429247158330553\\
-0.734	-0.00044797343018743\\
-0.732	-0.000465751338907004\\
-0.73	-0.000482045263511501\\
-0.728	-0.000496262780075944\\
-0.726	-0.000507760581960438\\
-0.724	-0.00051585189579976\\
-0.722	-0.000519815361290504\\
-0.72	-0.000518905316699608\\
-0.718	-0.000512363405699847\\
-0.716	-0.000499431394912251\\
-0.714	-0.000479365065970682\\
-0.712	-0.000451449021241096\\
-0.71	-0.000415012219066058\\
-0.708	-0.000369444032860303\\
-0.706	-0.000314210608975749\\
-0.704	-0.000248871281373614\\
-0.702	-0.000173094786997296\\
-0.7	-8.66750147986875e-05\\
-0.698	1.04539862217543e-05\\
-0.696	0.000118204018576897\\
-0.694	0.00023632004621401\\
-0.692	0.000364369077773951\\
-0.69	0.000501730890323422\\
-0.688	0.00064759084876742\\
-0.686	0.000800935060168149\\
-0.684	0.000960548082159567\\
-0.682	0.00112501338072694\\
-0.68	0.00129271670501126\\
-0.678	0.00146185251557295\\
-0.676	0.00163043356821678\\
-0.674	0.00179630371811322\\
-0.672	0.00195715396908699\\
-0.67	0.00211054175093707\\
-0.668	0.00225391336389917\\
-0.666	0.00238462948442354\\
-0.664	0.00249999358078422\\
-0.662	0.00259728304116189\\
-0.66	0.00267378277139501\\
-0.658	0.00272682097503596\\
-0.656	0.00275380678528886\\
-0.654	0.00275226937749181\\
-0.652	0.0027198981524423\\
-0.65	0.00265458354577178\\
-0.648	0.00255445798718733\\
-0.646	0.00241793650628306\\
-0.644	0.00224375645922478\\
-0.642	0.0020310158334134\\
-0.64	0.00177920957564193\\
-0.638	0.00148826338360221\\
-0.636	0.00115856440123582\\
-0.634	0.000790988265572822\\
-0.632	0.000386921966595023\\
-0.63	-5.17179976158646e-05\\
-0.628	-0.00052247266051837\\
-0.626	-0.00102233327702122\\
-0.624	-0.00154774007773995\\
-0.622	-0.00209458745352575\\
-0.62	-0.00265823588161581\\
-0.618	-0.00323353084398829\\
-0.616	-0.00381482892190793\\
-0.614	-0.00439603117876697\\
-0.612	-0.0049706238665413\\
-0.61	-0.0055317264102193\\
-0.608	-0.00607214654000808\\
-0.606	-0.00658444235370592\\
-0.604	-0.00706099100219718\\
-0.602	-0.00749406360035259\\
-0.6	-0.00787590587467172\\
-0.598	-0.00819882396865533\\
-0.596	-0.00845527473820927\\
-0.594	-0.00863795978329942\\
-0.592	-0.0087399223796824\\
-0.59	-0.0087546463968784\\
-0.588	-0.00867615621665886\\
-0.586	-0.00849911660128982\\
-0.584	-0.00821893140361219\\
-0.582	-0.00783183996276411\\
-0.58	-0.00733500999096781\\
-0.578	-0.00672662572919172\\
-0.576	-0.00600597013356059\\
-0.574	-0.00517349985091664\\
-0.572	-0.00423091175160497\\
-0.57	-0.00318119981100884\\
-0.568	-0.00202870116911556\\
-0.566	-0.000779130249754917\\
-0.564	0.000560400111543399\\
-0.562	0.0019813714998176\\
-0.56	0.00347386758457035\\
-0.558	0.00502659737474246\\
-0.556	0.00662693469784296\\
-0.554	0.00826097439606108\\
-0.552	0.00991360557886743\\
-0.55	0.0115686021061033\\
-0.548	0.0132087302989316\\
-0.546	0.014815873689645\\
-0.544	0.0163711744265414\\
-0.542	0.0178551907486471\\
-0.54	0.0192480697386881\\
-0.538	0.0205297343534177\\
-0.536	0.0216800835203138\\
-0.534	0.0226792038809506\\
-0.532	0.0235075915564756\\
-0.53	0.0241463821119478\\
-0.528	0.0245775867064319\\
-0.526	0.0247843322372339\\
-0.524	0.0247511031221008\\
-0.522	0.0244639822152345\\
-0.52	0.0239108882241201\\
-0.518	0.023081806886954\\
-0.516	0.0219690130872725\\
-0.514	0.0205672810255167\\
-0.512	0.0188740795387914\\
-0.51	0.0168897496619216\\
-0.508	0.0146176615567294\\
-0.506	0.0120643480036364\\
-0.504	0.00923961175138145\\
-0.502	0.00615660415752537\\
-0.5	0.00283187272501772\\
-0.498	-0.000714624651696806\\
-0.496	-0.00445954067397\\
-0.494	-0.00837619652454378\\
-0.492	-0.0124346780050502\\
-0.49	-0.0166019709361489\\
-0.488	-0.0208421364587353\\
-0.486	-0.0251165264755933\\
-0.484	-0.0293840390499583\\
-0.482	-0.0336014131343625\\
-0.48	-0.0377235615436241\\
-0.478	-0.0417039406143031\\
-0.476	-0.0454949545140182\\
-0.474	-0.049048391682747\\
-0.472	-0.0523158904098667\\
-0.47	-0.0552494300808275\\
-0.468	-0.0578018441717024\\
-0.466	-0.059927350634273\\
-0.464	-0.061582094904793\\
-0.462	-0.0627247003919735\\
-0.46	-0.0633168209600325\\
-0.458	-0.0633236896265098\\
-0.456	-0.0627146574475324\\
-0.454	-0.0614637163705329\\
-0.452	-0.0595499997009844\\
-0.45	-0.0569582537598778\\
-0.448	-0.0536792743064103\\
-0.446	-0.0497103013689442\\
-0.444	-0.0450553662694098\\
-0.442	-0.0397255848439149\\
-0.44	-0.0337393911565447\\
-0.438	-0.0271227063745041\\
-0.436	-0.0199090379203677\\
-0.434	-0.0121395045397838\\
-0.432	-0.00386278351816496\\
-0.43	0.00486502305563185\\
-0.428	0.0139806053511814\\
-0.426	0.0234137477672087\\
-0.424	0.0330877165515906\\
-0.422	0.0429197329216087\\
-0.42	0.0528215306073122\\
-0.418	0.0626999957827228\\
-0.416	0.0724578863709539\\
-0.414	0.0819946267201776\\
-0.412	0.0912071726584185\\
-0.41	0.0999909409574682\\
-0.408	0.108240796281281\\
-0.406	0.115852087773725\\
-0.404	0.122721726566285\\
-0.402	0.128749294670033\\
-0.4	0.133838174969545\\
-0.398	0.137896691370641\\
-0.396	0.140839247579759\\
-0.394	0.142587452520702\\
-0.392	0.143071220033799\\
-0.39	0.142229830261942\\
-0.388	0.14001294001494\\
-0.386	0.136381529424719\\
-0.384	0.131308772364184\\
-0.382	0.124780818405815\\
-0.38	0.116797474544654\\
-0.378	0.107372775504979\\
-0.376	0.0965354321896414\\
-0.374	0.0843291487134369\\
-0.372	0.0708127994824117\\
-0.37	0.0560604589338955\\
-0.368	0.0401612778293874\\
-0.366	0.0232192013848447\\
-0.364	0.00535252601931877\\
-0.362	-0.0133067069094683\\
-0.36	-0.0326134803492222\\
-0.358	-0.0524107320171791\\
-0.356	-0.0725304425534814\\
-0.354	-0.0927948760352802\\
-0.352	-0.113017965717228\\
-0.35	-0.133006836069949\\
-0.348	-0.152563450425806\\
-0.346	-0.171486371831391\\
-0.344	-0.189572623069269\\
-0.342	-0.206619630268091\\
-0.34	-0.222427233090447\\
-0.338	-0.236799743191315\\
-0.336	-0.249548031495408\\
-0.334	-0.260491623866748\\
-0.332	-0.269460783954773\\
-0.33	-0.276298561412995\\
-0.328	-0.280862783311595\\
-0.326	-0.283027966415631\\
-0.324	-0.282687128084391\\
-0.322	-0.279753473871638\\
-0.32	-0.274161940475052\\
-0.318	-0.265870573497634\\
-0.316	-0.254861720543064\\
-0.314	-0.241143021466991\\
-0.312	-0.224748179140391\\
-0.31	-0.20573749583981\\
-0.308	-0.184198162350366\\
-0.306	-0.160244289035724\\
-0.304	-0.134016670477304\\
-0.302	-0.10568227779264\\
-0.3	-0.0754334753875531\\
-0.298	-0.0434869616541319\\
-0.296	-0.0100824359697306\\
-0.294	0.0245190027467754\\
-0.292	0.0600367177362025\\
-0.29	0.0961726452461178\\
-0.288	0.132613640520112\\
-0.286	0.169034042100032\\
-0.284	0.20509843488492\\
-0.282	0.240464589793971\\
-0.28	0.274786555446393\\
-0.278	0.307717875029048\\
-0.276	0.338914899502247\\
-0.274	0.368040166522602\\
-0.272	0.394765812964705\\
-0.27	0.418776987723752\\
-0.268	0.439775230599391\\
-0.266	0.457481782514451\\
-0.264	0.4716407921251\\
-0.262	0.482022384041873\\
-0.26	0.488425554411311\\
-0.258	0.490680860508498\\
-0.256	0.488652872261203\\
-0.254	0.482242355261656\\
-0.252	0.471388156813475\\
-0.25	0.456068768896001\\
-0.248	0.436303544589455\\
-0.246	0.412153547470968\\
-0.244	0.383722016739232\\
-0.242	0.35115443432599\\
-0.24	0.314638183974431\\
-0.238	0.274401796173168\\
-0.236	0.230713776892586\\
-0.234	0.183881022238216\\
-0.232	0.134246825371821\\
-0.23	0.0821884863116141\\
-0.228	0.0281145394642759\\
-0.226	-0.0275383820821424\\
-0.224	-0.0843090224088398\\
-0.222	-0.141715291753375\\
-0.22	-0.19925833791396\\
-0.218	-0.256426865881463\\
-0.216	-0.312701668322722\\
-0.214	-0.36756032667386\\
-0.212	-0.420482040095631\\
-0.21	-0.470952537427168\\
-0.208	-0.518469025579397\\
-0.206	-0.562545126559981\\
-0.204	-0.602715754539186\\
-0.202	-0.638541884066789\\
-0.2	-0.669615160745775\\
-0.198	-0.695562306365401\\
-0.196	-0.716049271695619\\
-0.194	-0.73078509184263\\
-0.192	-0.739525401252118\\
-0.19	-0.742075568107567\\
-0.188	-0.738293410985996\\
-0.186	-0.728091464177167\\
-0.184	-0.711438762014728\\
-0.182	-0.688362116873911\\
-0.18	-0.658946870121242\\
-0.178	-0.623337100213874\\
-0.176	-0.581735277292879\\
-0.174	-0.534401358946182\\
-0.172	-0.481651327280096\\
-0.17	-0.423855172978994\\
-0.168	-0.361434337594201\\
-0.166	-0.294858630828481\\
-0.164	-0.224642645013918\\
-0.162	-0.15134169426126\\
-0.16	-0.0755473108313233\\
-0.158	0.00211766391098491\\
-0.156	0.0810043521050841\\
-0.154	0.160443429581949\\
-0.152	0.239750983930447\\
-0.15	0.318234587482769\\
-0.148	0.3951995294932\\
-0.146	0.469955149547344\\
-0.144	0.541821212559253\\
-0.142	0.610134264615085\\
-0.14	0.674253908421321\\
-0.138	0.733568937224702\\
-0.136	0.787503266794849\\
-0.134	0.83552160639793\\
-0.132	0.877134811632802\\
-0.13	0.911904864535886\\
-0.128	0.939449429466984\\
-0.126	0.959445936938934\\
-0.124	0.971635151716791\\
-0.122	0.975824186149056\\
-0.12	0.971888924761003\\
-0.118	0.959775831590135\\
-0.116	0.939503117523669\\
-0.114	0.911161250951374\\
-0.112	0.874912801314328\\
-0.11	0.830991611548978\\
-0.108	0.779701301931975\\
-0.106	0.721413114358951\\
-0.104	0.656563112573463\\
-0.102	0.585648760234488\\
-0.1	0.509224904906659\\
-0.098	0.427899202012912\\
-0.096	0.342327018442378\\
-0.094	0.253205860798266\\
-0.092	0.161269378145421\\
-0.09	0.0672809935236608\\
-0.088	-0.0279727776160573\\
-0.086	-0.123689260561994\\
-0.084	-0.219056741847806\\
-0.082	-0.313261710767653\\
-0.08	-0.405496154084853\\
-0.078	-0.494964835315319\\
-0.076	-0.580892489839126\\
-0.074	-0.662530867644811\\
-0.072	-0.739165556737904\\
-0.07	-0.8101225221373\\
-0.0679999999999999	-0.874774297923128\\
-0.0659999999999999	-0.932545772963656\\
-0.0639999999999999	-0.982919514706065\\
-0.0620000000000001	-1.02544057972982\\
-0.0600000000000001	-1.05972076458954\\
-0.0580000000000001	-1.08544225576867\\
-0.056	-1.10236064327362\\
-0.054	-1.11030726846264\\
-0.052	-1.10919088306463\\
-0.05	-1.09899860293635\\
-0.048	-1.07979614686499\\
-0.046	-1.0517273575809\\
-0.044	-1.015013009031\\
-0.042	-0.969948910810439\\
-0.04	-0.916903327387287\\
-0.038	-0.856313736316955\\
-0.036	-0.788682955962474\\
-0.034	-0.714574679251388\\
-0.032	-0.634608455649669\\
-0.03	-0.549454168761691\\
-0.028	-0.459826061721311\\
-0.026	-0.366476366775702\\
-0.024	-0.270188599139607\\
-0.022	-0.171770578277586\\
-0.02	-0.0720472422264003\\
-0.018	0.028146677623777\\
-0.016	0.127974052728318\\
-0.014	0.226602755537273\\
-0.012	0.323212990771405\\
-0.01	0.417004491879668\\
-0.00800000000000001	0.50720351627015\\
-0.00600000000000001	0.593069575085516\\
-0.004	0.67390183608999\\
-0.002	0.749045141617112\\
0	0.817895587459335\\
};
\node[anchor=north west] at (axis cs:-0.01,0.9) {\includegraphics[width=8.5cm]{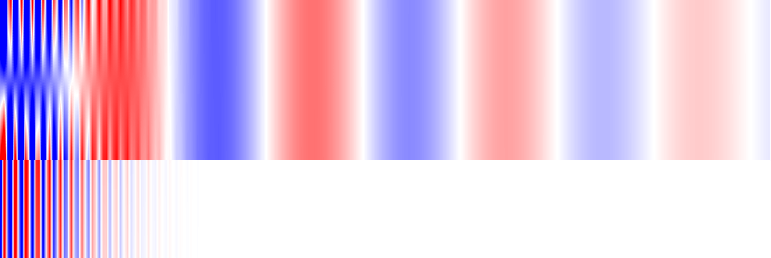}}; 
  \draw[line width=2.0pt, draw=black!50] (axis cs:0,-1) rectangle (axis cs:1,0.8);
  \draw[line width=2.0pt, draw=black] (axis cs:-1,-2) rectangle (axis cs:1,1.6);
  \end{axis}


\begin{axis}[
  name = Strain3,
  width=\figurewidth,
  height=0.3\figureheight,
  at={(0\figurewidth,0\figureheight)},
  scale only axis,
  point meta min=0,
  point meta max=1,
  xmin=-1,
  xmax=1,
  ymin=-2,
  ymax=1.6,
  xtick={-1,-0.5,0,0.5,1},
  ytick={-1,0,1},
xticklabels={{-2.5},{-1.25},{0},{125},{250}},
xlabel style={font=\color{white!15!black}},
xlabel={$x$\,[km]},
  ylabel={$\eta$\,[m]},
  axis background/.style={fill=white},
  axis y line*=left,
  ]
  \addplot [color=black, line width=3.0pt, forget plot]
  table[row sep=crcr]{%
-1	4.37845040139091e-05\\
-0.998	3.73913043532586e-05\\
-0.996	3.00682380583255e-05\\
-0.994	2.19736788522214e-05\\
-0.992	1.32803218198212e-05\\
-0.99	4.17225158660136e-06\\
-0.988	-5.15819446496164e-06\\
-0.986	-1.45137160846778e-05\\
-0.984	-2.36954384471196e-05\\
-0.982	-3.25063422920288e-05\\
-0.98	-4.07546783839666e-05\\
-0.978	-4.82573087863664e-05\\
-0.976	-5.4842918434445e-05\\
-0.974	-6.03550423623466e-05\\
-0.972	-6.46548568744229e-05\\
-0.97	-6.76236865839263e-05\\
-0.968	-6.91651838931192e-05\\
-0.966	-6.92071426916193e-05\\
-0.964	-6.77029140928181e-05\\
-0.962	-6.46323985078812e-05\\
-0.96	-6.00025953659987e-05\\
-0.958	-5.38476991184312e-05\\
-0.956	-4.62287377290168e-05\\
-0.954	-3.72327574905619e-05\\
-0.952	-2.6971565644088e-05\\
-0.95	-1.55800497388436e-05\\
-0.948	-3.21409986546786e-06\\
-0.946	9.95183336796661e-06\\
-0.944	2.37275059351879e-05\\
-0.942	3.7909927818241e-05\\
-0.94	5.22866410546185e-05\\
-0.938	6.66391746659614e-05\\
-0.936	8.07466194654088e-05\\
-0.934	9.43892635147047e-05\\
-0.932	0.000107352227709154\\
-0.93	0.00011942904067834\\
-0.928	0.000130425092939077\\
-0.926	0.000140160912018593\\
-0.924	0.000148475202947064\\
-0.922	0.000155227602256884\\
-0.92	0.000160301098156077\\
-0.918	0.000163604074941348\\
-0.916	0.00016507194581295\\
-0.914	0.000164668345001016\\
-0.912	0.000162385857350721\\
-0.91	0.000158246271177327\\
-0.908	0.000152300348118791\\
-0.906	0.000144627111765987\\
-0.904	0.000135332664930243\\
-0.902	0.000124548553359356\\
-0.9	0.000112429701318096\\
-0.898	9.91519518463505e-05\\
-0.896	8.49092511376773e-05\\
-0.894	6.99105227084978e-05\\
-0.892	5.43762823814603e-05\\
-0.89	3.85350496379013e-05\\
-0.888	2.26196146070288e-05\\
-0.886	6.86322253180705e-06\\
-0.884	-8.50426071272726e-06\\
-0.882	-2.32601379648451e-05\\
-0.88	-3.71925613685295e-05\\
-0.878	-5.01040421422835e-05\\
-0.876	-6.18147300984876e-05\\
-0.874	-7.21654071010284e-05\\
-0.872	-8.10201431809551e-05\\
-0.87	-8.82685695553515e-05\\
-0.868	-9.38277289791758e-05\\
-0.866	-9.76434708259876e-05\\
-0.864	-9.96913658027805e-05\\
-0.862	-9.99771231712706e-05\\
-0.86	-9.8536501636303e-05\\
-0.858	-9.54347135403294e-05\\
-0.856	-9.07653305094919e-05\\
-0.854	-8.46487071363632e-05\\
-0.852	-7.7229947427818e-05\\
-0.85	-6.86764465482556e-05\\
-0.848	-5.91750476943982e-05\\
-0.846	-4.89288605356115e-05\\
-0.844	-3.81537935761925e-05\\
-0.842	-2.70748578096404e-05\\
-0.84	-1.59223031419786e-05\\
-0.838	-4.92765211512956e-06\\
-0.836	5.68030251672308e-06\\
-0.834	1.5679468922364e-05\\
-0.832	2.48583288561904e-05\\
-0.83	3.30196460049409e-05\\
-0.828	3.99839467152706e-05\\
-0.826	4.55927132654722e-05\\
-0.824	4.97112345580764e-05\\
-0.822	5.22310645188237e-05\\
-0.82	5.30720449603361e-05\\
-0.818	5.21838567764717e-05\\
-0.816	4.95470711513137e-05\\
-0.814	4.51736807367822e-05\\
-0.812	3.91070994498586e-05\\
-0.81	3.1421628430889e-05\\
-0.808	2.22213946987832e-05\\
-0.806	1.16387779348849e-05\\
-0.804	-1.67650454268599e-07\\
-0.802	-1.30156435408464e-05\\
-0.8	-2.67022100636828e-05\\
-0.798	-4.10069131492766e-05\\
-0.796	-5.56954827083674e-05\\
-0.794	-7.05236821183296e-05\\
-0.792	-8.52413652587011e-05\\
-0.79	-9.95966564591294e-05\\
-0.788	-0.000113340183558835\\
-0.786	-0.000126229293035752\\
-0.784	-0.000138032176185431\\
-0.782	-0.00014853183646541\\
-0.78	-0.000157529830489722\\
-0.778	-0.000164849718689187\\
-0.776	-0.00017034016619304\\
-0.774	-0.000173877640172545\\
-0.772	-0.000175368656372609\\
-0.77	-0.000174751534997383\\
-0.768	-0.000171997634120821\\
-0.766	-0.000167112037514045\\
-0.764	-0.000160133682800406\\
-0.762	-0.000151134925240835\\
-0.76	-0.000140220541913677\\
-0.758	-0.000127526190482881\\
-0.756	-0.000113216346008688\\
-0.754	-9.74817481059261e-05\\
-0.752	-8.05363991898103e-05\\
-0.75	-6.26141621991295e-05\\
-0.748	-4.39650132300207e-05\\
-0.746	-2.48510104810151e-05\\
-0.744	-5.54204598888437e-06\\
-0.742	1.36885494333147e-05\\
-0.74	3.25684748772789e-05\\
-0.738	5.08309269127242e-05\\
-0.736	6.82189910145802e-05\\
-0.734	8.4489882804316e-05\\
-0.732	9.94189672373176e-05\\
-0.73	0.000112803487354325\\
-0.728	0.000124465938788775\\
-0.726	0.000134257031947625\\
-0.724	0.000142058190465683\\
-0.722	0.000147783542164846\\
-0.72	0.000151381367096695\\
-0.718	0.000152834976275686\\
-0.716	0.000152163004181768\\
-0.714	0.000149419107943119\\
-0.712	0.000144691076082192\\
-0.71	0.00013809935968682\\
-0.708	0.000129795048651381\\
-0.706	0.000119957325126822\\
-0.704	0.000108790435227128\\
-0.702	9.6520228387466e-05\\
-0.7	8.33903212526383e-05\\
-0.698	6.96579495451458e-05\\
-0.696	5.55895769139144e-05\\
-0.694	4.14563341422245e-05\\
-0.692	2.75293652654278e-05\\
-0.69	1.40751590550789e-05\\
-0.688	1.35094487715986e-06\\
-0.686	-1.03997688101338e-05\\
-0.684	-2.09514371309679e-05\\
-0.682	-3.01004298679808e-05\\
-0.68	-3.76686227373263e-05\\
-0.678	-4.35065665184497e-05\\
-0.676	-4.74961785242779e-05\\
-0.674	-4.95529087798913e-05\\
-0.672	-4.96273419393215e-05\\
-0.67	-4.77062053618654e-05\\
-0.668	-4.38127637013805e-05\\
-0.666	-3.80065906480812e-05\\
-0.664	-3.03827189793049e-05\\
-0.662	-2.10701805928723e-05\\
-0.66	-1.02299585968337e-05\\
-0.658	1.94761645127932e-06\\
-0.656	1.52459844681785e-05\\
-0.654	2.94259373923286e-05\\
-0.652	4.42296057185107e-05\\
-0.65	5.9384784266255e-05\\
-0.648	7.46095254048422e-05\\
-0.646	8.96169230923162e-05\\
-0.644	0.000104120007524295\\
-0.642	0.00011783666794407\\
-0.64	0.000130494520332617\\
-0.638	0.000141835637241235\\
-0.636	0.000151621058997628\\
-0.634	0.000159635008851997\\
-0.632	0.000165688739292907\\
-0.63	0.000169623942672302\\
-0.628	0.000171315666358436\\
-0.626	0.000170674680739768\\
-0.624	0.000167649257445244\\
-0.622	0.000162226324915788\\
-0.62	0.000154431978863303\\
-0.618	0.000144331335962337\\
-0.616	0.000132027730172578\\
-0.614	0.000117661262222106\\
-0.612	0.000101406723746136\\
-0.61	8.34709282779428e-05\\
-0.608	6.40894914414449e-05\\
-0.606	4.35231121919345e-05\\
-0.604	2.20534156447562e-05\\
-0.602	-2.15743185177702e-08\\
-0.6	-2.23922579156529e-05\\
-0.598	-4.47424977203023e-05\\
-0.596	-6.67547882511461e-05\\
-0.594	-8.81154702694009e-05\\
-0.592	-0.000108519902011978\\
-0.59	-0.000127677499983476\\
-0.588	-0.000145316563736782\\
-0.586	-0.000161188802412478\\
-0.584	-0.000175073485485956\\
-0.582	-0.000186781146228781\\
-0.58	-0.000196156773683332\\
-0.578	-0.000203082437332581\\
-0.576	-0.000207479298044046\\
-0.574	-0.000209308969080927\\
-0.572	-0.000208574201834093\\
-0.57	-0.000205318882281487\\
-0.568	-0.000199627335804512\\
-0.566	-0.000191622949716066\\
-0.564	-0.00018146613446104\\
-0.562	-0.000169351655755645\\
-0.56	-0.000155505380716424\\
-0.558	-0.000140180491175642\\
-0.556	-0.000123653226590437\\
-0.554	-0.000106218227214403\\
-0.552	-8.81835552474903e-05\\
-0.55	-6.9865477446402e-05\\
-0.548	-5.15830970572997e-05\\
-0.546	-3.3652925817686e-05\\
-0.544	-1.63834881488051e-05\\
-0.542	-7.00494405584361e-08\\
-0.54	1.50104414177687e-05\\
-0.538	2.86041084369552e-05\\
-0.536	4.04845218223037e-05\\
-0.534	5.04566123671885e-05\\
-0.532	5.83600586806852e-05\\
-0.53	6.40720877300307e-05\\
-0.528	6.75096389605108e-05\\
-0.526	6.86308529573419e-05\\
-0.524	6.74358569773659e-05\\
-0.522	6.39668316191295e-05\\
-0.52	5.83073551183571e-05\\
-0.518	5.05810341049892e-05\\
-0.516	4.09494419040131e-05\\
-0.514	2.96093973992167e-05\\
-0.512	1.67896289133367e-05\\
-0.51	2.7468782491831e-06\\
-0.508	-1.22384901250845e-05\\
-0.506	-2.78673010331674e-05\\
-0.504	-4.38267204601467e-05\\
-0.502	-5.97956599395992e-05\\
-0.5	-7.54503413767722e-05\\
-0.498	-9.04699271141198e-05\\
-0.496	-0.000104542118525021\\
-0.494	-0.000117368626578442\\
-0.492	-0.000128670419632799\\
-0.49	-0.000138192657162326\\
-0.488	-0.000145709223154158\\
-0.486	-0.000151026779474895\\
-0.484	-0.000153988267433563\\
-0.482	-0.000154475795019079\\
-0.48	-0.000152412857623676\\
-0.478	-0.000147765851376864\\
-0.476	-0.00014054485028199\\
-0.474	-0.000130803630948493\\
-0.472	-0.000118638941732232\\
-0.47	-0.000104189026135561\\
-0.468	-8.76314233590986e-05\\
-0.466	-6.91800815300946e-05\\
-0.464	-4.90818312847527e-05\\
-0.462	-2.76122787294743e-05\\
-0.46	-5.07118721413118e-06\\
-0.458	1.82225733721655e-05\\
-0.456	4.19364231435965e-05\\
-0.454	6.57297214452675e-05\\
-0.452	8.92595707007425e-05\\
-0.45	0.000112186673591141\\
-0.448	0.000134181140746792\\
-0.446	0.000154928146442111\\
-0.444	0.000174133331842593\\
-0.442	0.000191527859178727\\
-0.44	0.000206873025791534\\
-0.438	0.000219964354148189\\
-0.436	0.000230635082635961\\
-0.434	0.000238758992013526\\
-0.432	0.000244252513697009\\
-0.43	0.000247076078373515\\
-0.428	0.000247234676603203\\
-0.426	0.000244777616815716\\
-0.424	0.000239797480265943\\
-0.422	0.000232428286767611\\
-0.42	0.000222842899199066\\
-0.418	0.000211249708583149\\
-0.416	0.000197888654746163\\
-0.414	0.000183026649958264\\
-0.412	0.000166952484269165\\
-0.41	0.000149971301346578\\
-0.408	0.000132398742226653\\
-0.406	0.00011455486142605\\
-0.404	9.6757925135671e-05\\
-0.402	7.93182046218242e-05\\
-0.4	6.2531879444198e-05\\
-0.398	4.66751646145621e-05\\
-0.396	3.19987732835533e-05\\
-0.394	1.87228220868388e-05\\
-0.392	7.03227985439635e-06\\
-0.39	-2.9269478677546e-06\\
-0.388	-1.10512160254296e-05\\
-0.386	-1.72815480912944e-05\\
-0.384	-2.160519701353e-05\\
-0.382	-2.40563153272411e-05\\
-0.38	-2.47157058604826e-05\\
-0.378	-2.37096403445722e-05\\
-0.376	-2.12077496611722e-05\\
-0.374	-1.74200060285417e-05\\
-0.372	-1.25928340106851e-05\\
-0.37	-7.00440342242569e-06\\
-0.368	-9.59172820572877e-07\\
-0.366	5.21823300283185e-06\\
-0.364	1.1189714538978e-05\\
-0.362	1.66108488710595e-05\\
-0.36	2.11383809748456e-05\\
-0.358	2.44379263870432e-05\\
-0.356	2.61917498126733e-05\\
-0.354	2.61064795353596e-05\\
-0.352	2.39206150295396e-05\\
-0.35	1.94116850333744e-05\\
-0.348	1.24029155396394e-05\\
-0.346	2.76927175778751e-06\\
-0.344	-9.55725501727953e-06\\
-0.342	-2.45832354757917e-05\\
-0.34	-4.22503738500209e-05\\
-0.338	-6.24331580912734e-05\\
-0.336	-8.49376686162719e-05\\
-0.334	-0.000109501608664316\\
-0.332	-0.000135795601038268\\
-0.33	-0.00016342577652629\\
-0.328	-0.000191937658963331\\
-0.326	-0.000220821330940791\\
-0.324	-0.000249517842960586\\
-0.322	-0.000277426807617198\\
-0.32	-0.000303915099554924\\
-0.318	-0.00032832656175684\\
-0.316	-0.000349992599536721\\
-0.314	-0.000368243525682977\\
-0.312	-0.000382420503877261\\
-0.31	-0.000391887923008231\\
-0.308	-0.000396046022627219\\
-0.306	-0.000394343579690667\\
-0.304	-0.000386290459184237\\
-0.302	-0.000371469826317751\\
-0.3	-0.000349549815870608\\
-0.298	-0.000320294455052939\\
-0.296	-0.000283573639950104\\
-0.294	-0.000239371972274533\\
-0.292	-0.000187796272713514\\
-0.29	-0.000129081599578736\\
-0.288	-6.35956166136469e-05\\
-0.286	8.15882843477274e-06\\
-0.284	8.55430327080093e-05\\
-0.282	0.000167783770444922\\
-0.28	0.00025397670004666\\
-0.278	0.000343091891510088\\
-0.276	0.000433981492345109\\
-0.274	0.000525389521623936\\
-0.272	0.000615963752589343\\
-0.27	0.000704269614675896\\
-0.268	0.000788806016159712\\
-0.266	0.000868022959390226\\
-0.264	0.000940340791966953\\
-0.262	0.00100417090972226\\
-0.26	0.00105793770127696\\
-0.258	0.00110010149961059\\
-0.256	0.00112918228386002\\
-0.254	0.00114378385473247\\
-0.252	0.00114261818978925\\
-0.25	0.00112452967069432\\
-0.248	0.00108851886353999\\
-0.246	0.00103376552580248\\
-0.244	0.000959650509500827\\
-0.242	0.000865776229859261\\
-0.24	0.000751985372355755\\
-0.238	0.000618377518498852\\
-0.236	0.000465323382085419\\
-0.234	0.000293476363010571\\
-0.232	0.000103781144929351\\
-0.23	-0.000102520913937636\\
-0.228	-0.000323889820838427\\
-0.226	-0.000558488613158929\\
-0.224	-0.000804188807947885\\
-0.222	-0.00105857975989149\\
-0.22	-0.00131898201184229\\
-0.218	-0.00158246468000576\\
-0.216	-0.00184586687167357\\
-0.214	-0.00210582308734473\\
-0.212	-0.00235879251158613\\
-0.21	-0.00260109204845312\\
-0.208	-0.00282893290818778\\
-0.206	-0.00303846050267545\\
-0.204	-0.00322579735826885\\
-0.202	-0.0033870887065888\\
-0.2	-0.00351855036726792\\
-0.198	-0.00361651849188346\\
-0.196	-0.00367750069603048\\
-0.194	-0.00369822806715901\\
-0.192	-0.0036757074999756\\
-0.19	-0.00360727377942682\\
-0.188	-0.00349064080404091\\
-0.186	-0.00332395132022254\\
-0.184	-0.00310582452145594\\
-0.182	-0.00283540085573298\\
-0.18	-0.00251238338029975\\
-0.178	-0.0021370750054255\\
-0.176	-0.00171041097865349\\
-0.174	-0.00123398597822638\\
-0.172	-0.000710075209295793\\
-0.17	-0.000141648929339264\\
-0.168	0.000467620130001536\\
-0.166	0.00111335692862326\\
-0.164	0.00179049229836641\\
-0.162	0.00249327950063406\\
-0.16	0.00321531945145899\\
-0.158	0.00394959482466095\\
-0.156	0.00468851315726408\\
-0.154	0.00542395898847729\\
-0.152	0.00614735496493631\\
-0.15	0.00684973174125963\\
-0.148	0.00752180639713617\\
-0.146	0.00815406898104262\\
-0.144	0.00873687667734854\\
-0.142	0.00926055497908898\\
-0.14	0.00971550513432331\\
-0.138	0.0100923170209949\\
-0.136	0.010381886494977\\
-0.134	0.0105755361498972\\
-0.132	0.0106651383268894\\
-0.13	0.0106432391190916\\
-0.128	0.0105031820310204\\
-0.126	0.0102392298784198\\
-0.124	0.00984668345127261\\
-0.122	0.00932199541286332\\
-0.12	0.00866287787244029\\
-0.118	0.00786840204948642\\
-0.116	0.00693908844504648\\
-0.114	0.00587698595107279\\
-0.112	0.00468573836325203\\
-0.11	0.00337063681703909\\
-0.108	0.00193865674120439\\
-0.106	0.00039847801847129\\
-0.104	-0.00123951284110347\\
-0.102	-0.00296323957156825\\
-0.1	-0.00475897296779442\\
-0.098	-0.00661138837113979\\
-0.096	-0.00850364435217653\\
-0.094	-0.0104174829055109\\
-0.092	-0.0123333512425687\\
-0.09	-0.0142305450248207\\
-0.088	-0.0160873726243143\\
-0.086	-0.0178813397329266\\
-0.084	-0.0195893533690542\\
-0.082	-0.0211879440533498\\
-0.08	-0.0226535046466774\\
-0.078	-0.0239625440669553\\
-0.076	-0.0250919538303969\\
-0.074	-0.0260192851004384\\
-0.072	-0.0267230336779752\\
-0.07	-0.0271829301331631\\
-0.0679999999999999	-0.0273802320656584\\
-0.0659999999999999	-0.0272980152904767\\
-0.0639999999999999	-0.0269214605842283\\
-0.0620000000000001	-0.0262381324947878\\
-0.0600000000000001	-0.0252382466197639\\
-0.0580000000000001	-0.023914921698454\\
-0.056	-0.0222644128410561\\
-0.054	-0.0202863222401566\\
-0.052	-0.0179837837748973\\
-0.05	-0.0153636180293561\\
-0.048	-0.0124364544045927\\
-0.046	-0.00921681720911614\\
-0.044	-0.00572317286520994\\
-0.042	-0.00197793566805478\\
-0.04	0.00199257012028099\\
-0.038	0.00615819367092747\\
-0.036	0.0104850878809388\\
-0.034	0.0149358874884741\\
-0.032	0.0194699354368424\\
-0.03	0.0240435573800451\\
-0.028	0.0286103836631754\\
-0.026	0.0331217175317148\\
-0.024	0.037526947730336\\
-0.022	0.0417740030506885\\
-0.02	0.04580984578564\\
-0.018	0.0495810004517023\\
-0.016	0.0530341135593205\\
-0.014	0.0561165396499545\\
-0.012	0.0587769482871189\\
-0.01	0.0609659461935069\\
-0.00800000000000001	0.0626367082756051\\
-0.00600000000000001	0.0637456108782688\\
-0.004	0.0642528602717007\\
-0.002	0.0641231090989415\\
0	0.0633260533095864\\
  };
  \node[anchor=north west] at (axis cs:-0.01,0.9) {\includegraphics[width=8.5cm]{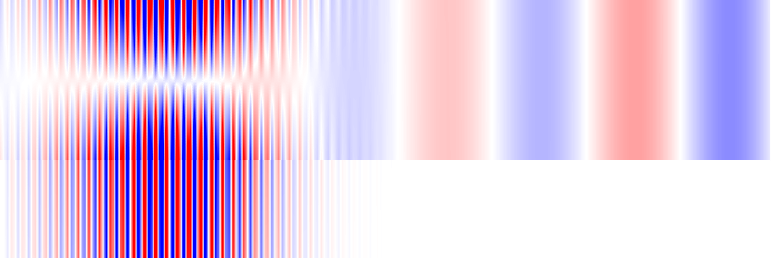}}; 
  \draw[line width=2.0pt, draw=black!50] (axis cs:0,-1) rectangle (axis cs:1,0.8);
  \draw[line width=2.0pt, draw=black] (axis cs:-1,-2) rectangle (axis cs:1,1.6);
  \end{axis}

\node[align=center,anchor=south,text opacity=1,text opacity=1,yshift=20pt,xshift=135pt,rotate=90]
at (Strain1.north)(bar) {\includegraphics[width=0.65cm]{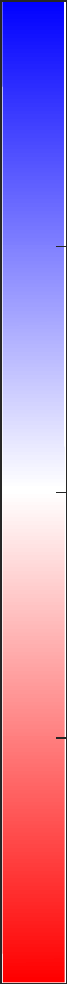}}; 
\node[align=center,anchor=east,text opacity=1,xshift=30pt,yshift=20pt]
at (bar.east) {$\epsilon\,[\times 10^7]$}; 
\node[align=center,anchor=east,text opacity=1,xshift=-130pt,yshift=5pt]
at (bar.east) {$-6$}; 
\node[align=center,anchor=east,text opacity=1,xshift=-62pt,yshift=5pt]
at (bar.east) {$-3$};
\node[align=center,anchor=east,text opacity=1,xshift=6pt,yshift=5pt]
at (bar.east) {$0$};
\node[align=center,anchor=east,text opacity=1,xshift=73pt,yshift=5pt]
at (bar.east) {$3$};
\node[align=center,anchor=east,text opacity=1,xshift=140pt,yshift=5pt]
at (bar.east) {$6$};

\node[align=left,anchor=east,text opacity=1,rounded corners,fill=none,opacity=0.6,text opacity=1,xshift=70pt,yshift=-10pt]
at (Strain1.north west) {(a)~$t = \phantom{0}20\,\text{s}$};
\node[align=left,anchor=east,text opacity=1,rounded corners,fill=none,opacity=0.6,text opacity=1,xshift=70pt,yshift=-10pt]
at (Strain2.north west) {(b)~$t = 100\,\text{s}$};
\node[align=left,anchor=east,text opacity=1,rounded corners,fill=none,opacity=0.6,text opacity=1,xshift=70pt,yshift=-10pt]
at (Strain3.north west) {(c)~$t = 200\,\text{s}$};
\end{tikzpicture}%


%% file: tikz/Fig4_Strain.tex
%
\pgfplotsset{compat=newest,plot coordinates/math parser=false,every axis legend/.append style={
at={(0.5,1.05)},
anchor=south,legend cell align=left, align=left, draw=white!15!black}}
%
%
\begin{tikzpicture}

\begin{axis}[%
grid=both, 
name = period,
width=\figurewidth,
height=0.4\figureheight,
at={(0\figurewidth,0.5\figureheight)},
scale only axis,
xmin=10,
xmax=40,
xlabel style={font=\color{white!15!black}},
xlabel={$T_{\text{peak}}\,[\text{s}]$},
xtick={10,15,20,25,30,35,40},
ymin=0,
ymax=1.2,
ytick={0,0.2,0.4,0.6,0.8,1},
ylabel style={font=\color{white!15!black}},
ylabel={{$\hat{\epsilon}^{\max}_{\bullet}$}},
axis background/.style={fill=white},
legend style={legend cell align=left, align=left, draw=white!15!black},
legend columns=3
]
\addplot [color=blue, line width=1.5pt]
  table[row sep=crcr]{%
  10	1.01895637009693\\
  11.56	1.02043754218482\\
  11.74	1.01992328715032\\
  11.86	1.01921895354449\\
  12.1	1.01722366038898\\
  12.28	1.01586305566939\\
  12.4	1.01533387768829\\
  12.52	1.015246222071\\
  12.7	1.01568172734852\\
  12.94	1.01644419865804\\
  13.06	1.01653590834098\\
  13.42	1.01640920699738\\
  13.54	1.01698638714076\\
  13.66	1.01823675592457\\
  13.9	1.02161034046652\\
  13.96	1.02220332510365\\
  14.02	1.02254546980787\\
  14.08	1.02256143691235\\
  14.14	1.02217588875033\\
  14.2	1.02131348765499\\
  14.26	1.01993759135273\\
  14.32	1.01816633914267\\
  14.56	1.01026446507068\\
  14.62	1.00886895370628\\
  14.68	1.00801909458795\\
  14.74	1.00787172240799\\
  14.8	1.00858367185871\\
  14.86	1.01026579238474\\
  14.92	1.01284499244002\\
  14.98	1.01620219523083\\
  15.04	1.02021832396343\\
  15.1	1.02477430184407\\
  15.16	1.02975105207904\\
  15.28	1.04049056243697\\
  15.4	1.05148424068737\\
  15.46	1.05678148396426\\
  15.52	1.0618017378912\\
  15.58	1.06644262473262\\
  15.64	1.07060176675291\\
  15.7	1.07417678621649\\
  15.76	1.07706530538778\\
  15.82	1.07916494653117\\
  15.88	1.08037333191109\\
  15.94	1.08058808379194\\
  16	1.07970682443813\\
  16.06	1.07768747057862\\
  16.12	1.07472911680051\\
  16.18	1.07109115215542\\
  16.3	1.06281394647084\\
  16.36	1.05869348353463\\
  16.42	1.05493096593798\\
  16.48	1.05178578273252\\
  16.54	1.0495173229699\\
  16.6	1.04838497570174\\
  16.66	1.04856734562153\\
  16.72	1.04991989999014\\
  16.78	1.05221732171031\\
  16.84	1.05523429368476\\
  16.9	1.05874549881621\\
  17.08	1.06999134217988\\
  17.14	1.07322630896663\\
  17.2	1.07582892342403\\
  17.26	1.07763129100701\\
  17.32	1.07869520737943\\
  17.38	1.07913989075733\\
  17.44	1.07908455935679\\
  17.5	1.07864843139385\\
  17.62	1.07711065864508\\
  17.74	1.07548031823948\\
  17.8	1.07492848070553\\
  17.86	1.07468632661603\\
  17.92	1.0747489277394\\
  17.98	1.07508652655453\\
  18.04	1.0756693655403\\
  18.16	1.07745173393936\\
  18.28	1.07985797276767\\
  18.46	1.08411927417855\\
  18.88	1.09442179671511\\
  19	1.09700804825913\\
  19.12	1.09928250905472\\
  19.24	1.10127089058044\\
  19.42	1.10382691080078\\
  19.66	1.10670009372835\\
  19.84	1.10853545776668\\
  20.02	1.10992007831778\\
  20.14	1.11049681775747\\
  20.26	1.1107319545054\\
  20.38	1.11063080141617\\
  20.56	1.10999276573202\\
  20.8	1.10859223798559\\
  21.16	1.10608073777785\\
  21.46	1.10362240479237\\
  22	1.0989467959846\\
  22.36	1.09629579234434\\
  22.48	1.09506275291106\\
  22.6	1.09343786614987\\
  22.9	1.08844290718257\\
  22.96	1.08773730729795\\
  23.02	1.08727480776841\\
  23.08	1.08711768133057\\
  23.14	1.08732820072105\\
  23.2	1.08796863867649\\
  23.26	1.08906941806987\\
  23.32	1.09053356231973\\
  23.44	1.09403663960854\\
  23.56	1.09744725898214\\
  23.62	1.09879583083801\\
  23.68	1.09973480887979\\
  23.74	1.10013536666239\\
  23.8	1.09986867774073\\
  23.86	1.09885383513198\\
  23.92	1.09720160970249\\
  23.98	1.09507069178085\\
  24.1	1.09000753977563\\
  24.22	1.08493390174522\\
  24.28	1.08278987629209\\
  24.34	1.08111930031851\\
  24.4	1.08008086415309\\
  24.46	1.07978963586065\\
  24.52	1.08018619445092\\
  24.58	1.08116749666983\\
  24.64	1.08263049926331\\
  24.7	1.08447215897731\\
  24.82	1.08887927675057\\
  24.94	1.0935645039571\\
  25	1.09575380046269\\
  25.06	1.09772405345828\\
  25.12	1.09947501415922\\
  25.24	1.10240029911407\\
  25.36	1.10469213620106\\
  25.48	1.106513006294\\
  25.66	1.10871434541781\\
  25.96	1.11183319453944\\
  26.26	1.11498616620746\\
  26.56	1.11831946122906\\
  26.74	1.11988318606424\\
  26.86	1.12055023033897\\
  26.98	1.12089651605247\\
  27.16	1.12099134600573\\
  28	1.12029715892456\\
  28.36	1.11952585175577\\
  28.66	1.11898816534203\\
  28.96	1.1185038569585\\
  29.08	1.11801128885201\\
  29.2	1.11719123642568\\
  29.32	1.11597397477868\\
  29.44	1.11442878844561\\
  29.62	1.11172373086402\\
  29.98	1.10613572690801\\
  32.14	1.07467386996251\\
  32.98	1.0622640729493\\
  33.34	1.05697175014582\\
  33.64	1.0530446761823\\
  33.94	1.04907483639712\\
  34.12	1.04633284335311\\
  34.84	1.03468244313732\\
  35.2	1.02930401062882\\
  35.5	1.02463548374491\\
  35.62	1.02315219717322\\
  35.74	1.0221409092781\\
  35.86	1.02176039687158\\
  35.98	1.02194964989965\\
  36.34	1.02308980858761\\
  36.46	1.02270054019169\\
  36.58	1.02163187239608\\
  36.7	1.01995637423961\\
  36.82	1.01775433650388\\
  36.94	1.01510604997048\\
  37.06	1.01209682377152\\
  37.3	1.00591389921448\\
  37.42	1.00337752571843\\
  37.48	1.00238862374377\\
  37.54	1.00163944021907\\
  37.6	1.00117012159509\\
  37.66	1.00100655416699\\
  37.72	1.00111758360728\\
  37.84	1.00198177516047\\
  37.96	1.00339938038959\\
  38.2	1.00644156841535\\
  38.32	1.00739450766203\\
  38.44	1.00777752620556\\
  38.56	1.00755723726208\\
  38.68	1.00670025404771\\
  38.8	1.00517318977862\\
  38.92	1.00298875726929\\
  39.04	1.00034406772771\\
  39.34	0.993314790274454\\
  39.46	0.990974486220111\\
  39.52	0.990026955366176\\
  39.58	0.989268373762137\\
  39.64	0.988729142758778\\
  39.7	0.988439663706892\\
  39.76	0.988430337957283\\
  39.82	0.988731566860743\\
  39.88	0.989373751768056\\
  39.94	0.990387294030036\\
  40	0.991802594997459\\
};
\addlegendentry{$\hat{\epsilon}^{\max}$}

\addplot [color=red, line width=1.5pt]
  table[row sep=crcr]{%
  10	0.623814687540779\\
  10.3	0.617533848461051\\
  10.6	0.610921453340048\\
  10.9	0.603984137758779\\
  11.2	0.596728537298254\\
  11.44	0.59064203314351\\
  11.62	0.585741424705553\\
  11.8	0.580393433693025\\
  11.98	0.574560419107527\\
  12.22	0.566714767665822\\
  12.34	0.563090948278443\\
  12.46	0.559854755854346\\
  12.58	0.557002206028123\\
  12.82	0.551851291023162\\
  13	0.547867154379077\\
  13.24	0.541901064230416\\
  13.36	0.539007099556343\\
  13.48	0.536462052038388\\
  13.6	0.534483069343516\\
  13.72	0.533159532917161\\
  13.9	0.531440830560619\\
  13.96	0.530664298886187\\
  14.02	0.529671483087689\\
  14.08	0.528393701957363\\
  14.14	0.526762274287449\\
  14.2	0.524708518870185\\
  14.26	0.522202725559296\\
  14.38	0.516378692716948\\
  14.5	0.51057236598843\\
  14.56	0.508088705319835\\
  14.62	0.506104906663275\\
  14.68	0.504786115179932\\
  14.74	0.504297476031006\\
  14.8	0.504804134377679\\
  14.86	0.506423083600531\\
  14.92	0.50907870995767\\
  14.98	0.512647247926587\\
  15.04	0.517004931984786\\
  15.1	0.522027996609758\\
  15.16	0.527592676278992\\
  15.22	0.533575205469987\\
  15.34	0.546298750327232\\
  15.46	0.559212724234527\\
  15.52	0.565457538828312\\
  15.58	0.57142821660581\\
  15.64	0.577026295443005\\
  15.7	0.582153313215898\\
  15.76	0.586710807800465\\
  15.82	0.590600317072706\\
  15.88	0.593723378908599\\
  15.94	0.595981531184144\\
  16	0.59727631177531\\
  16.06	0.59756648375037\\
  16.12	0.597039710946611\\
  16.18	0.595940882393585\\
  16.36	0.591660952534539\\
  16.42	0.590722791280058\\
  16.48	0.590437019424115\\
  16.54	0.591048525996257\\
  16.6	0.59280220002605\\
  16.66	0.595864977492937\\
  16.72	0.600091982175854\\
  16.78	0.605260384803643\\
  16.84	0.611147356105128\\
  16.9	0.617530066809138\\
  17.08	0.637424342624627\\
  17.14	0.643561718227041\\
  17.2	0.649080686876118\\
  17.26	0.65381511238953\\
  17.32	0.657825630940202\\
  17.38	0.661229571789903\\
  17.44	0.664144264200388\\
  17.5	0.666687037433427\\
  17.62	0.671126143414199\\
  17.74	0.67548552382631\\
  17.8	0.677928640098528\\
  17.86	0.680679213472061\\
  17.92	0.683731576749622\\
  17.98	0.687055463442128\\
  18.04	0.690620607060488\\
  18.16	0.698353599118406\\
  18.28	0.706688421010675\\
  18.46	0.719792369735764\\
  18.82	0.746161939498499\\
  18.94	0.754616840995752\\
  19.06	0.762739604721077\\
  19.18	0.770501925917671\\
  19.3	0.777944786227138\\
  19.48	0.788606324607997\\
  19.66	0.798794683234448\\
  19.84	0.808544910505901\\
  19.96	0.814734480961924\\
  20.08	0.82062277902471\\
  20.2	0.82616454108291\\
  20.32	0.831330781283697\\
  20.44	0.836157624808457\\
  20.56	0.840697474597143\\
  20.74	0.847083767456148\\
  20.92	0.853109451501226\\
  21.1	0.858829688662546\\
  21.28	0.864227345033356\\
  21.46	0.869285048623134\\
  21.7	0.875585237111949\\
  22.24	0.889236270126297\\
  22.36	0.892007971728901\\
  22.48	0.89444095597942\\
  22.6	0.896406343417013\\
  22.72	0.897880013817691\\
  22.9	0.900065167568307\\
  22.96	0.90103662963849\\
  23.02	0.902233729610366\\
  23.08	0.903718926978954\\
  23.14	0.905554681239281\\
  23.2	0.907803451886366\\
  23.26	0.910495844678103\\
  23.32	0.913535050423924\\
  23.56	0.926586744393951\\
  23.62	0.929430070994641\\
  23.68	0.931848232983199\\
  23.74	0.933712567431925\\
  23.8	0.934894411413119\\
  23.86	0.935313022209044\\
  23.92	0.935079337941808\\
  23.98	0.934352216943495\\
  24.1	0.932053098081916\\
  24.22	0.929686532280904\\
  24.28	0.928875102608302\\
  24.34	0.928523386197064\\
  24.4	0.928790241379275\\
  24.46	0.929790899418556\\
  24.52	0.931466083304763\\
  24.58	0.933712888959306\\
  24.64	0.936428412303584\\
  24.7	0.939509749259017\\
  24.82	0.946358247688948\\
  24.94	0.953435151620361\\
  25	0.956801995452643\\
  25.06	0.959937786236161\\
  25.12	0.962842408950564\\
  25.18	0.965536306387129\\
  25.3	0.970373696591942\\
  25.42	0.974613499180904\\
  25.54	0.978419256484337\\
  25.72	0.983653126243901\\
  25.96	0.990178079231711\\
  26.62	1.00764032206939\\
  26.74	1.01041847606285\\
  26.86	1.01286346424286\\
  26.98	1.0149694712329\\
  27.16	1.01765786578226\\
  27.46	1.02153495551907\\
  27.88	1.02667042865463\\
  28.12	1.0292085641786\\
  28.54	1.0330855547186\\
  28.96	1.03685161613306\\
  29.08	1.03757934636572\\
  29.2	1.03797531812251\\
  29.32	1.03797510440421\\
  29.44	1.03764342243731\\
  29.62	1.03673659120219\\
  30.22	1.03327013693419\\
  30.64	1.03077478673007\\
  31.12	1.02747558859519\\
  31.66	1.02343119045041\\
  32.26	1.0185178467989\\
  32.74	1.01429874830522\\
  33.52	1.00709158718902\\
  33.88	1.00415101622981\\
  34.06	1.00234240587982\\
  34.3	0.999453009773831\\
  34.66	0.995003858846253\\
  35.62	0.98437205304959\\
  35.74	0.983771329204181\\
  35.86	0.983793263217144\\
  35.98	0.984377015897465\\
  36.28	0.986431790530588\\
  36.4	0.986730108558397\\
  36.52	0.986355717225379\\
  36.64	0.985327281986429\\
  36.76	0.983725239406347\\
  36.88	0.981630026049956\\
  37	0.979122078482042\\
  37.36	0.971010089985853\\
  37.42	0.969976466238627\\
  37.48	0.969140907207255\\
  37.54	0.968543575624253\\
  37.6	0.968224634222118\\
  37.66	0.968209985308008\\
  37.72	0.968468489487584\\
  37.84	0.969623357849066\\
  37.96	0.971326040749098\\
  38.2	0.974922055934805\\
  38.32	0.976143977246657\\
  38.44	0.976790835713942\\
  38.56	0.97682934962603\\
  38.68	0.976226237272279\\
  38.8	0.974948216942053\\
  38.92	0.973008104327754\\
  39.04	0.970603104733875\\
  39.34	0.964153399231712\\
  39.46	0.962037304031128\\
  39.52	0.961200290713236\\
  39.58	0.960551184034863\\
  39.64	0.960120396834455\\
  39.7	0.959938341950448\\
  39.76	0.960035432221282\\
  39.82	0.960442080485407\\
  39.88	0.961188699581264\\
  39.94	0.962305702347287\\
  40	0.963823501621917\\
};
\addlegendentry{$\hat{\epsilon}_{\text{flex}}^{\max}$}

\addplot [color=green, line width=1.5pt]
  table[row sep=crcr]{%
  10	0.395141682556158\\
  10.3	0.401736671540924\\
  10.6	0.408688687346235\\
  10.9	0.415979867686119\\
  11.26	0.425151338902509\\
  11.92	0.442223327963049\\
  12.82	0.464255270838521\\
  13.3	0.475870014010255\\
  14.02	0.492873986720177\\
  14.2	0.496604968784808\\
  14.32	0.498795270688191\\
  14.44	0.500674362260547\\
  14.56	0.502175759750841\\
  14.68	0.503232979408018\\
  14.8	0.50377953748103\\
  14.92	0.503766282482353\\
  15.04	0.50321339197864\\
  15.16	0.502158375800043\\
  15.28	0.50063874377674\\
  15.4	0.498692005738903\\
  15.52	0.496344199062897\\
  15.64	0.493575471309903\\
  15.76	0.490354497587305\\
  15.88	0.486649953002484\\
  16	0.482430512662823\\
  16.12	0.477689405853901\\
  16.24	0.472518078574119\\
  16.42	0.464208174657919\\
  17.08	0.432566999555256\\
  17.38	0.417910318967429\\
  17.98	0.3880310631124\\
  18.46	0.364326904442791\\
  18.82	0.346870873687372\\
  19.18	0.329807395335429\\
  19.48	0.315983953500492\\
  19.72	0.305250642097874\\
  20.02	0.29220096029961\\
  20.44	0.274316816048447\\
  20.74	0.26189194264164\\
  20.98	0.252334310031507\\
  21.22	0.243154787005359\\
  21.46	0.234337356169235\\
  21.7	0.225867761791598\\
  21.94	0.217735890991186\\
  22.18	0.209931987141054\\
  22.42	0.202445251496037\\
  22.72	0.19351556778291\\
  23.02	0.185041078158044\\
  23.32	0.17699851189581\\
  23.62	0.169365759843366\\
  23.92	0.16212227176068\\
  24.22	0.155247369464313\\
  24.52	0.148720111146154\\
  24.82	0.142521029061619\\
  25.18	0.135490686287611\\
  25.54	0.128878347506436\\
  25.9	0.122655974590614\\
  26.26	0.116799801772622\\
  26.68	0.110360078584499\\
  27.22	0.102484030424101\\
  27.58	0.0975897190194743\\
  28	0.0923074662219321\\
  28.42	0.0873994498973829\\
  28.84	0.0828845775246947\\
  29.68	0.074397122405685\\
  30.1	0.0704644429304224\\
  30.58	0.0663526452576875\\
  31.12	0.0621184510425081\\
  31.66	0.0582741659872994\\
  32.26	0.0544071497922829\\
  32.86	0.0509221759050433\\
  33.52	0.0474811387031338\\
  34.18	0.0444082962834145\\
  34.9	0.0414240941170974\\
  35.68	0.0385738362527448\\
  36.52	0.0358913610374643\\
  37.42	0.0334010594798144\\
  38.38	0.0311179742528935\\
  39.46	0.028937182188983\\
  40	0.0279790933755422\\
};
\addlegendentry{$\hat{\epsilon}_{\text{ext}}^{\max}$}

\end{axis}

\begin{axis}[%
grid = both,
name = width,
width=\figurewidth,
height=0.4\figureheight,
at={(0\figurewidth,0\figureheight)},
scale only axis,
xmin=5,
xmax=35,
xlabel style={font=\color{white!15!black}},
xlabel={$\sigma\,[\times 10^4$\,m$^{-1}]$},
xtick = {5,10,15,20,25,30,35},
ymin=0.4,
ymax=1.4,
ytick = {0.4,0.6,...,1.2},
ylabel style={font=\color{white!15!black}},
ylabel={{$\hat{\epsilon}^{\max}_{\bullet}$}},
axis background/.style={fill=white}
]
\addplot [color=blue, line width=1.5pt, forget plot]
  table[row sep=crcr]{%
3.5	1.00695761013848\\
3.563	1.01862475066382\\
3.626	1.02722819043901\\
3.689	1.03303469027989\\
3.752	1.03631101100225\\
3.815	1.03732391342193\\
3.878	1.03634015835473\\
3.941	1.03362650661646\\
4.004	1.02944971902295\\
4.067	1.02407655639001\\
4.13	1.01777377953344\\
4.193	1.01080814926907\\
4.445	0.981652310449853\\
4.508	0.975373825383684\\
4.571	0.970033051804606\\
4.634	0.965896750528422\\
4.697	0.963231682370953\\
4.76	0.962304608148017\\
4.823	0.963281328042115\\
4.886	0.96592379970258\\
4.949	0.969893020145406\\
5.012	0.974849986386623\\
5.138	0.986371144328253\\
5.201	0.992257330060696\\
5.264	0.997775249655568\\
5.327	1.0025859001289\\
5.39	1.00635027849668\\
5.453	1.00882102730093\\
5.516	1.01011737118759\\
5.579	1.0104501803286\\
5.642	1.01003032489589\\
5.768	1.00777610099706\\
5.894	1.0050416608666\\
5.957	1.00402153514435\\
6.02	1.00351396587999\\
6.083	1.00367436346166\\
6.146	1.00443629914226\\
6.209	1.00567788439089\\
6.335	1.00911244946859\\
6.524	1.01481445557268\\
6.587	1.01637427908042\\
6.65	1.01756053243985\\
6.713	1.01828419022405\\
6.776	1.01858767942207\\
6.902	1.01823531243175\\
7.091	1.01643774987602\\
7.217	1.01527175121088\\
7.343	1.0148111942882\\
7.532	1.01537931296974\\
7.721	1.01597080697267\\
7.847	1.01552863928666\\
7.91	1.01486547259231\\
8.036	1.01250399191933\\
8.162	1.00911958829203\\
8.54	0.997786419420798\\
8.666	0.995211594440043\\
8.792	0.993754780736595\\
8.855	0.993494972594171\\
8.918	0.993573880647453\\
8.981	0.994011242688558\\
9.044	0.994826796509614\\
9.107	0.996040279902743\\
9.17	0.997671430660063\\
9.233	0.999721238920166\\
9.359	1.00476207839273\\
9.674	1.01852639826922\\
9.737	1.02073651513048\\
9.8	1.0225493039857\\
9.863	1.02389898292038\\
9.926	1.02482763487859\\
10.052	1.02569805450042\\
10.178	1.02571295612088\\
10.43	1.02538577843639\\
10.556	1.02601681564059\\
10.745	1.02794667846702\\
11.123	1.03241215689965\\
11.312	1.03374558426285\\
11.564	1.03472774485949\\
12.068	1.03577833203583\\
12.446	1.03599495268365\\
12.887	1.03546183521422\\
13.391	1.03419739488592\\
13.895	1.03225383790963\\
14.84	1.0279304879741\\
15.218	1.02463172257666\\
15.344	1.02421209473302\\
15.47	1.02458927998535\\
15.596	1.02592767772877\\
15.785	1.0291030126919\\
16.037	1.03350003547641\\
16.163	1.03491787529443\\
16.289	1.03563398384986\\
16.478	1.0359543346157\\
16.73	1.03638417987008\\
16.919	1.03765775282519\\
17.171	1.04030756632289\\
17.486	1.04368589185825\\
17.864	1.04687672746198\\
18.746	1.0536636537069\\
19.061	1.05569568551733\\
19.25	1.05755476753475\\
19.439	1.0602233504872\\
19.754	1.06568887535834\\
20.069	1.07093659053275\\
20.447	1.07633499140317\\
21.203	1.08627244198399\\
21.77	1.09297459880993\\
22.463	1.1004891515865\\
23.282	1.10865833989688\\
24.416	1.11921290581662\\
26.432	1.13783299785858\\
26.81	1.14135292661223\\
27.251	1.14517066131366\\
27.44	1.14767937907843\\
27.629	1.15117548858561\\
27.881	1.15686506545644\\
28.322	1.16705965165209\\
29.204	1.18598756305385\\
30.212	1.20739744395899\\
30.527	1.21430792204412\\
30.779	1.22047311592219\\
31.22	1.23214389013278\\
33.047	1.28081615198219\\
33.236	1.2868850050866\\
33.488	1.29593110554043\\
35	1.35289453944409\\
};
\addplot [color=red, line width=1.5pt, forget plot]
  table[row sep=crcr]{%
3.5	0.528597640812897\\
3.563	0.540256299621383\\
3.626	0.548851105925458\\
3.689	0.554648820510899\\
3.752	0.55791620416349\\
3.815	0.558920017669017\\
3.878	0.557927021813263\\
3.941	0.555203977382007\\
4.004	0.551017645161032\\
4.067	0.545634785936123\\
4.13	0.539322160493057\\
4.193	0.532346529617627\\
4.445	0.50314916750785\\
4.508	0.496859921257304\\
4.571	0.491508234289086\\
4.634	0.487360867388972\\
4.697	0.484684581342755\\
4.76	0.483746136936212\\
4.823	0.484711334320892\\
4.886	0.487342131111447\\
4.949	0.491299524288287\\
5.012	0.496244510831829\\
5.138	0.507741251940686\\
5.201	0.513615000466828\\
5.264	0.519120330281339\\
5.327	0.523918238364637\\
5.39	0.527669721697123\\
5.453	0.530127422784531\\
5.516	0.531410566233824\\
5.579	0.531730022177257\\
5.642	0.531296660747103\\
5.768	0.529014966295108\\
5.894	0.526252443935974\\
5.957	0.525218047621898\\
6.02	0.524696054727833\\
6.083	0.524841875601503\\
6.146	0.525589081452416\\
6.209	0.52681578370553\\
6.335	0.530220123118191\\
6.524	0.535875640877656\\
6.587	0.537419661468107\\
6.65	0.538589958435466\\
6.713	0.539297506307967\\
6.776	0.539584732026889\\
6.902	0.539199377182243\\
7.091	0.53735117987739\\
7.217	0.536150655502048\\
7.343	0.535654957047271\\
7.532	0.536169207872668\\
7.721	0.536705446082593\\
7.847	0.536225669420467\\
7.91	0.535543466545256\\
8.036	0.53314344985489\\
8.162	0.529719891623223\\
8.54	0.518265542841007\\
8.666	0.515649084326178\\
8.792	0.514150016179087\\
8.855	0.513868847819943\\
8.918	0.513926240244537\\
8.981	0.514341931183466\\
9.044	0.515135658367342\\
9.107	0.516327159526746\\
9.17	0.517936172392304\\
9.233	0.519963687040331\\
9.359	0.524959471866772\\
9.674	0.538608425428407\\
9.737	0.540795000551689\\
9.8	0.542584091357043\\
9.863	0.543909915863445\\
9.926	0.544814556945596\\
10.052	0.545636485191793\\
10.178	0.545602268804821\\
10.43	0.545174972968219\\
10.556	0.545755008901189\\
10.745	0.547607190434164\\
11.123	0.551913053136708\\
11.312	0.553164542011338\\
11.564	0.554035237044772\\
12.068	0.554855282445303\\
12.446	0.554892316912735\\
12.887	0.554142419039763\\
13.391	0.552620615456192\\
13.895	0.550409396200735\\
14.777	0.546087435402207\\
15.029	0.543671011641237\\
15.218	0.542035316441712\\
15.344	0.541540322357527\\
15.47	0.54184148579737\\
15.596	0.543103205354754\\
15.785	0.546162290553589\\
16.037	0.550402008714045\\
16.163	0.551740206546498\\
16.289	0.552376012234852\\
16.478	0.552574667724976\\
16.73	0.552839931750121\\
16.919	0.553988324481793\\
17.171	0.556468899845498\\
17.486	0.55963192162713\\
17.864	0.562558866460691\\
18.746	0.568706478639591\\
19.061	0.570502143715082\\
19.25	0.572217365401563\\
19.439	0.574740554654625\\
19.754	0.579960343276248\\
20.069	0.584958046769415\\
20.447	0.590050771743378\\
21.14	0.598624568271866\\
21.707	0.60491060578638\\
22.4	0.611869055550052\\
23.156	0.61878316493312\\
24.164	0.627292211855533\\
27.377	0.653746293549986\\
27.503	0.655607157061496\\
27.692	0.659173338541876\\
28.007	0.666168485059607\\
28.322	0.672996649239934\\
29.204	0.690897106939737\\
30.212	0.71110513932242\\
30.527	0.717635114059142\\
30.779	0.72349445823027\\
31.283	0.736222030040203\\
32.984	0.779272155296063\\
33.173	0.784852279336555\\
33.362	0.791146646382131\\
33.677	0.802509206410527\\
34.244	0.822971342435146\\
34.811	0.843562885080033\\
35	0.850688292582085\\
};
\addplot [color=green, line width=1.5pt, forget plot]
  table[row sep=crcr]{%
3.5	0.478359969325581\\
7.658	0.479246787989112\\
11.753	0.480777770414335\\
15.785	0.482940722138309\\
19.754	0.485728532082092\\
23.66	0.489134375558699\\
27.566	0.493203729383879\\
32.354	0.498910597486919\\
35	0.502206246862009\\
};
\end{axis}
\node[align=center,anchor=east,text opacity=1,rounded corners,fill=none,opacity=0.6,text opacity=1,xshift=-15pt,yshift=5pt]
at (period.north west) {(a)};
\node[align=center,anchor=east,text opacity=1,rounded corners,fill=none,opacity=0.6,text opacity=1,xshift=-15pt,yshift=5pt]
at (width.north west) {(b)};
\end{tikzpicture}%


%% file: tikz/Fig5_boxplot_period.tex
\pgfplotsset{compat=newest,plot coordinates/math parser=false,every axis legend/.append style={
at={(0.5,0.88)},
anchor=south,legend cell align=left, align=left, draw=white!15!black}}

%
\definecolor{mycolor1}{rgb}{0.00000,0.44700,0.74100}%
\begin{tikzpicture}

\begin{axis}[%
grid = both,
name = Thick100,
width=\figurewidth,
height=0.3\figureheight,
at={(0\figurewidth,0.7\figureheight)},
scale only axis,
xmin=10,
xmax=40,
xtick={14.8,20.2,25,29.8,35.2},
xticklabels=\empty,
ymin=0.4,
ymax=1.2,
ylabel style={font=\color{white!15!black}},
ylabel={{$\max_{x}\vert\hat{\epsilon}\vert$}},
axis background/.style={fill=white}
]
\addplot [color=mycolor1, line width=1.5pt, forget plot]
  table[row sep=crcr]{%
10	0.749861323749229\\
10.12	0.757972344353696\\
10.24	0.766600726133937\\
10.66	0.797520033509805\\
10.78	0.805601470768877\\
10.9	0.81292246039834\\
10.96	0.816225184578926\\
11.02	0.819250673524472\\
11.08	0.821969886125757\\
11.14	0.824353781273565\\
11.2	0.826373317858696\\
11.26	0.828013226439332\\
11.32	0.829313324243252\\
11.44	0.831108443101712\\
11.56	0.832187385595574\\
11.8	0.833911586135535\\
11.92	0.835331629215531\\
12.04	0.837254516842414\\
12.16	0.839613136443141\\
12.34	0.843821254541716\\
12.52	0.848655917649552\\
12.64	0.852251933455122\\
12.76	0.856231993866061\\
12.88	0.860670924057679\\
13	0.865643549205274\\
13.3	0.878890023755034\\
13.36	0.881036762166204\\
13.42	0.882812568829884\\
13.48	0.884130266782691\\
13.54	0.884902679061234\\
13.6	0.885042628702131\\
13.66	0.884501521007465\\
13.72	0.883385090341193\\
13.78	0.881837653332738\\
13.9	0.878027026807025\\
14.02	0.874224174465738\\
14.08	0.872686455187825\\
14.14	0.8715836293443\\
14.2	0.871060013564595\\
14.26	0.871218876101729\\
14.32	0.871999291703034\\
14.38	0.873299286739467\\
14.44	0.875016887581964\\
14.56	0.87929701216887\\
14.74	0.886299901868107\\
14.8	0.88838169133664\\
14.86	0.890188507973193\\
14.98	0.893015300307553\\
15.1	0.894962617812581\\
15.22	0.896234036195082\\
15.4	0.89731917477269\\
15.76	0.89892456969055\\
15.88	0.899946201797135\\
16	0.901504091343938\\
16.12	0.903675653348856\\
16.36	0.908496836601898\\
16.48	0.910317394825391\\
16.54	0.910898808049737\\
16.6	0.911191888763767\\
16.66	0.911164702601454\\
16.78	0.910359787424838\\
17.08	0.907118212801144\\
17.14	0.90679897100879\\
17.2	0.906746864258288\\
17.32	0.907520174611037\\
17.5	0.909827121288743\\
17.62	0.911221338734286\\
17.68	0.911641734499526\\
17.74	0.911773837690575\\
17.8	0.911539785355544\\
17.86	0.910890039154133\\
17.98	0.908626830187885\\
18.28	0.901644414641076\\
18.34	0.900757570450153\\
18.4	0.900255470239813\\
18.46	0.900204941965001\\
18.52	0.900571786529945\\
18.58	0.901296548076196\\
18.7	0.903581998678831\\
18.82	0.90658564690532\\
19.06	0.912857214662679\\
19.24	0.91667208813481\\
19.48	0.92096816767905\\
19.78	0.926287887183598\\
20.02	0.931074536799166\\
20.26	0.936507963751474\\
20.5	0.942046051548175\\
20.62	0.944274072877562\\
20.74	0.945801013164349\\
20.86	0.946382306617686\\
20.98	0.946020262698752\\
21.1	0.944964066120718\\
21.28	0.9426468065612\\
21.52	0.939507775061571\\
21.7	0.937653515668664\\
21.94	0.935801051629639\\
22.18	0.934547740608984\\
22.48	0.933593759050666\\
23.02	0.932128724104231\\
23.2	0.930819333340587\\
23.38	0.928753193345713\\
23.68	0.924931393060568\\
23.8	0.923791310563566\\
23.92	0.923101920392597\\
24.04	0.922845162067716\\
24.22	0.92314564762404\\
24.4	0.924096681974014\\
24.88	0.927349190258916\\
25	0.927542756365774\\
25.12	0.927242971680933\\
25.3	0.926158772558964\\
25.6	0.924159603943316\\
25.78	0.923711513926726\\
25.96	0.923922556720534\\
26.14	0.924660906699202\\
26.44	0.926615684485462\\
26.68	0.928005108714586\\
26.86	0.928392669793467\\
27.04	0.928081006318763\\
27.88	0.925365675887114\\
28.36	0.924816276460845\\
28.6	0.925031873616135\\
28.9	0.926005868298901\\
29.2	0.926886640250871\\
29.38	0.926938748909102\\
29.62	0.926368747806151\\
29.86	0.925142532064676\\
30.16	0.923472394132524\\
30.28	0.923243383506708\\
30.4	0.923505915674269\\
30.52	0.924344158179593\\
30.94	0.928121614115817\\
31.06	0.92822668532412\\
31.18	0.927616625412789\\
31.3	0.92649706486602\\
31.84	0.92058129204456\\
32.38	0.916478853259434\\
32.56	0.915292432286172\\
32.74	0.914666967589184\\
32.86	0.914733584206374\\
33.04	0.915516098760698\\
33.52	0.918323372160202\\
33.7	0.918561335701881\\
33.88	0.918117706177789\\
34.06	0.916913201484405\\
34.48	0.913092165103102\\
34.6	0.912641322171716\\
34.72	0.912773943110373\\
34.9	0.913641619558689\\
35.08	0.914528751607889\\
35.2	0.914695960281442\\
35.32	0.914299289127186\\
35.5	0.912976101562144\\
35.74	0.911002429561648\\
35.86	0.910410642272588\\
36.04	0.910151277393616\\
36.4	0.909901071430859\\
36.58	0.908983810685818\\
36.94	0.906845417618108\\
37.06	0.906634292057532\\
37.24	0.906918664079598\\
37.54	0.907540505920352\\
37.66	0.90732806006492\\
37.84	0.906336092410015\\
38.5	0.901532960429684\\
38.92	0.899720925270124\\
39.28	0.897967050439497\\
39.52	0.896196471160984\\
39.7	0.894350381841669\\
39.88	0.891926671699181\\
40	0.889930636442138\\
};
\addplot [color=red, line width=1.5pt, forget plot]
  table[row sep=crcr]{%
14.8	0.940687425693406\\
14.8	0.98913403174784\\
};
\addplot [color=red, line width=1.5pt, forget plot]
  table[row sep=crcr]{%
20.2	0.964995803781644\\
20.2	1.002653179533\\
};
\addplot [color=red, line width=1.5pt, forget plot]
  table[row sep=crcr]{%
25	0.941080031525477\\
25	1.00847609895477\\
};
\addplot [color=red, line width=1.5pt, forget plot]
  table[row sep=crcr]{%
29.8	0.93741494439637\\
29.8	0.986623110274483\\
};
\addplot [color=red, line width=1.5pt, forget plot]
  table[row sep=crcr]{%
35.2	0.922944456028837\\
35.2	0.981464575819409\\
};
\addplot [color=red, line width=1.5pt, forget plot]
  table[row sep=crcr]{%
14.8	0.794483223655483\\
14.8	0.824116861414447\\
};
\addplot [color=red, line width=1.5pt, forget plot]
  table[row sep=crcr]{%
20.2	0.902926006967643\\
20.2	0.921716908201944\\
};
\addplot [color=red, line width=1.5pt, forget plot]
  table[row sep=crcr]{%
25	0.903066699166622\\
25	0.915211922050961\\
};
\addplot [color=red, line width=1.5pt, forget plot]
  table[row sep=crcr]{%
29.8	0.853031966112756\\
29.8	0.913194452745646\\
};
\addplot [color=red, line width=1.5pt, forget plot]
  table[row sep=crcr]{%
35.2	0.863753300510581\\
35.2	0.900997656625236\\
};
\addplot [color=red, forget plot]
  table[row sep=crcr]{%
13.8	0.98913403174784\\
15.8	0.98913403174784\\
};
\addplot [color=red, forget plot]
  table[row sep=crcr]{%
19.2	1.002653179533\\
21.2	1.002653179533\\
};
\addplot [color=red, forget plot]
  table[row sep=crcr]{%
24	1.00847609895477\\
26	1.00847609895477\\
};
\addplot [color=red, forget plot]
  table[row sep=crcr]{%
28.8	0.986623110274483\\
30.8	0.986623110274483\\
};
\addplot [color=red, forget plot]
  table[row sep=crcr]{%
34.2	0.981464575819409\\
36.2	0.981464575819409\\
};
\addplot [color=red, forget plot]
  table[row sep=crcr]{%
13.8	0.794483223655483\\
15.8	0.794483223655483\\
};
\addplot [color=red, forget plot]
  table[row sep=crcr]{%
19.2	0.902926006967643\\
21.2	0.902926006967643\\
};
\addplot [color=red, forget plot]
  table[row sep=crcr]{%
24	0.903066699166622\\
26	0.903066699166622\\
};
\addplot [color=red, forget plot]
  table[row sep=crcr]{%
28.8	0.853031966112756\\
30.8	0.853031966112756\\
};
\addplot [color=red, forget plot]
  table[row sep=crcr]{%
34.2	0.863753300510581\\
36.2	0.863753300510581\\
};
\addplot [color=red, forget plot]
  table[row sep=crcr]{%
13.8	0.824116861414447\\
13.8	0.940687425693406\\
15.8	0.940687425693406\\
15.8	0.824116861414447\\
13.8	0.824116861414447\\
};
\addplot [color=red, forget plot]
  table[row sep=crcr]{%
19.2	0.921716908201944\\
19.2	0.964995803781644\\
21.2	0.964995803781644\\
21.2	0.921716908201944\\
19.2	0.921716908201944\\
};
\addplot [color=red, forget plot]
  table[row sep=crcr]{%
24	0.915211922050961\\
24	0.941080031525477\\
26	0.941080031525477\\
26	0.915211922050961\\
24	0.915211922050961\\
};
\addplot [color=red, forget plot]
  table[row sep=crcr]{%
28.8	0.913194452745646\\
28.8	0.93741494439637\\
30.8	0.93741494439637\\
30.8	0.913194452745646\\
28.8	0.913194452745646\\
};
\addplot [color=red, forget plot]
  table[row sep=crcr]{%
34.2	0.900997656625236\\
34.2	0.922944456028837\\
36.2	0.922944456028837\\
36.2	0.900997656625236\\
34.2	0.900997656625236\\
};
\addplot [color=red, forget plot]
  table[row sep=crcr]{%
13.8	0.888381691336638\\
13.8	0.888381691336638\\
};
\addplot [color=red, forget plot]
  table[row sep=crcr]{%
19.2	0.935084171615468\\
19.2	0.935084171615468\\
};
\addplot [color=red, forget plot]
  table[row sep=crcr]{%
24	0.927542756365771\\
24	0.927542756365771\\
};
\addplot [color=red, forget plot]
  table[row sep=crcr]{%
28.8	0.925504281469799\\
28.8	0.925504281469799\\
};
\addplot [color=red, forget plot]
  table[row sep=crcr]{%
34.2	0.914695960281442\\
34.2	0.914695960281442\\
};

\addplot[area legend, draw=black, fill=red, fill opacity=0.5, forget plot]
table[row sep=crcr] {%
x	y\\
13.8	0.824116861414447\\
13.8	0.940687425693406\\
15.8	0.940687425693406\\
15.8	0.824116861414447\\
13.8	0.824116861414447\\
}--cycle;

\addplot[area legend, draw=black, fill=red, fill opacity=0.5, forget plot]
table[row sep=crcr] {%
x	y\\
19.2	0.921716908201943\\
19.2	0.964995803781642\\
21.2	0.964995803781642\\
21.2	0.921716908201943\\
19.2	0.921716908201943\\
}--cycle;

\addplot[area legend, draw=black, fill=red, fill opacity=0.5, forget plot]
table[row sep=crcr] {%
x	y\\
24	0.915211922050961\\
24	0.941080031525478\\
26	0.941080031525478\\
26	0.915211922050961\\
24	0.915211922050961\\
}--cycle;

\addplot[area legend, draw=black, fill=red, fill opacity=0.5, forget plot]
table[row sep=crcr] {%
x	y\\
28.8	0.913194452745645\\
28.8	0.937414944396369\\
30.8	0.937414944396369\\
30.8	0.913194452745645\\
28.8	0.913194452745645\\
}--cycle;

\addplot[area legend, draw=black, fill=red, fill opacity=0.5, forget plot]
table[row sep=crcr] {%
x	y\\
34.2	0.900997656625236\\
34.2	0.92294445602884\\
36.2	0.92294445602884\\
36.2	0.900997656625236\\
34.2	0.900997656625236\\
}--cycle;
\addplot[only marks, mark=*, mark options={}, mark size=1.2247pt, draw=white!70!black, fill=white!70!black, forget plot] table[row sep=crcr]{%
x	y\\
14.8	0.888381691336639\\
20.2	0.935084171615468\\
25	0.927542756365771\\
29.8	0.925504281469799\\
35.2	0.914695960281445\\
};
\end{axis}

\begin{axis}[%
grid = both,
name = Thick200,
width=\figurewidth,
height=0.3\figureheight,
at={(0\figurewidth,0.35\figureheight)},
scale only axis,
xmin=10,
xmax=40,
xtick={14.8,20.2,25,29.8,35.2},
xticklabels=\empty,
ymin=0.4,
ymax=1.2,
ylabel style={font=\color{white!15!black}},
ylabel={{$\max_{x}\vert\hat{\epsilon}\vert$}},
axis background/.style={fill=white}
]
\addplot [color=mycolor1, line width=1.5pt, forget plot]
  table[row sep=crcr]{%
10	0.658103653809363\\
10.06	0.660282014930722\\
10.12	0.661967962055421\\
10.18	0.66320380258481\\
10.24	0.664031843920242\\
10.3	0.664494393463066\\
10.36	0.664633758614642\\
10.48	0.664112165349444\\
10.6	0.662805523335464\\
10.84	0.65919092990444\\
10.96	0.657559896909035\\
11.08	0.656497652008113\\
11.14	0.656285593596472\\
11.2	0.656342654412512\\
11.32	0.657312352810763\\
11.44	0.659140153802326\\
11.86	0.666712974159829\\
11.98	0.667969393886501\\
12.34	0.671276058123837\\
12.46	0.673161352344358\\
12.58	0.675685733803007\\
12.7	0.678700189805063\\
12.88	0.683794647355683\\
13.3	0.696435981927877\\
13.42	0.70055101157493\\
13.54	0.705191284565394\\
13.66	0.710521015954221\\
13.78	0.716611092961671\\
13.9	0.723439074973285\\
14.02	0.73097846364265\\
14.14	0.739202760623371\\
14.26	0.748068547052142\\
14.44	0.761503168457473\\
14.5	0.76560505335712\\
14.56	0.769344051481205\\
14.62	0.772615327435027\\
14.68	0.775314045823883\\
14.74	0.777335371253074\\
14.8	0.778574468327903\\
14.86	0.778978795549932\\
14.92	0.778704987005824\\
15.04	0.776958674550876\\
15.16	0.775006954826438\\
15.22	0.77447638719547\\
15.28	0.774521251695901\\
15.34	0.775350476310649\\
15.4	0.777172989022652\\
15.46	0.780119925186483\\
15.52	0.784011249643392\\
15.58	0.788589134606283\\
15.82	0.808609723775966\\
15.88	0.812752994462457\\
15.94	0.816035858932395\\
16	0.818200489398698\\
16.06	0.819071380921251\\
16.12	0.81880231994792\\
16.18	0.817629415773524\\
16.24	0.815788777692916\\
16.36	0.81104873699239\\
16.42	0.808621552962158\\
16.48	0.806471072205049\\
16.54	0.804833404015916\\
16.6	0.803944657689584\\
16.66	0.803978431357166\\
16.72	0.804858278494834\\
16.78	0.806445241415013\\
16.84	0.808600362430141\\
16.9	0.811184683852659\\
17.02	0.817085097169588\\
17.14	0.823034819865292\\
17.2	0.825680778011268\\
17.26	0.827952965299531\\
17.32	0.829866298343951\\
17.38	0.831466468618693\\
17.5	0.833910086755807\\
17.62	0.835649351504166\\
18.04	0.840897196431015\\
18.7	0.850159815759035\\
19	0.854441534606181\\
19.24	0.858477636892076\\
19.72	0.866842172085676\\
20.02	0.871447999702013\\
20.98	0.885700933007485\\
21.4	0.892296859365246\\
22.06	0.901803996948743\\
22.42	0.907940203791796\\
22.54	0.909567531068895\\
22.66	0.910738845027304\\
22.84	0.911748078736281\\
23.14	0.913094227251463\\
23.26	0.914097207497846\\
23.44	0.916213154321568\\
23.68	0.919098379191759\\
23.8	0.920096327150816\\
23.98	0.920762068720961\\
24.16	0.921311740702151\\
24.28	0.92219660092195\\
24.4	0.923904217663747\\
24.52	0.926626266419881\\
24.82	0.934632518912025\\
24.88	0.935841044455728\\
24.94	0.936744060372646\\
25	0.937266737530535\\
25.06	0.937359654125117\\
25.12	0.937075017663972\\
25.24	0.935683544916579\\
25.54	0.931030723059742\\
25.6	0.930519796231735\\
25.66	0.930311422906996\\
25.78	0.930725829933444\\
25.9	0.93202879566757\\
26.02	0.933956744677126\\
26.26	0.938638855661516\\
26.8	0.949084070355362\\
27.04	0.95346324828558\\
27.22	0.956222943250921\\
27.34	0.957625097921671\\
27.46	0.958564637485935\\
27.58	0.959045705796107\\
27.76	0.959136744379535\\
28.42	0.958478217003574\\
28.54	0.959064728132269\\
28.66	0.960103364220657\\
28.9	0.963048721816207\\
29.08	0.965021538242368\\
29.2	0.965743970094842\\
29.32	0.965766753358025\\
29.44	0.965200873467708\\
29.62	0.963651717470206\\
29.98	0.960185880588732\\
30.7	0.954431951324182\\
30.88	0.953422860859341\\
31	0.953145697560146\\
31.18	0.953403662894125\\
31.48	0.954125913400205\\
31.6	0.953919237768673\\
31.72	0.953189935856543\\
31.9	0.951433129357433\\
32.14	0.948920556840712\\
32.26	0.948041739886641\\
32.44	0.947347773544443\\
32.92	0.946135715572481\\
33.16	0.944698400389171\\
33.7	0.94123113059009\\
33.94	0.940227164177294\\
34.24	0.939594491742341\\
34.48	0.93894218656493\\
34.66	0.937871792485488\\
34.84	0.936201719781138\\
35.32	0.931352350313766\\
35.5	0.930165345065177\\
35.68	0.929431128365096\\
35.86	0.929155835690615\\
36.34	0.929036542452693\\
36.52	0.928215568320695\\
36.76	0.92635674708464\\
37.06	0.923987622597977\\
37.3	0.922731146244146\\
37.66	0.921503562533665\\
38.02	0.920216782437222\\
38.26	0.918852995789543\\
38.74	0.91566733648223\\
38.92	0.915255953090771\\
39.22	0.915447031378349\\
39.46	0.915350697605625\\
39.58	0.914900163089293\\
39.7	0.914026570316679\\
39.82	0.912613809410551\\
39.94	0.910545770493691\\
40	0.909229737444605\\
};
\addplot [color=red, line width=1.5pt, forget plot]
  table[row sep=crcr]{%
14.8	0.891634405593788\\
14.8	1.02021832396343\\
};
\addplot [color=red, line width=1.5pt, forget plot]
  table[row sep=crcr]{%
20.2	0.981669618897556\\
20.2	1.10992007831778\\
};
\addplot [color=red, line width=1.5pt, forget plot]
  table[row sep=crcr]{%
25	0.968998997550614\\
25	1.09575380046269\\
};
\addplot [color=red, line width=1.5pt, forget plot]
  table[row sep=crcr]{%
29.8	0.993835615829159\\
29.8	1.10613572690801\\
};
\addplot [color=red, line width=1.5pt, forget plot]
  table[row sep=crcr]{%
35.2	0.957167393523449\\
35.2	1.02930401062882\\
};
\addplot [color=red, line width=1.5pt, forget plot]
  table[row sep=crcr]{%
14.8	0.561278283307555\\
14.8	0.648247533614082\\
};
\addplot [color=red, line width=1.5pt, forget plot]
  table[row sep=crcr]{%
20.2	0.67165101649764\\
20.2	0.800944207511286\\
};
\addplot [color=red, line width=1.5pt, forget plot]
  table[row sep=crcr]{%
25	0.79713801886522\\
25	0.900175001762527\\
};
\addplot [color=red, line width=1.5pt, forget plot]
  table[row sep=crcr]{%
29.8	0.857156615361177\\
29.8	0.945333261247093\\
};
\addplot [color=red, line width=1.5pt, forget plot]
  table[row sep=crcr]{%
35.2	0.848041378688436\\
35.2	0.916856462525608\\
};
\addplot [color=red, forget plot]
  table[row sep=crcr]{%
13.8	1.02021832396343\\
15.8	1.02021832396343\\
};
\addplot [color=red, forget plot]
  table[row sep=crcr]{%
19.2	1.10992007831778\\
21.2	1.10992007831778\\
};
\addplot [color=red, forget plot]
  table[row sep=crcr]{%
24	1.09575380046269\\
26	1.09575380046269\\
};
\addplot [color=red, forget plot]
  table[row sep=crcr]{%
28.8	1.10613572690801\\
30.8	1.10613572690801\\
};
\addplot [color=red, forget plot]
  table[row sep=crcr]{%
34.2	1.02930401062882\\
36.2	1.02930401062882\\
};
\addplot [color=red, forget plot]
  table[row sep=crcr]{%
13.8	0.561278283307555\\
15.8	0.561278283307555\\
};
\addplot [color=red, forget plot]
  table[row sep=crcr]{%
19.2	0.67165101649764\\
21.2	0.67165101649764\\
};
\addplot [color=red, forget plot]
  table[row sep=crcr]{%
24	0.79713801886522\\
26	0.79713801886522\\
};
\addplot [color=red, forget plot]
  table[row sep=crcr]{%
28.8	0.857156615361177\\
30.8	0.857156615361177\\
};
\addplot [color=red, forget plot]
  table[row sep=crcr]{%
34.2	0.848041378688436\\
36.2	0.848041378688436\\
};
\addplot [color=red, forget plot]
  table[row sep=crcr]{%
13.8	0.648247533614082\\
13.8	0.891634405593788\\
15.8	0.891634405593788\\
15.8	0.648247533614082\\
13.8	0.648247533614082\\
};
\addplot [color=red, forget plot]
  table[row sep=crcr]{%
19.2	0.800944207511286\\
19.2	0.981669618897556\\
21.2	0.981669618897556\\
21.2	0.800944207511286\\
19.2	0.800944207511286\\
};
\addplot [color=red, forget plot]
  table[row sep=crcr]{%
24	0.900175001762527\\
24	0.968998997550614\\
26	0.968998997550614\\
26	0.900175001762527\\
24	0.900175001762527\\
};
\addplot [color=red, forget plot]
  table[row sep=crcr]{%
28.8	0.945333261247093\\
28.8	0.993835615829159\\
30.8	0.993835615829159\\
30.8	0.945333261247093\\
28.8	0.945333261247093\\
};
\addplot [color=red, forget plot]
  table[row sep=crcr]{%
34.2	0.916856462525608\\
34.2	0.957167393523449\\
36.2	0.957167393523449\\
36.2	0.916856462525608\\
34.2	0.916856462525608\\
};
\addplot [color=red, forget plot]
  table[row sep=crcr]{%
13.8	0.778574468327902\\
13.8	0.778574468327902\\
};
\addplot [color=red, forget plot]
  table[row sep=crcr]{%
19.2	0.87405404248528\\
19.2	0.87405404248528\\
};
\addplot [color=red, forget plot]
  table[row sep=crcr]{%
24	0.937266737530532\\
24	0.937266737530532\\
};
\addplot [color=red, forget plot]
  table[row sep=crcr]{%
28.8	0.961815402307877\\
28.8	0.961815402307877\\
};
\addplot [color=red, forget plot]
  table[row sep=crcr]{%
34.2	0.932391045776285\\
34.2	0.932391045776285\\
};

\addplot[area legend, draw=black, fill=red, fill opacity=0.5, forget plot]
table[row sep=crcr] {%
x	y\\
13.8	0.648247533614082\\
13.8	0.891634405593788\\
15.8	0.891634405593788\\
15.8	0.648247533614082\\
13.8	0.648247533614082\\
}--cycle;

\addplot[area legend, draw=black, fill=red, fill opacity=0.5, forget plot]
table[row sep=crcr] {%
x	y\\
19.2	0.800944207511284\\
19.2	0.981669618897556\\
21.2	0.981669618897556\\
21.2	0.800944207511284\\
19.2	0.800944207511284\\
}--cycle;

\addplot[area legend, draw=black, fill=red, fill opacity=0.5, forget plot]
table[row sep=crcr] {%
x	y\\
24	0.900175001762527\\
24	0.968998997550613\\
26	0.968998997550613\\
26	0.900175001762527\\
24	0.900175001762527\\
}--cycle;

\addplot[area legend, draw=black, fill=red, fill opacity=0.5, forget plot]
table[row sep=crcr] {%
x	y\\
28.8	0.945333261247094\\
28.8	0.993835615829159\\
30.8	0.993835615829159\\
30.8	0.945333261247094\\
28.8	0.945333261247094\\
}--cycle;

\addplot[area legend, draw=black, fill=red, fill opacity=0.5, forget plot]
table[row sep=crcr] {%
x	y\\
34.2	0.91685646252561\\
34.2	0.957167393523447\\
36.2	0.957167393523447\\
36.2	0.91685646252561\\
34.2	0.91685646252561\\
}--cycle;
\addplot[only marks, mark=*, mark options={}, mark size=1.2247pt, draw=white!70!black, fill=white!70!black, forget plot] table[row sep=crcr]{%
x	y\\
14.8	0.778574468327901\\
20.2	0.874054042485279\\
25	0.937266737530532\\
29.8	0.961815402307877\\
35.2	0.932391045776283\\
};
\end{axis}

\begin{axis}[%
grid = both,
name = Thick400,
width=\figurewidth,
height=0.3\figureheight,
at={(0\figurewidth,0\figureheight)},
scale only axis,
xmin=10,
xmax=40,
xtick={10,14.8,20.2,25,29.8,35.2,40},
xticklabels={10,15,20,25,30,35,40},
xlabel style={font=\color{white!15!black}},
xlabel={$T_{\text{peak}}\,[\text{s}]$},
ymin=0.4,
ymax=1.2,
ylabel style={font=\color{white!15!black}},
ylabel={{$\max_{x}\vert\hat{\epsilon}\vert$}},
axis background/.style={fill=white}
]
\addplot [color=mycolor1, line width=1.5pt, forget plot]
  table[row sep=crcr]{%
10	0.671396553232917\\
10.18	0.670879672040464\\
10.42	0.669565829651042\\
11.02	0.665793892408196\\
11.2	0.665260870641745\\
11.38	0.665245955359801\\
11.98	0.665877508972088\\
12.16	0.665311182906137\\
12.34	0.664075000168232\\
12.52	0.662044691598219\\
13	0.65572288773032\\
13.12	0.654808373911358\\
13.3	0.65414728648129\\
13.48	0.654155802878016\\
13.72	0.654857624433532\\
14.02	0.656357056116512\\
14.5	0.658939875408493\\
14.68	0.65934384511911\\
14.8	0.65923784198101\\
14.98	0.658459883014402\\
15.34	0.656488394934811\\
15.46	0.656304155366172\\
15.64	0.656722645516261\\
15.82	0.657733930237036\\
16.18	0.66047085608097\\
16.48	0.663145048222709\\
16.72	0.665813168138854\\
16.96	0.669121658707226\\
17.14	0.672076692668703\\
17.32	0.675466692461377\\
17.62	0.681754528713647\\
18.04	0.690522662922241\\
19.18	0.713532559352828\\
19.3	0.71642961973204\\
19.42	0.720195229530049\\
19.48	0.72252521390795\\
19.54	0.725218103776761\\
19.6	0.728322488515218\\
19.66	0.731862183776421\\
19.78	0.739921615143658\\
20.02	0.757159157955108\\
20.08	0.761117959652694\\
20.14	0.764736421798005\\
20.2	0.767914491415958\\
20.26	0.770578508497721\\
20.32	0.772760384899442\\
20.38	0.774518425443532\\
20.44	0.775910934952378\\
20.56	0.777832580153962\\
20.68	0.778991759083411\\
20.92	0.780877437124232\\
21.04	0.782470601483162\\
21.1	0.78360624357947\\
21.16	0.78503463006507\\
21.22	0.786805789273494\\
21.28	0.788969749538296\\
21.34	0.791576539193017\\
21.4	0.794676186571195\\
21.46	0.798290668308447\\
21.52	0.802329754248646\\
21.64	0.811208611321625\\
21.82	0.824754382101645\\
21.88	0.82885717432422\\
21.94	0.832556597771308\\
22	0.835734370588845\\
22.06	0.838303223092844\\
22.12	0.840299934279571\\
22.18	0.841792295315351\\
22.24	0.84284809736652\\
22.3	0.84353513159941\\
22.42	0.844074061275698\\
22.6	0.843811476311366\\
22.84	0.843579734018512\\
23.08	0.84391611545017\\
23.32	0.844816857932308\\
23.44	0.845685662753873\\
23.56	0.847101149760476\\
23.68	0.849273058919884\\
23.74	0.850708215546817\\
23.8	0.852411130199869\\
23.92	0.856604678809745\\
24.1	0.863978049365265\\
24.22	0.868867815646126\\
24.34	0.873138653975438\\
24.4	0.874881722476132\\
24.46	0.876296419972995\\
24.58	0.878276245918812\\
24.7	0.879442540753118\\
25.06	0.881877370808525\\
25.3	0.88477870738604\\
25.48	0.886686831315352\\
25.6	0.88732568626218\\
25.72	0.887260296786636\\
26.02	0.886340033361158\\
26.14	0.886803629555487\\
26.2	0.887429728932517\\
26.26	0.888370923211461\\
26.38	0.89099889440137\\
26.74	0.900183679039884\\
26.8	0.901224050804025\\
26.92	0.902460360208217\\
27.04	0.902880479323379\\
27.34	0.903216128657398\\
27.46	0.904195182365108\\
27.58	0.906021065281621\\
27.7	0.908573452829714\\
27.82	0.911720200163543\\
27.94	0.915329162437288\\
28.12	0.921317428318076\\
28.48	0.933569392590059\\
28.6	0.937189862708152\\
28.72	0.940343435617791\\
28.84	0.942945084854472\\
28.96	0.94492252406026\\
29.08	0.946203466877208\\
29.2	0.946715626947359\\
29.32	0.946498617789906\\
29.5	0.94590665896164\\
29.62	0.946208096000937\\
29.68	0.946792304737336\\
29.74	0.947765359102377\\
29.8	0.949202148209146\\
29.86	0.951146394736689\\
29.92	0.953517155627736\\
30.04	0.95908978253496\\
30.22	0.967844843338533\\
30.28	0.970420391092418\\
30.34	0.972637686770419\\
30.4	0.974384620881196\\
30.46	0.975583693334073\\
30.52	0.976295841641019\\
30.58	0.976616612714686\\
30.7	0.976466210812738\\
30.88	0.975693951526246\\
30.94	0.975672944365698\\
31	0.97592938836037\\
31.06	0.976533190148764\\
31.18	0.978626609000671\\
31.3	0.981512483338648\\
31.54	0.987796167045204\\
31.66	0.990275896389591\\
31.78	0.992048542032535\\
31.96	0.993870876235178\\
32.26	0.996637620174972\\
32.44	0.998975960789188\\
32.8	1.003932896974\\
32.98	1.00583467640157\\
33.22	1.00772007979124\\
33.58	1.00984582728963\\
34.18	1.01339456162076\\
34.36	1.01429568064395\\
34.48	1.01456534751379\\
34.6	1.01443378834282\\
34.78	1.01349842123266\\
34.96	1.01262253588137\\
35.08	1.01267670741682\\
35.14	1.01303135789374\\
35.2	1.01366488981161\\
35.32	1.01577020682112\\
35.56	1.02100633679716\\
35.62	1.02200663453399\\
35.68	1.02270433673997\\
35.74	1.02300989260112\\
35.8	1.02283375130347\\
35.86	1.02211961234347\\
35.92	1.02094417645948\\
36.04	1.01764921811458\\
36.22	1.01199582988678\\
36.28	1.01036158448319\\
36.34	1.00903569949708\\
36.4	1.00812812597724\\
36.46	1.00771405106358\\
36.52	1.00772960626063\\
36.64	1.00865507736945\\
36.88	1.01132175512431\\
36.94	1.0115830137746\\
37	1.01148484330082\\
37.06	1.01096121700105\\
37.18	1.00892778705698\\
37.42	1.00363226847549\\
37.48	1.00261826216168\\
37.54	1.00192220880084\\
37.6	1.00164111020314\\
37.66	1.00183610642597\\
37.72	1.00242489051556\\
37.84	1.0043111474688\\
37.96	1.00635453141044\\
38.02	1.00713972423554\\
38.08	1.00760969268808\\
38.14	1.00764626806151\\
38.2	1.00713128164928\\
38.26	1.00598789083342\\
38.32	1.0043045573503\\
38.38	1.00221106902489\\
38.5	0.997312779147016\\
38.62	0.992331323799483\\
38.68	0.990133878636996\\
38.74	0.988305005581978\\
38.8	0.986974492459382\\
38.86	0.986237276943612\\
38.92	0.986048896106865\\
38.98	0.986330036870754\\
39.04	0.987001386156919\\
39.16	0.98919745798257\\
39.46	0.995964371200458\\
39.52	0.996902117843419\\
39.58	0.997516255303303\\
39.64	0.997727470501751\\
39.7	0.997456450360389\\
39.76	0.996623881800829\\
39.82	0.99515045174472\\
39.88	0.992956847113682\\
39.94	0.98996375482934\\
40	0.986091861813328\\
};
\addplot [color=red, line width=1.5pt, forget plot]
  table[row sep=crcr]{%
14.8	0.845723794162506\\
14.8	0.998719411569507\\
};
\addplot [color=red, line width=1.5pt, forget plot]
  table[row sep=crcr]{%
20.2	0.871286217156754\\
20.2	0.989685170288361\\
};
\addplot [color=red, line width=1.5pt, forget plot]
  table[row sep=crcr]{%
25	0.957933598037759\\
25	1.03105980909527\\
};
\addplot [color=red, line width=1.5pt, forget plot]
  table[row sep=crcr]{%
29.8	1.02873892977675\\
29.8	1.08350160257272\\
};
\addplot [color=red, line width=1.5pt, forget plot]
  table[row sep=crcr]{%
35.2	1.05038735607727\\
35.2	1.12342209611429\\
};
\addplot [color=red, line width=1.5pt, forget plot]
  table[row sep=crcr]{%
14.8	0.46448962941964\\
14.8	0.583151609801547\\
};
\addplot [color=red, line width=1.5pt, forget plot]
  table[row sep=crcr]{%
20.2	0.530260426695598\\
20.2	0.629613169283335\\
};
\addplot [color=red, line width=1.5pt, forget plot]
  table[row sep=crcr]{%
25	0.719348174349747\\
25	0.779402549635829\\
};
\addplot [color=red, line width=1.5pt, forget plot]
  table[row sep=crcr]{%
29.8	0.877973573434662\\
29.8	0.914826069551921\\
};
\addplot [color=red, line width=1.5pt, forget plot]
  table[row sep=crcr]{%
35.2	0.952925858010545\\
35.2	0.988357757037797\\
};
\addplot [color=red, forget plot]
  table[row sep=crcr]{%
13.8	0.998719411569507\\
15.8	0.998719411569507\\
};
\addplot [color=red, forget plot]
  table[row sep=crcr]{%
19.2	0.989685170288361\\
21.2	0.989685170288361\\
};
\addplot [color=red, forget plot]
  table[row sep=crcr]{%
24	1.03105980909527\\
26	1.03105980909527\\
};
\addplot [color=red, forget plot]
  table[row sep=crcr]{%
28.8	1.08350160257272\\
30.8	1.08350160257272\\
};
\addplot [color=red, forget plot]
  table[row sep=crcr]{%
34.2	1.12342209611429\\
36.2	1.12342209611429\\
};
\addplot [color=red, forget plot]
  table[row sep=crcr]{%
13.8	0.46448962941964\\
15.8	0.46448962941964\\
};
\addplot [color=red, forget plot]
  table[row sep=crcr]{%
19.2	0.530260426695598\\
21.2	0.530260426695598\\
};
\addplot [color=red, forget plot]
  table[row sep=crcr]{%
24	0.719348174349747\\
26	0.719348174349747\\
};
\addplot [color=red, forget plot]
  table[row sep=crcr]{%
28.8	0.877973573434662\\
30.8	0.877973573434662\\
};
\addplot [color=red, forget plot]
  table[row sep=crcr]{%
34.2	0.952925858010545\\
36.2	0.952925858010545\\
};
\addplot [color=red, forget plot]
  table[row sep=crcr]{%
13.8	0.583151609801547\\
13.8	0.845723794162506\\
15.8	0.845723794162506\\
15.8	0.583151609801547\\
13.8	0.583151609801547\\
};
\addplot [color=red, forget plot]
  table[row sep=crcr]{%
19.2	0.629613169283335\\
19.2	0.871286217156754\\
21.2	0.871286217156754\\
21.2	0.629613169283335\\
19.2	0.629613169283335\\
};
\addplot [color=red, forget plot]
  table[row sep=crcr]{%
24	0.779402549635829\\
24	0.957933598037759\\
26	0.957933598037759\\
26	0.779402549635829\\
24	0.779402549635829\\
};
\addplot [color=red, forget plot]
  table[row sep=crcr]{%
28.8	0.914826069551921\\
28.8	1.02873892977675\\
30.8	1.02873892977675\\
30.8	0.914826069551921\\
28.8	0.914826069551921\\
};
\addplot [color=red, forget plot]
  table[row sep=crcr]{%
34.2	0.988357757037797\\
34.2	1.05038735607727\\
36.2	1.05038735607727\\
36.2	0.988357757037797\\
34.2	0.988357757037797\\
};
\addplot [color=red, forget plot]
  table[row sep=crcr]{%
13.8	0.583151609801547\\
13.8	0.583151609801547\\
};
\addplot [color=red, forget plot]
  table[row sep=crcr]{%
19.2	0.767914491415958\\
19.2	0.767914491415958\\
};
\addplot [color=red, forget plot]
  table[row sep=crcr]{%
24	0.88131299672596\\
24	0.88131299672596\\
};
\addplot [color=red, forget plot]
  table[row sep=crcr]{%
28.8	0.949202148209149\\
28.8	0.949202148209149\\
};
\addplot [color=red, forget plot]
  table[row sep=crcr]{%
34.2	1.01366488981161\\
34.2	1.01366488981161\\
};

\addplot[area legend, draw=black, fill=red, fill opacity=0.5, forget plot]
table[row sep=crcr] {%
x	y\\
13.8	0.583151609801547\\
13.8	0.845723794162506\\
15.8	0.845723794162506\\
15.8	0.583151609801547\\
13.8	0.583151609801547\\
}--cycle;

\addplot[area legend, draw=black, fill=red, fill opacity=0.5, forget plot]
table[row sep=crcr] {%
x	y\\
19.2	0.629613169283333\\
19.2	0.871286217156755\\
21.2	0.871286217156755\\
21.2	0.629613169283333\\
19.2	0.629613169283333\\
}--cycle;

\addplot[area legend, draw=black, fill=red, fill opacity=0.5, forget plot]
table[row sep=crcr] {%
x	y\\
24	0.77940254963583\\
24	0.957933598037757\\
26	0.957933598037757\\
26	0.77940254963583\\
24	0.77940254963583\\
}--cycle;

\addplot[area legend, draw=black, fill=red, fill opacity=0.5, forget plot]
table[row sep=crcr] {%
x	y\\
28.8	0.914826069551923\\
28.8	1.02873892977675\\
30.8	1.02873892977675\\
30.8	0.914826069551923\\
28.8	0.914826069551923\\
}--cycle;

\addplot[area legend, draw=black, fill=red, fill opacity=0.5, forget plot]
table[row sep=crcr] {%
x	y\\
34.2	0.988357757037794\\
34.2	1.05038735607726\\
36.2	1.05038735607726\\
36.2	0.988357757037794\\
34.2	0.988357757037794\\
}--cycle;
\addplot[only marks, mark=*, mark options={}, mark size=1.2247pt, draw=white!70!black, fill=white!70!black, forget plot] table[row sep=crcr]{%
x	y\\
14.8	0.659237841981012\\
20.2	0.767914491415959\\
25	0.881312996725962\\
29.8	0.949202148209148\\
35.2	1.01366488981161\\
};
\end{axis}
\node[align=center,anchor=east,text opacity=1,rounded corners,fill=none,opacity=0.6,text opacity=1,xshift=-20pt,yshift=0pt]
at (Thick100.north west) {(a)};
\node[align=center,anchor=east,text opacity=1,rounded corners,fill=none,opacity=0.6,text opacity=1,xshift=65pt,yshift=-10pt]
at (Thick100.north west) {$D = 100 $\,m};
\node[align=center,anchor=east,text opacity=1,rounded corners,fill=none,opacity=0.6,text opacity=1,xshift=-20pt,yshift=0pt]
at (Thick200.north west) {(b)};
\node[align=center,anchor=east,text opacity=1,rounded corners,fill=none,opacity=0.6,text opacity=1,xshift=65pt,yshift=-10pt]
at (Thick200.north west) {$D = 200 $\,m};
\node[align=center,anchor=east,text opacity=1,rounded corners,fill=none,opacity=0.6,text opacity=1,xshift=-20pt,yshift=0pt]
at (Thick400.north west) {(c)};
\node[align=center,anchor=east,text opacity=1,rounded corners,fill=none,opacity=0.6,text opacity=1,xshift=65pt,yshift=-10pt]
at (Thick400.north west) {$D = 400 $\,m};
\end{tikzpicture}%

%% file: tikz/Fig5_boxplot_width.tex
\pgfplotsset{compat=newest,plot coordinates/math parser=false,every axis legend/.append style={
at={(0.5,0.88)},
anchor=south,legend cell align=left, align=left, draw=white!15!black}}

%
\definecolor{mycolor1}{rgb}{0.00000,0.44700,0.74100}%
\begin{tikzpicture}

\begin{axis}[%
grid = both,
name = Thick100,
width=\figurewidth,
height=0.3\figureheight,
at={(0\figurewidth,0.7\figureheight)},
scale only axis,
xmin=5,
xmax=35,
xtick={9.8,14.84,19.88,24.92,29.96},
xticklabels=\empty,
ymin=0.4,
ymax=1.4,
ylabel style={font=\color{white!15!black}},
ylabel={{$\max_{x}\vert\hat{\epsilon}\vert$}},
axis background/.style={fill=white}
]
\addplot [color=mycolor1, line width=1.5pt, forget plot]
  table[row sep=crcr]{%
3.5	0.891866061592921\\
3.563	0.896057770476936\\
3.626	0.899338205143131\\
3.689	0.901780497489867\\
3.752	0.903457779415497\\
3.815	0.904443182818376\\
3.878	0.904809839596844\\
3.941	0.904630881649268\\
4.004	0.903979440873997\\
4.13	0.90155163843378\\
4.256	0.898111487463012\\
4.508	0.890534360677137\\
4.634	0.887567495235672\\
4.697	0.886545448647041\\
4.76	0.885928502010955\\
4.823	0.885763439646048\\
4.949	0.88652819614925\\
5.138	0.889138230287791\\
5.264	0.890862394203616\\
5.39	0.891780970965641\\
5.516	0.891425033233077\\
5.642	0.890168458859314\\
5.894	0.887291716795986\\
6.02	0.886840707410322\\
6.146	0.887635897121662\\
6.272	0.889304460734031\\
6.524	0.893004485399572\\
6.65	0.893907334321611\\
6.776	0.893593854434954\\
6.902	0.892337347436815\\
7.28	0.887317896271028\\
7.406	0.886607479994723\\
7.532	0.886617952877089\\
7.658	0.887249938189257\\
7.847	0.889145865368995\\
8.099	0.892820013969825\\
8.288	0.895442112297147\\
8.414	0.896603172903497\\
8.54	0.896968105824534\\
8.666	0.896341692415426\\
8.792	0.894867349380476\\
8.918	0.892773152261277\\
9.17	0.887637497936481\\
9.359	0.88387646875556\\
9.485	0.881893460711765\\
9.611	0.880640643810246\\
9.737	0.880384345491699\\
9.863	0.881372672498998\\
9.989	0.883434655524951\\
10.304	0.88937482773796\\
10.43	0.890468284087817\\
10.556	0.890120955502979\\
10.682	0.88874725391188\\
10.934	0.885396521811458\\
11.06	0.884657386352409\\
11.186	0.88519199333178\\
11.312	0.886740916995294\\
11.564	0.891141312734128\\
11.69	0.89312258398914\\
11.816	0.894453634302188\\
11.942	0.895078683335591\\
12.068	0.895017814766199\\
12.194	0.894291112270849\\
12.32	0.892918659526366\\
12.509	0.889922808037248\\
12.698	0.886965046979597\\
12.824	0.885728362795241\\
12.95	0.885579600446441\\
13.076	0.886775189774916\\
13.202	0.888883691581476\\
13.454	0.893425806791122\\
13.58	0.894652607275354\\
13.706	0.894595910414189\\
14.084	0.892420435932671\\
14.147	0.892684010168054\\
14.21	0.893359781892755\\
14.273	0.894500059048553\\
14.399	0.897785330975609\\
14.588	0.903371282421467\\
14.651	0.90491441105592\\
14.714	0.906092139824032\\
14.777	0.906780885360241\\
14.84	0.906857064298983\\
14.903	0.906242267398305\\
14.966	0.905038781910612\\
15.092	0.901455590678253\\
15.281	0.895352173827469\\
15.344	0.89371324571399\\
15.407	0.892517858645583\\
15.47	0.891913473998272\\
15.533	0.892001753914641\\
15.596	0.892701163603562\\
15.659	0.893884369040499\\
15.785	0.897192831060202\\
15.974	0.902600641586169\\
16.037	0.904012606995671\\
16.1	0.905017029981359\\
16.163	0.90552080036165\\
16.226	0.905567703326838\\
16.352	0.904602943144965\\
16.478	0.902746821701832\\
16.667	0.899656495254405\\
16.793	0.898286529506379\\
16.919	0.89781552252888\\
17.045	0.898126952736781\\
17.171	0.899088541228949\\
17.36	0.901460610654041\\
17.549	0.904599411098445\\
17.801	0.909628158954661\\
18.431	0.923021946551614\\
18.557	0.924663198719195\\
18.683	0.925519976774936\\
18.872	0.925675820767239\\
19.061	0.925725422048799\\
19.187	0.926406240445374\\
19.313	0.928078634588168\\
19.439	0.930653856995455\\
19.691	0.936170649231208\\
19.817	0.937909099608284\\
19.88	0.938260155981638\\
19.943	0.938194401637119\\
20.069	0.937118478309145\\
20.321	0.933937893077328\\
20.447	0.93340404259061\\
20.51	0.933744842877033\\
20.573	0.934591216054613\\
20.636	0.935888945769605\\
20.762	0.939499695742917\\
21.14	0.952356246770144\\
21.266	0.955198313377224\\
21.392	0.956961094504472\\
21.518	0.957808890142175\\
21.644	0.957906000280644\\
21.77	0.957416724910175\\
22.022	0.955441425467008\\
22.211	0.954022404733365\\
22.337	0.953562141878571\\
22.463	0.953765094071578\\
22.589	0.954655444910095\\
22.778	0.956870745104922\\
23.282	0.963600814006007\\
23.66	0.967576788797388\\
23.975	0.97050654547143\\
24.164	0.971779701792883\\
24.353	0.972405896249313\\
24.542	0.97234617208634\\
24.92	0.971231450232374\\
25.109	0.970851125385309\\
25.235	0.971128681139895\\
25.361	0.972147282745439\\
25.487	0.974177925626655\\
25.613	0.977471377270824\\
25.739	0.981813162603451\\
25.991	0.990902635851704\\
26.054	0.992749190221062\\
26.117	0.994250432619921\\
26.18	0.995318869851566\\
26.243	0.995898012109421\\
26.369	0.995889510110089\\
26.558	0.994437697986164\\
26.684	0.99365993383573\\
26.747	0.993623666372713\\
26.81	0.993953844843404\\
26.873	0.994717940239561\\
26.999	0.997289811183926\\
27.377	1.0070325617645\\
27.44	1.00802997519353\\
27.503	1.00864018549898\\
27.629	1.00889210921126\\
27.755	1.00830416535839\\
28.007	1.00677569079559\\
28.133	1.00690721241691\\
28.259	1.00807949261205\\
28.385	1.01029587270568\\
28.511	1.01354823843445\\
28.637	1.01782847553505\\
28.763	1.02311302926466\\
29.078	1.03745149919146\\
29.141	1.0397697679206\\
29.204	1.04168636152453\\
29.267	1.04310762284332\\
29.33	1.04393989471706\\
29.393	1.04412845656504\\
29.456	1.04377433412343\\
29.582	1.04199788563298\\
29.771	1.0387621386099\\
29.834	1.03809111905375\\
29.897	1.03785715141569\\
29.96	1.03820019801108\\
30.023	1.03921476747941\\
30.086	1.04081355375663\\
30.212	1.04523273777785\\
30.464	1.05524027896863\\
30.527	1.05721131018913\\
30.59	1.05870448029884\\
30.653	1.05962713433066\\
30.716	1.06004703640905\\
30.842	1.05981006029484\\
31.094	1.05804931610749\\
31.157	1.05797028363693\\
31.22	1.05825145039153\\
31.283	1.05897033388257\\
31.409	1.06151534755631\\
31.535	1.0650417606926\\
31.787	1.07266312394643\\
31.913	1.07558979151744\\
32.039	1.07761090106126\\
32.228	1.07966135790056\\
32.48	1.08237179294756\\
32.606	1.08441614401864\\
32.795	1.08841396343963\\
32.984	1.09320022462209\\
33.425	1.10499515349893\\
33.551	1.10760101940278\\
33.677	1.10941606937461\\
33.803	1.11020637509463\\
33.929	1.11005493089655\\
34.37	1.10797282340185\\
34.496	1.10864098884988\\
34.559	1.10944414060348\\
34.622	1.11062895441564\\
34.685	1.11224714231547\\
34.748	1.11435041633213\\
34.811	1.11699048849473\\
34.874	1.12021907083239\\
34.937	1.12408787537425\\
35	1.12864861414943\\
};
\addplot [color=red, line width=1.5pt, forget plot]
  table[row sep=crcr]{%
9.8	0.933906466376202\\
9.8	1.00860319676062\\
};
\addplot [color=red, line width=1.5pt, forget plot]
  table[row sep=crcr]{%
14.84	0.957842173683567\\
14.84	1.0348989206812\\
};
\addplot [color=red, line width=1.5pt, forget plot]
  table[row sep=crcr]{%
19.88	0.986097775874477\\
19.88	1.07171729388073\\
};
\addplot [color=red, line width=1.5pt, forget plot]
  table[row sep=crcr]{%
24.92	1.04510749718168\\
24.92	1.14370697216931\\
};
\addplot [color=red, line width=1.5pt, forget plot]
  table[row sep=crcr]{%
29.96	1.1142055086061\\
29.96	1.24327903611487\\
};
\addplot [color=red, line width=1.5pt, forget plot]
  table[row sep=crcr]{%
9.8	0.807301311512836\\
9.8	0.842934598590242\\
};
\addplot [color=red, line width=1.5pt, forget plot]
  table[row sep=crcr]{%
14.84	0.822487552239943\\
14.84	0.856384032300861\\
};
\addplot [color=red, line width=1.5pt, forget plot]
  table[row sep=crcr]{%
19.88	0.860356234633581\\
19.88	0.893879707809678\\
};
\addplot [color=red, line width=1.5pt, forget plot]
  table[row sep=crcr]{%
24.92	0.897332044691332\\
24.92	0.937533363422055\\
};
\addplot [color=red, line width=1.5pt, forget plot]
  table[row sep=crcr]{%
29.96	0.931588546832732\\
29.96	1.00413949345292\\
};
\addplot [color=red, forget plot]
  table[row sep=crcr]{%
8.8	1.00860319676062\\
10.8	1.00860319676062\\
};
\addplot [color=red, forget plot]
  table[row sep=crcr]{%
13.84	1.0348989206812\\
15.84	1.0348989206812\\
};
\addplot [color=red, forget plot]
  table[row sep=crcr]{%
18.88	1.07171729388073\\
20.88	1.07171729388073\\
};
\addplot [color=red, forget plot]
  table[row sep=crcr]{%
23.92	1.14370697216931\\
25.92	1.14370697216931\\
};
\addplot [color=red, forget plot]
  table[row sep=crcr]{%
28.96	1.24327903611487\\
30.96	1.24327903611487\\
};
\addplot [color=red, forget plot]
  table[row sep=crcr]{%
8.8	0.807301311512836\\
10.8	0.807301311512836\\
};
\addplot [color=red, forget plot]
  table[row sep=crcr]{%
13.84	0.822487552239943\\
15.84	0.822487552239943\\
};
\addplot [color=red, forget plot]
  table[row sep=crcr]{%
18.88	0.860356234633581\\
20.88	0.860356234633581\\
};
\addplot [color=red, forget plot]
  table[row sep=crcr]{%
23.92	0.897332044691332\\
25.92	0.897332044691332\\
};
\addplot [color=red, forget plot]
  table[row sep=crcr]{%
28.96	0.931588546832732\\
30.96	0.931588546832732\\
};
\addplot [color=red, forget plot]
  table[row sep=crcr]{%
8.8	0.842934598590242\\
8.8	0.933906466376202\\
10.8	0.933906466376202\\
10.8	0.842934598590242\\
8.8	0.842934598590242\\
};
\addplot [color=red, forget plot]
  table[row sep=crcr]{%
13.84	0.856384032300861\\
13.84	0.957842173683567\\
15.84	0.957842173683567\\
15.84	0.856384032300861\\
13.84	0.856384032300861\\
};
\addplot [color=red, forget plot]
  table[row sep=crcr]{%
18.88	0.893879707809678\\
18.88	0.986097775874477\\
20.88	0.986097775874477\\
20.88	0.893879707809678\\
18.88	0.893879707809678\\
};
\addplot [color=red, forget plot]
  table[row sep=crcr]{%
23.92	0.937533363422055\\
23.92	1.04510749718168\\
25.92	1.04510749718168\\
25.92	0.937533363422055\\
23.92	0.937533363422055\\
};
\addplot [color=red, forget plot]
  table[row sep=crcr]{%
28.96	1.00413949345292\\
28.96	1.1142055086061\\
30.96	1.1142055086061\\
30.96	1.00413949345292\\
28.96	1.00413949345292\\
};
\addplot [color=red, forget plot]
  table[row sep=crcr]{%
8.8	0.880713118126259\\
8.8	0.880713118126259\\
};
\addplot [color=red, forget plot]
  table[row sep=crcr]{%
13.84	0.906857064298986\\
13.84	0.906857064298986\\
};
\addplot [color=red, forget plot]
  table[row sep=crcr]{%
18.88	0.938260155981634\\
18.88	0.938260155981634\\
};
\addplot [color=red, forget plot]
  table[row sep=crcr]{%
23.92	0.97123145023237\\
23.92	0.97123145023237\\
};
\addplot [color=red, forget plot]
  table[row sep=crcr]{%
28.96	1.03820019801108\\
28.96	1.03820019801108\\
};

\addplot[area legend, draw=black, fill=red, fill opacity=0.5, forget plot]
table[row sep=crcr] {%
x	y\\
8.8	0.842934598590241\\
8.8	0.933906466376202\\
10.8	0.933906466376202\\
10.8	0.842934598590241\\
8.8	0.842934598590241\\
}--cycle;

\addplot[area legend, draw=black, fill=red, fill opacity=0.5, forget plot]
table[row sep=crcr] {%
x	y\\
13.84	0.856384032300861\\
13.84	0.957842173683566\\
15.84	0.957842173683566\\
15.84	0.856384032300861\\
13.84	0.856384032300861\\
}--cycle;

\addplot[area legend, draw=black, fill=red, fill opacity=0.5, forget plot]
table[row sep=crcr] {%
x	y\\
18.88	0.893879707809676\\
18.88	0.986097775874478\\
20.88	0.986097775874478\\
20.88	0.893879707809676\\
18.88	0.893879707809676\\
}--cycle;

\addplot[area legend, draw=black, fill=red, fill opacity=0.5, forget plot]
table[row sep=crcr] {%
x	y\\
23.92	0.937533363422056\\
23.92	1.04510749718168\\
25.92	1.04510749718168\\
25.92	0.937533363422056\\
23.92	0.937533363422056\\
}--cycle;

\addplot[area legend, draw=black, fill=red, fill opacity=0.5, forget plot]
table[row sep=crcr] {%
x	y\\
28.96	1.00413949345292\\
28.96	1.1142055086061\\
30.96	1.1142055086061\\
30.96	1.00413949345292\\
28.96	1.00413949345292\\
}--cycle;
\addplot[only marks, mark=*, mark options={}, mark size=1.2247pt, draw=white!70!black, fill=white!70!black, forget plot] table[row sep=crcr]{%
x	y\\
9.8	0.880713118126259\\
14.84	0.906857064298985\\
19.88	0.938260155981635\\
24.92	0.97123145023237\\
29.96	1.03820019801108\\
};
\end{axis}

\begin{axis}[%
grid = both,
name = Thick200,
width=\figurewidth,
height=0.3\figureheight,
at={(0\figurewidth,0.35\figureheight)},
scale only axis,
xmin=5,
xmax=35,
xtick={9.8,14.84,19.88,24.92,29.96},
xticklabels=\empty,
ymin=0.4,
ymax=1.4,
ylabel style={font=\color{white!15!black}},
ylabel={{$\max_{x}\vert\hat{\epsilon}\vert$}},
axis background/.style={fill=white}
]
\addplot [color=mycolor1, line width=1.5pt, forget plot]
  table[row sep=crcr]{%
3.5	0.766407698841974\\
3.563	0.770645511557348\\
3.626	0.774248053369483\\
3.689	0.777255980247155\\
3.752	0.779709948159123\\
3.815	0.781650613074177\\
3.878	0.783118630961091\\
3.941	0.784154657788626\\
4.004	0.784799349525564\\
4.13	0.785077351602744\\
4.256	0.784277884942824\\
4.382	0.782726197296\\
4.886	0.775406520883273\\
5.327	0.77122317452617\\
5.453	0.769225824624066\\
5.768	0.763413215074877\\
5.894	0.762054681635853\\
5.957	0.761832006620935\\
6.02	0.761996886854838\\
6.083	0.762586293676435\\
6.209	0.764754785265154\\
6.524	0.771747974123535\\
6.587	0.772695891625062\\
6.65	0.773283510963431\\
6.713	0.773450385520256\\
6.839	0.772782944407659\\
7.217	0.769067119976739\\
7.28	0.769116100016696\\
7.343	0.769561631195693\\
7.469	0.771413090146964\\
7.658	0.775421359242934\\
7.784	0.778004328453079\\
7.91	0.779787985006934\\
8.036	0.780297049076133\\
8.162	0.779650931255595\\
8.288	0.778117714746067\\
8.477	0.774741011941551\\
8.918	0.765977118270804\\
9.107	0.76318383147769\\
9.233	0.762002028102266\\
9.359	0.761413633343103\\
9.548	0.761447403144636\\
9.737	0.762298785524493\\
9.989	0.764246762884774\\
10.178	0.766397405623458\\
10.367	0.769380095662015\\
10.556	0.773343343726609\\
10.808	0.778908114831623\\
10.934	0.781013517694021\\
11.06	0.782192228489073\\
11.186	0.782190718068762\\
11.312	0.781167451514825\\
11.438	0.779383892466434\\
11.627	0.775854957177472\\
12.005	0.768465096441581\\
12.194	0.765332254698201\\
12.383	0.762892702303624\\
12.635	0.76058104397994\\
12.887	0.758313417609756\\
13.076	0.755981090113366\\
13.517	0.749886135899779\\
13.643	0.748738250368298\\
13.769	0.74817851664406\\
13.895	0.74840130607253\\
14.021	0.749601509170958\\
14.147	0.751974016456586\\
14.21	0.753660787456248\\
14.336	0.757989708120931\\
14.588	0.767801928489703\\
14.714	0.771665457768876\\
14.777	0.772982252401711\\
14.84	0.77375408616129\\
14.903	0.773919193603625\\
14.966	0.773573690114922\\
15.092	0.771893192719503\\
15.281	0.768932504487907\\
15.344	0.768370632882203\\
15.407	0.768247249501862\\
15.47	0.768697939940665\\
15.533	0.76980513272219\\
15.659	0.773415069214664\\
15.848	0.779568498867057\\
15.911	0.781083154752196\\
15.974	0.782023972258969\\
16.037	0.782207594560049\\
16.1	0.781450664828107\\
16.163	0.779641130728074\\
16.226	0.776952157893881\\
16.289	0.773628216451712\\
16.541	0.758872167127102\\
16.604	0.756040434537894\\
16.667	0.754040554097841\\
16.73	0.753116995933134\\
16.793	0.753438243522218\\
16.856	0.754868833752582\\
16.919	0.75719731686398\\
16.982	0.760212243096156\\
17.108	0.767455625881865\\
17.234	0.774907384027706\\
17.297	0.778182779460053\\
17.36	0.780875919451688\\
17.423	0.782824900730326\\
17.486	0.784066005975518\\
17.549	0.784685064354804\\
17.612	0.7847679050357\\
17.675	0.784400357185731\\
17.801	0.78265741256331\\
18.305	0.773142133342994\\
18.494	0.770751373918792\\
18.683	0.769266457854407\\
18.872	0.768725049074703\\
19.061	0.769035880163628\\
19.25	0.77009659283712\\
19.817	0.774443932998203\\
19.943	0.774539534796958\\
20.447	0.773659326950096\\
20.573	0.774737199627694\\
20.888	0.778587857136095\\
20.951	0.778820652066258\\
21.014	0.778657014875115\\
21.077	0.778003808449455\\
21.14	0.776767895676088\\
21.203	0.7748986259029\\
21.266	0.772515294322147\\
21.392	0.766853572351309\\
21.518	0.761076982990332\\
21.581	0.758550067171896\\
21.644	0.756479779465963\\
21.707	0.755027901525871\\
21.77	0.754356215004961\\
21.833	0.754573560570677\\
21.896	0.75557701494683\\
21.959	0.757210713871338\\
22.085	0.761745388317046\\
22.274	0.769377627545985\\
22.337	0.771519644256713\\
22.4	0.773200855681161\\
22.463	0.774304841415514\\
22.526	0.774872956489013\\
22.589	0.774985999789187\\
22.715	0.774170066619618\\
22.904	0.771553100753252\\
23.03	0.769822399788715\\
23.156	0.768732344573678\\
23.282	0.768306168419379\\
23.408	0.76844247307929\\
23.597	0.769479772931525\\
23.786	0.771233424808536\\
23.975	0.773628043163392\\
24.164	0.776746757420547\\
24.353	0.780665606338438\\
24.605	0.786104915062339\\
24.731	0.787966913315657\\
24.794	0.788490232695956\\
24.857	0.788668140542796\\
24.92	0.788445418079263\\
24.983	0.787790245927042\\
25.109	0.785453064819521\\
25.298	0.780657865370912\\
25.424	0.777595186051705\\
25.55	0.775434794053524\\
25.613	0.774885120293305\\
25.739	0.774815555372761\\
25.865	0.77585570295367\\
25.991	0.777689565877964\\
26.243	0.782478817456784\\
26.495	0.787188269393106\\
26.684	0.79007893689387\\
26.81	0.791501265016052\\
26.999	0.792780364060704\\
27.44	0.794812133482239\\
27.944	0.798718486713248\\
28.07	0.798386885920607\\
28.196	0.79703052623239\\
28.637	0.790609839175218\\
28.763	0.790420689603742\\
28.889	0.79144359051616\\
29.015	0.793295509912348\\
29.393	0.799820944251088\\
29.708	0.804191114128692\\
29.834	0.806925266544994\\
29.897	0.808732101587964\\
29.96	0.810908260861083\\
30.023	0.813490629261615\\
30.149	0.819677642448561\\
30.275	0.826799294866348\\
30.59	0.845318854436712\\
30.716	0.85170224759662\\
30.842	0.856940012845456\\
30.905	0.859041471206204\\
30.968	0.860751146032776\\
31.031	0.86203391180635\\
31.094	0.862854643008127\\
31.157	0.863178214119301\\
31.22	0.862969499621073\\
31.283	0.862220486630036\\
31.409	0.859530428438035\\
31.661	0.852953591437142\\
31.724	0.851804108771503\\
31.787	0.851107617327052\\
31.85	0.850991667397459\\
31.913	0.851540517768889\\
31.976	0.852665261197465\\
32.102	0.856113632220534\\
32.354	0.864104345891555\\
32.417	0.865558775681421\\
32.48	0.866531508517362\\
32.543	0.866938444199668\\
32.669	0.866536198683022\\
32.858	0.865238426584874\\
32.921	0.865246231767308\\
32.984	0.865734889155092\\
33.047	0.866860791306891\\
33.11	0.86878033078137\\
33.173	0.87159358595008\\
33.236	0.875175378436111\\
33.299	0.879344215675424\\
33.425	0.888717054157802\\
33.551	0.898260160884966\\
33.614	0.902641833430273\\
33.677	0.906521595344685\\
33.74	0.909717954064185\\
33.803	0.912095542194493\\
33.866	0.913703493020449\\
33.929	0.914637064996626\\
33.992	0.914991516577615\\
34.055	0.914862106218003\\
34.181	0.913532733495337\\
34.496	0.908409809549333\\
34.559	0.907838808030093\\
34.622	0.907641270661543\\
34.685	0.90791245589827\\
34.748	0.908747622194866\\
34.811	0.910242028005918\\
34.874	0.91249093178601\\
34.937	0.915589591989736\\
35	0.919633267071681\\
};
\addplot [color=red, line width=1.5pt, forget plot]
  table[row sep=crcr]{%
9.8	0.896819067570775\\
9.8	0.97683530616329\\
};
\addplot [color=red, line width=1.5pt, forget plot]
  table[row sep=crcr]{%
14.84	0.871564152756182\\
14.84	0.981403047669374\\
};
\addplot [color=red, line width=1.5pt, forget plot]
  table[row sep=crcr]{%
19.88	0.891634405593788\\
19.88	1.05337372715129\\
};
\addplot [color=red, line width=1.5pt, forget plot]
  table[row sep=crcr]{%
24.92	0.898106187210502\\
24.92	1.11\\
};
\addplot [color=red, line width=1.5pt, forget plot]
  table[row sep=crcr]{%
29.96	0.953181743471131\\
29.96	1.2\\
};
\addplot [color=red, line width=1.5pt, forget plot]
  table[row sep=crcr]{%
9.8	0.548096984768598\\
9.8	0.621407165557315\\
};
\addplot [color=red, line width=1.5pt, forget plot]
  table[row sep=crcr]{%
14.84	0.543148148722398\\
14.84	0.645962009006515\\
};
\addplot [color=red, line width=1.5pt, forget plot]
  table[row sep=crcr]{%
19.88	0.561278283307555\\
19.88	0.648247533614082\\
};
\addplot [color=red, line width=1.5pt, forget plot]
  table[row sep=crcr]{%
24.92	0.601858942737667\\
24.92	0.684573163269768\\
};
\addplot [color=red, line width=1.5pt, forget plot]
  table[row sep=crcr]{%
29.96	0.629390705537148\\
29.96	0.72558697385487\\
};
\addplot [color=red, forget plot]
  table[row sep=crcr]{%
8.8	0.97683530616329\\
10.8	0.97683530616329\\
};
\addplot [color=red, forget plot]
  table[row sep=crcr]{%
13.84	0.981403047669374\\
15.84	0.981403047669374\\
};
\addplot [color=red, forget plot]
  table[row sep=crcr]{%
18.88	1.05337372715129\\
20.88	1.05337372715129\\
};
\addplot [color=red, forget plot]
  table[row sep=crcr]{%
23.92	1.11\\
25.92	1.11\\
};
\addplot [color=red, forget plot]
  table[row sep=crcr]{%
28.96	1.2\\
30.96	1.2\\
};
\addplot [color=red, forget plot]
  table[row sep=crcr]{%
8.8	0.548096984768598\\
10.8	0.548096984768598\\
};
\addplot [color=red, forget plot]
  table[row sep=crcr]{%
13.84	0.543148148722398\\
15.84	0.543148148722398\\
};
\addplot [color=red, forget plot]
  table[row sep=crcr]{%
18.88	0.561278283307555\\
20.88	0.561278283307555\\
};
\addplot [color=red, forget plot]
  table[row sep=crcr]{%
23.92	0.601858942737667\\
25.92	0.601858942737667\\
};
\addplot [color=red, forget plot]
  table[row sep=crcr]{%
28.96	0.629390705537148\\
30.96	0.629390705537148\\
};
\addplot [color=red, forget plot]
  table[row sep=crcr]{%
8.8	0.621407165557315\\
8.8	0.896819067570775\\
10.8	0.896819067570775\\
10.8	0.621407165557315\\
8.8	0.621407165557315\\
};
\addplot [color=red, forget plot]
  table[row sep=crcr]{%
13.84	0.645962009006515\\
13.84	0.871564152756182\\
15.84	0.871564152756182\\
15.84	0.645962009006515\\
13.84	0.645962009006515\\
};
\addplot [color=red, forget plot]
  table[row sep=crcr]{%
18.88	0.648247533614082\\
18.88	0.891634405593788\\
20.88	0.891634405593788\\
20.88	0.648247533614082\\
18.88	0.648247533614082\\
};
\addplot [color=red, forget plot]
  table[row sep=crcr]{%
23.92	0.684573163269768\\
23.92	0.898106187210502\\
25.92	0.898106187210502\\
25.92	0.684573163269768\\
23.92	0.684573163269768\\
};
\addplot [color=red, forget plot]
  table[row sep=crcr]{%
28.96	0.72558697385487\\
28.96	0.953181743471131\\
30.96	0.953181743471131\\
30.96	0.72558697385487\\
28.96	0.72558697385487\\
};
\addplot [color=red, forget plot]
  table[row sep=crcr]{%
8.8	0.762709433772676\\
8.8	0.762709433772676\\
};
\addplot [color=red, forget plot]
  table[row sep=crcr]{%
13.84	0.773754086161292\\
13.84	0.773754086161292\\
};
\addplot [color=red, forget plot]
  table[row sep=crcr]{%
18.88	0.774564467013192\\
18.88	0.774564467013192\\
};
\addplot [color=red, forget plot]
  table[row sep=crcr]{%
23.92	0.788445418079263\\
23.92	0.788445418079263\\
};
\addplot [color=red, forget plot]
  table[row sep=crcr]{%
28.96	0.810908260861087\\
28.96	0.810908260861087\\
};

\addplot[area legend, draw=black, fill=red, fill opacity=0.5, forget plot]
table[row sep=crcr] {%
x	y\\
8.8	0.621407165557315\\
8.8	0.896819067570774\\
10.8	0.896819067570774\\
10.8	0.621407165557315\\
8.8	0.621407165557315\\
}--cycle;

\addplot[area legend, draw=black, fill=red, fill opacity=0.5, forget plot]
table[row sep=crcr] {%
x	y\\
13.84	0.645962009006516\\
13.84	0.871564152756182\\
15.84	0.871564152756182\\
15.84	0.645962009006516\\
13.84	0.645962009006516\\
}--cycle;

\addplot[area legend, draw=black, fill=red, fill opacity=0.5, forget plot]
table[row sep=crcr] {%
x	y\\
18.88	0.648247533614081\\
18.88	0.891634405593789\\
20.88	0.891634405593789\\
20.88	0.648247533614081\\
18.88	0.648247533614081\\
}--cycle;

\addplot[area legend, draw=black, fill=red, fill opacity=0.5, forget plot]
table[row sep=crcr] {%
x	y\\
23.92	0.684573163269769\\
23.92	0.898106187210503\\
25.92	0.898106187210503\\
25.92	0.684573163269769\\
23.92	0.684573163269769\\
}--cycle;

\addplot[area legend, draw=black, fill=red, fill opacity=0.5, forget plot]
table[row sep=crcr] {%
x	y\\
28.96	0.725586973854871\\
28.96	0.953181743471131\\
30.96	0.953181743471131\\
30.96	0.725586973854871\\
28.96	0.725586973854871\\
}--cycle;
\addplot[only marks, mark=*, mark options={}, mark size=1.2247pt, draw=white!70!black, fill=white!70!black, forget plot] table[row sep=crcr]{%
x	y\\
9.8	0.762709433772676\\
14.84	0.773754086161292\\
19.88	0.774564467013192\\
24.92	0.788445418079262\\
29.96	0.810908260861086\\
};
\end{axis}

\begin{axis}[%
grid = both,
name = Thick400,
width=\figurewidth,
height=0.3\figureheight,
at={(0\figurewidth,0\figureheight)},
scale only axis,
xmin=5,
xmax=35,
xtick={5,9.8,14.84,19.88,24.92,29.96,35},
xticklabels={{5},{10},{15},{20},{25},{30},{35}},
xlabel style={font=\color{white!15!black}},
xlabel={$\sigma\,[\times{10^4}\,\text{m}^{-1}]$},
ymin=0.4,
ymax=1.4,
ylabel style={font=\color{white!15!black}},
ylabel={{$\max_{x}\vert\hat{\epsilon}\vert$}},
axis background/.style={fill=white}
]
\addplot [color=mycolor1, line width=1.5pt, forget plot]
  table[row sep=crcr]{%
3.5	0.65849034266121\\
3.752	0.657965390618671\\
4.508	0.655335376220378\\
4.697	0.655654225730039\\
4.886	0.656782351003557\\
5.327	0.660091371692722\\
5.453	0.660315865093018\\
5.642	0.659700244306066\\
5.831	0.65828206547625\\
6.587	0.651516243344489\\
6.776	0.650657468551358\\
7.028	0.650337466283432\\
7.28	0.65083962718866\\
7.532	0.652046863507728\\
7.784	0.653910088150283\\
8.288	0.658303143985734\\
8.414	0.658447058366008\\
8.54	0.657684217567059\\
8.666	0.655890612950813\\
9.044	0.649274666045827\\
9.107	0.648752278581469\\
9.17	0.64858879856375\\
9.233	0.648832056587295\\
9.359	0.650312115273529\\
9.548	0.653889428247361\\
9.674	0.656302851129148\\
9.8	0.658038096812668\\
9.926	0.658632771790089\\
10.052	0.658178389382911\\
10.178	0.656904939619935\\
10.367	0.653962116871398\\
10.682	0.648413960806295\\
10.808	0.646955046093844\\
10.934	0.646492813943269\\
10.997	0.646752722561104\\
11.06	0.647402524298315\\
11.186	0.649867958447032\\
11.312	0.653306983054676\\
11.501	0.658598725847504\\
11.627	0.661087174144235\\
11.69	0.661729479726638\\
11.753	0.661868194933355\\
11.816	0.661539738265716\\
11.942	0.65973347931282\\
12.068	0.656815042878854\\
12.509	0.644992447932147\\
12.635	0.642834653437248\\
12.761	0.64175597964649\\
12.824	0.64170609778423\\
12.887	0.642027737820889\\
12.95	0.642754813664084\\
13.013	0.643899757883332\\
13.139	0.647127798980165\\
13.391	0.65467721641798\\
13.517	0.6574790068223\\
13.58	0.658304263022167\\
13.643	0.658647183093301\\
13.706	0.658553069752209\\
13.832	0.657347175169484\\
13.958	0.655273443946207\\
14.21	0.650869930266957\\
14.336	0.649555076165242\\
14.588	0.648139875014657\\
14.777	0.64670027407665\\
14.903	0.644891287957364\\
15.092	0.641008056909193\\
15.281	0.637038281330383\\
15.407	0.63509242126014\\
15.533	0.634216516890959\\
15.659	0.634507347099145\\
15.785	0.635723913249834\\
15.911	0.637610530729397\\
16.163	0.642373695686203\\
16.415	0.647058129853782\\
16.604	0.649825594544744\\
16.73	0.651062292377787\\
16.856	0.651694859209819\\
16.982	0.651788207511842\\
17.171	0.651146648700085\\
17.423	0.649382793137598\\
17.864	0.646062140396339\\
18.053	0.645386468176675\\
18.494	0.64452567771896\\
18.62	0.64347278750386\\
18.746	0.641748106622082\\
19.124	0.635873448998446\\
19.25	0.634999300727621\\
19.376	0.635168814937444\\
19.754	0.637404568789876\\
19.88	0.636899614022454\\
20.006	0.635056123958023\\
20.195	0.630827677034922\\
20.321	0.627980356553486\\
20.447	0.625868046327163\\
20.51	0.62529047406359\\
20.573	0.625119031170371\\
20.699	0.625874731940847\\
20.825	0.627789705539669\\
20.951	0.63049734086978\\
21.392	0.641163469582438\\
21.518	0.643471606765054\\
21.644	0.645096363328207\\
21.77	0.645851234402514\\
21.896	0.64563851503771\\
22.022	0.644715699959946\\
22.4	0.641149626901601\\
22.526	0.64076799489002\\
22.715	0.641150727269171\\
22.967	0.642698600559982\\
23.282	0.644691593613466\\
23.471	0.645027792192451\\
23.597	0.644579262137995\\
23.723	0.643425515481383\\
23.849	0.641632959678141\\
24.29	0.634354365242913\\
24.416	0.633108053146564\\
24.542	0.632472776249983\\
24.668	0.632430355143626\\
24.794	0.632962610417955\\
24.92	0.634051362663435\\
25.172	0.637315059476165\\
25.298	0.638751596413279\\
25.424	0.639523105538004\\
25.55	0.639228471725673\\
25.676	0.637629095969523\\
25.802	0.635136443734417\\
26.054	0.629767236158365\\
26.18	0.628038643982926\\
26.306	0.6275486482213\\
26.432	0.628050963264897\\
26.621	0.629767401925101\\
26.81	0.631408382758508\\
26.936	0.63186176998731\\
27.062	0.631767089384113\\
27.251	0.630818329685354\\
27.503	0.6285784743962\\
27.818	0.625546914700635\\
28.007	0.624463487452672\\
28.133	0.624343882326869\\
28.322	0.62495684617943\\
28.637	0.626162252047976\\
28.763	0.626030627515938\\
28.952	0.624937038657905\\
29.645	0.619454177559113\\
29.834	0.619097805884522\\
30.023	0.619514014349889\\
30.212	0.620764011291293\\
30.401	0.622825012751548\\
30.59	0.62566700962995\\
30.842	0.630401403252463\\
31.094	0.635092052605572\\
31.22	0.636990088476111\\
31.409	0.638954237684537\\
31.598	0.640114407905813\\
32.102	0.642002773614649\\
32.48	0.643130410647153\\
32.795	0.643352163910215\\
33.11	0.642816939793697\\
33.614	0.641508739737532\\
33.803	0.641876287329538\\
33.992	0.643006709668711\\
34.307	0.645818101167535\\
34.622	0.648547286602557\\
34.811	0.649549467482217\\
35	0.649697001373283\\
};
\addplot [color=red, line width=1.5pt, forget plot]
  table[row sep=crcr]{%
9.8	0.765721431662547\\
9.8	1.00060358947047\\
};
\addplot [color=red, line width=1.5pt, forget plot]
  table[row sep=crcr]{%
14.84	0.757137768954747\\
14.84	1.00379843127069\\
};
\addplot [color=red, line width=1.5pt, forget plot]
  table[row sep=crcr]{%
19.88	0.745605737828615\\
19.88	1.00837045226813\\
};
\addplot [color=red, line width=1.5pt, forget plot]
  table[row sep=crcr]{%
24.92	0.733877801475707\\
24.92	1.01447160147949\\
};
\addplot [color=red, line width=1.5pt, forget plot]
  table[row sep=crcr]{%
29.96	0.726622546421556\\
29.96	1.02268919870025\\
};
\addplot [color=red, line width=1.5pt, forget plot]
  table[row sep=crcr]{%
9.8	0.467061387333089\\
9.8	0.547900770735643\\
};
\addplot [color=red, line width=1.5pt, forget plot]
  table[row sep=crcr]{%
14.84	0.458197622823104\\
14.84	0.548298942687003\\
};
\addplot [color=red, line width=1.5pt, forget plot]
  table[row sep=crcr]{%
19.88	0.468831013395469\\
19.88	0.531515585299406\\
};
\addplot [color=red, line width=1.5pt, forget plot]
  table[row sep=crcr]{%
24.92	0.445392028783107\\
24.92	0.529697133941518\\
};
\addplot [color=red, line width=1.5pt, forget plot]
  table[row sep=crcr]{%
29.96	0.445593092772416\\
29.96	0.529593570049524\\
};
\addplot [color=red, forget plot]
  table[row sep=crcr]{%
8.8	1.00060358947047\\
10.8	1.00060358947047\\
};
\addplot [color=red, forget plot]
  table[row sep=crcr]{%
13.84	1.00379843127069\\
15.84	1.00379843127069\\
};
\addplot [color=red, forget plot]
  table[row sep=crcr]{%
18.88	1.00837045226813\\
20.88	1.00837045226813\\
};
\addplot [color=red, forget plot]
  table[row sep=crcr]{%
23.92	1.01447160147949\\
25.92	1.01447160147949\\
};
\addplot [color=red, forget plot]
  table[row sep=crcr]{%
28.96	1.02268919870025\\
30.96	1.02268919870025\\
};
\addplot [color=red, forget plot]
  table[row sep=crcr]{%
8.8	0.467061387333089\\
10.8	0.467061387333089\\
};
\addplot [color=red, forget plot]
  table[row sep=crcr]{%
13.84	0.458197622823104\\
15.84	0.458197622823104\\
};
\addplot [color=red, forget plot]
  table[row sep=crcr]{%
18.88	0.468831013395469\\
20.88	0.468831013395469\\
};
\addplot [color=red, forget plot]
  table[row sep=crcr]{%
23.92	0.445392028783107\\
25.92	0.445392028783107\\
};
\addplot [color=red, forget plot]
  table[row sep=crcr]{%
28.96	0.445593092772416\\
30.96	0.445593092772416\\
};
\addplot [color=red, forget plot]
  table[row sep=crcr]{%
8.8	0.547900770735643\\
8.8	0.765721431662547\\
10.8	0.765721431662547\\
10.8	0.547900770735643\\
8.8	0.547900770735643\\
};
\addplot [color=red, forget plot]
  table[row sep=crcr]{%
13.84	0.548298942687003\\
13.84	0.757137768954747\\
15.84	0.757137768954747\\
15.84	0.548298942687003\\
13.84	0.548298942687003\\
};
\addplot [color=red, forget plot]
  table[row sep=crcr]{%
18.88	0.531515585299406\\
18.88	0.745605737828615\\
20.88	0.745605737828615\\
20.88	0.531515585299406\\
18.88	0.531515585299406\\
};
\addplot [color=red, forget plot]
  table[row sep=crcr]{%
23.92	0.529697133941518\\
23.92	0.733877801475707\\
25.92	0.733877801475707\\
25.92	0.529697133941518\\
23.92	0.529697133941518\\
};
\addplot [color=red, forget plot]
  table[row sep=crcr]{%
28.96	0.529593570049524\\
28.96	0.726622546421556\\
30.96	0.726622546421556\\
30.96	0.529593570049524\\
28.96	0.529593570049524\\
};
\addplot [color=red, forget plot]
  table[row sep=crcr]{%
8.8	0.658038096812664\\
8.8	0.658038096812664\\
};
\addplot [color=red, forget plot]
  table[row sep=crcr]{%
13.84	0.645904408141853\\
13.84	0.645904408141853\\
};
\addplot [color=red, forget plot]
  table[row sep=crcr]{%
18.88	0.636899614022454\\
18.88	0.636899614022454\\
};
\addplot [color=red, forget plot]
  table[row sep=crcr]{%
23.92	0.634051362663435\\
23.92	0.634051362663435\\
};
\addplot [color=red, forget plot]
  table[row sep=crcr]{%
28.96	0.619282758687309\\
28.96	0.619282758687309\\
};

\addplot[area legend, draw=black, fill=red, fill opacity=0.5, forget plot]
table[row sep=crcr] {%
x	y\\
8.8	0.547900770735643\\
8.8	0.765721431662547\\
10.8	0.765721431662547\\
10.8	0.547900770735643\\
8.8	0.547900770735643\\
}--cycle;

\addplot[area legend, draw=black, fill=red, fill opacity=0.5, forget plot]
table[row sep=crcr] {%
x	y\\
13.84	0.548298942687004\\
13.84	0.757137768954748\\
15.84	0.757137768954748\\
15.84	0.548298942687004\\
13.84	0.548298942687004\\
}--cycle;

\addplot[area legend, draw=black, fill=red, fill opacity=0.5, forget plot]
table[row sep=crcr] {%
x	y\\
18.88	0.531515585299406\\
18.88	0.745605737828615\\
20.88	0.745605737828615\\
20.88	0.531515585299406\\
18.88	0.531515585299406\\
}--cycle;

\addplot[area legend, draw=black, fill=red, fill opacity=0.5, forget plot]
table[row sep=crcr] {%
x	y\\
23.92	0.529697133941519\\
23.92	0.733877801475707\\
25.92	0.733877801475707\\
25.92	0.529697133941519\\
23.92	0.529697133941519\\
}--cycle;

\addplot[area legend, draw=black, fill=red, fill opacity=0.5, forget plot]
table[row sep=crcr] {%
x	y\\
28.96	0.529593570049526\\
28.96	0.726622546421557\\
30.96	0.726622546421557\\
30.96	0.529593570049526\\
28.96	0.529593570049526\\
}--cycle;
\addplot[only marks, mark=*, mark options={}, mark size=1.2247pt, draw=white!70!black, fill=white!70!black, forget plot] table[row sep=crcr]{%
x	y\\
9.8	0.658038096812665\\
14.84	0.645904408141853\\
19.88	0.636899614022453\\
24.92	0.634051362663433\\
29.96	0.61928275868731\\
};
\end{axis}

\node[align=center,anchor=east,text opacity=1,rounded corners,fill=none,opacity=0.6,text opacity=1,xshift=-20pt,yshift=0pt]
at (Thick100.north west) {(a)};
\node[align=center,anchor=east,text opacity=1,rounded corners,fill=none,opacity=0.6,text opacity=1,xshift=65pt,yshift=-10pt]
at (Thick100.north west) {$D = 100 $\,m};
\node[align=center,anchor=east,text opacity=1,rounded corners,fill=none,opacity=0.6,text opacity=1,xshift=-20pt,yshift=0pt]
at (Thick200.north west) {(b)};
\node[align=center,anchor=east,text opacity=1,rounded corners,fill=none,opacity=0.6,text opacity=1,xshift=65pt,yshift=-10pt]
at (Thick200.north west) {$D = 200 $\,m};
\node[align=center,anchor=east,text opacity=1,rounded corners,fill=none,opacity=0.6,text opacity=1,xshift=-20pt,yshift=0pt]
at (Thick400.north west) {(c)};
\node[align=center,anchor=east,text opacity=1,rounded corners,fill=none,opacity=0.6,text opacity=1,xshift=65pt,yshift=-10pt]
at (Thick400.north west) {$D = 400 $\,m};

\end{tikzpicture}%